\documentclass[prd,nopacs,preprintnumbers,twocolumn,amsmath,nofootinbib,amssymb]{revtex4}
\usepackage{rotating}
\usepackage{natbib}
\usepackage{longtable,lscape}
\usepackage{graphicx,color,dcolumn,booktabs,bm}
\usepackage{epsfig,dsfont,amssymb,amsmath,amsfonts,amsbsy,mathrsfs}
\usepackage{feynmf}   
\usepackage{slashed}  
\usepackage{cases}
\usepackage{xcolor}
\usepackage{float}
\usepackage{makecell}
\usepackage{multirow}
\usepackage{array}
\usepackage{gensymb}
\usepackage{txfonts}
\usepackage{overpic}
\usepackage{epstopdf}
\usepackage{indentfirst}
\usepackage{threeparttable}

\usepackage[margin=1in,letterpaper]{geometry} 

\graphicspath{{Figures/}} %

\usepackage{hyperref}
\hypersetup{colorlinks,citecolor=blue,anchorcolor=red,menucolor=red, linkcolor=red,filecolor=red,runcolor=red,urlcolor=blue,frenchlinks=red}

\makeatletter
\@addtoreset{equation}{section}
\makeatother
\renewcommand{\theequation}{\arabic{section}.\arabic{equation}}
\allowdisplaybreaks

\newcolumntype{I}{!{\vrule width 0.9pt}}

\begin{document}

\title{The nonleptonic decays of double-charmed baryon $\Omega_{cc}^{+}$ within the nonrelativistic quark model}
\author{Yu-Shuai Li$^{1}$}\email{liysh@pku.edu.cn}
\affiliation{$^1$ School of Physics and Center of High Energy Physics, Peking University, Beijing 100871, China}

\begin{abstract}
In this work, we investigate the two-body nonleptonic decays of double-charmed baryon $\Omega_{cc}^{+}$ within a nonrelativistic quark model, in which the nonfactorizable amplitudes contributed by $W$-exchange diagrams are evaluated under pole model assumption. To reduce sensitivity of decay amplitudes to arbitrary choice of baryon wave functions, we adopted charmed baryon wave functions obtained by solving the Schr\"{o}dinger equation with a nonrelativistic quark potential model. Our results show that the branching fractions of CF decays $\Omega_{cc}^{+}\to\Omega_{c}^{0}\pi^{+}$ and $\Xi_{c}^{(\prime)+}\pi^{+}$, as well as the SCS decays $\Omega_{cc}^{+}\to\Omega_{c}^{0}K^{+}$, $\Sigma_{cc}^{++}K^{-}$, $\Xi_{c}^{\prime+}\eta$ and $\Lambda_{c}^{+}\bar{K}^{0}$ can reach up to a few percents. Particularly for the $\Omega_{cc}^{+}\to\Omega_{c}^{0}K^{+}$ mode, owing to the large factorizable amplitude and constructive pole contributions, the branching fraction is comparable to CF decays. These decay modes can serve as the discovery channels for the double-charmed baryon $\Omega_{cc}^{+}$ in future experiments like LHCb and Belle II.
\end{abstract}

\maketitle

\section{Introduction}
\label{sec1}

To complete the charmed-baryon family, the establishment of double-charmed baryons is of particular important. As predicted by quark model, the double-charmed baryons $\Xi_{cc}^{++}$ and $\Xi_{cc}^{+}$ with quark contents being $ccu$ and $ccd$ respectively, form an isodoublet ($I=1/2$), while the $\Omega_{cc}^{+}$ forms an isosinglet ($I=0$) with quark contents of $ccs$. In contrast to the well-established state $\Xi_{cc}^{++}$, which has been observed in the decay modes $\Lambda_{c}^{+}K^{-}\pi^{+}\pi^{+}$~\cite{LHCb:2017iph}, $\Xi_{c}^{+}\pi^{(\prime)+}$~\cite{LHCb:2018pcs,LHCb:2022rpd}, and $\Xi_{cc}^{++}\to\Xi_{c}^{0}\pi^{+}\pi^{+}$~\cite{LHCb:2025shu}, the evidences of $\Xi_{cc}^{+}$ was only reported by SELEX in the decay modes $\Lambda_{c}^{+}K^{-}\pi^{+}$~\cite{SELEX:2002wqn} and $pD^{+}K^{-}$~\cite{SELEX:2004lln}. None of them were confirmed by FOCUS~\cite{Ratti:2003ez}, BaBar~\cite{BaBar:2006bab}, Belle~\cite{Belle:2013htj} and LHCb~\cite{LHCb:2013hvt,LHCb:2019gqy} collaborations. Moreover, the mass of $\Xi_{cc}^{+}$ was measured to be about $3519~\text{MeV}$ by SELEX, which differs from $m_{\Xi_{cc}^{++}}=(3621.55\pm0.23\pm0.30)~\text{MeV}$~\cite{LHCb:2019epo} by about $100~\text{MeV}$. More experimental measurements of the $\Xi_{cc}^{+}$ are needed to establish its existence and precisely measure its properties. The $\Omega_{cc}^{+}$ has been searched in the $\Xi_{c}^{+}K^{-}\pi^{+}$ final state at the LHCb experiment~\cite{LHCb:2021rkb}, while no significant signal was observed. The future experimental measurement is of key importance in completing the double-charmed baryon family. The ground double-charmed baryons with $J^{P}=1/2^{+}$ can decay only via the weak interaction; therefore, the study of their weak decays is essential to facilitate experimental discovery~\cite{Yu:2017zst}. In this work, we focus on the two-body nonleptonic decays of double-charmed baryon $\Omega_{cc}^{+}$, aiming to identify the highest-potential processes for probing the $\Omega_{cc}^{+}$ state.

\begin{figure*}[htbp]\centering
  \includegraphics[width=100mm]{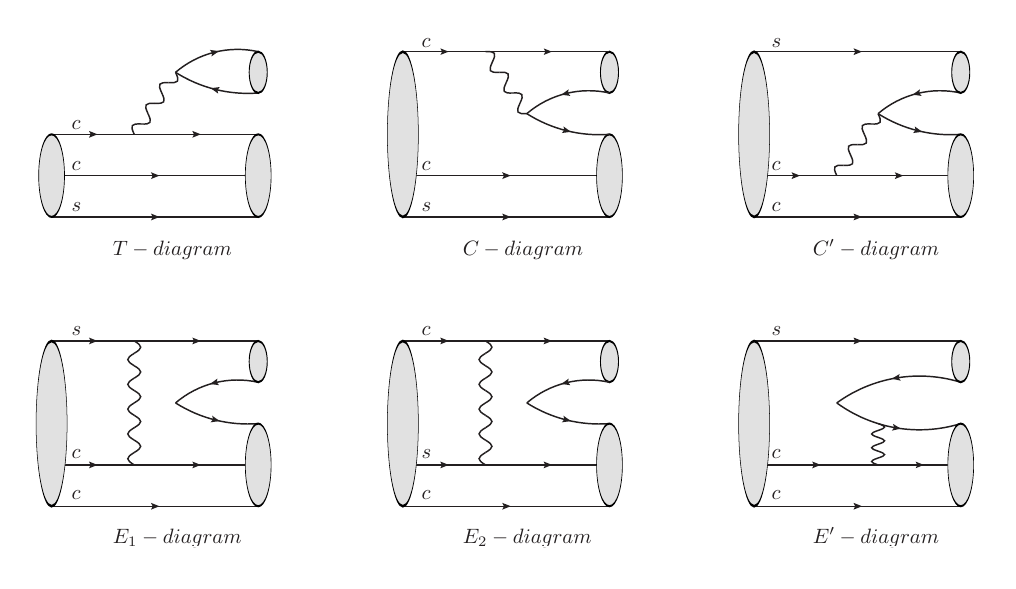}\\
  \caption{The Feynman diagrams for nonleptonic decays of double-charmed baryon $\Omega_{cc}^{+}$.}
  \label{fig:FeynmanDiagrams}
\end{figure*}

The study of double-charmed baryon weak decays also provides valuable insights into weak decay mechanisms. Based on the Cabibbo-Kobayashi-Maskawa (CKM) matrix elements, the concerned nonleptonic decays of double-charmed baryon $\Omega_{cc}^{+}$ can be divided into three groups: (i) the Cabibbo-favored (CF) decays induced by $c\to{su\bar{d}}$ and $cd\to{su}$, (ii) the single Cabibbo-suppressed (SCS) decays induced by $c\to{du\bar{d}}$ and $cd\to{du}$, as well as $c\to{su\bar{s}}$ and $cs\to{su}$, (iii) the doubly Cabibbo-suppressed (DCS) decays induced by $c\to{du\bar{s}}$ and $cs\to{du}$. From another perspective, in theoretical study of charmed-baryon nonleptonic decay, the decay amplitudes are dominated by following Feynman diagrams: the external $W$-emission diagrams ($T$) and internal $W$-emission diagram ($C$), which are factorizable, and the inner $W$-emission diagram ($C^{\prime}$) and $W$-exchange diagram ($E_{1,2},E^{\prime}$), which are nonfactorizable~\cite{Cheng:2021qpd}. The schematic Feynman diagrams are displayed in Fig.~\ref{fig:FeynmanDiagrams}, and the corresponding Feynman diagrams to relevant decay channels are summarized in Table~\ref{tab:FeynmanDiagrams}.

\begin{table*}[htbp]\centering
\caption{The Feynman diagrams of two-body nonleptonic decays of double-charmed baryon $\Omega_{cc}^{+}$.}
\label{tab:FeynmanDiagrams}
\renewcommand\arraystretch{1.05}
\begin{tabular*}{130mm}{c@{\extracolsep{\fill}}ccccc}
\toprule[1pt]
\toprule[0.5pt]
$\text{CF}$   &$\text{Diagrams}$   &$\text{SCS}$   &$\text{Diagrams}$   &$\text{DCS}$   &$\text{Diagrams}$\\
\midrule[0.5pt]
$\Omega_{cc}^{+}\to\Omega_{c}^{0}\pi^{+}$  &$T$  
&$\Omega_{cc}^{+}\to\Xi_{c}^{(\prime)0}\pi^{+}$  &$T,E_{1}$  
&$\Omega_{cc}^{+}\to\Xi_{c}^{(\prime)0}K^{+}$  &$T,E_{1}$\\
$\Omega_{cc}^{+}\to\Xi_{c}^{(\prime)+}\bar{K}^{0}$  &$C,C^{\prime},E^{\prime}$  
&$\Omega_{cc}^{+}\to\Omega_{c}^{0}K^{+}$  &$T,E_{1}$  
&$\Omega_{cc}^{+}\to\Xi_{c}^{(\prime)+}K^{0}$  &$C,E_{2}$\\
&  
&$\Omega_{cc}^{+}\to\Xi_{c}^{(\prime)+}\eta^{(\prime)}$  &$C,C^{\prime},E_{1},E_{2},E^{\prime}$  
&$\Omega_{cc}^{+}\to\Sigma_{c}^{+}\pi^{0}$  &$E_{1},E_{2}$\\
&  
&$\Omega_{cc}^{+}\to\Xi_{c}^{(\prime)+}\pi^{0}$  &$C,E_{1}$  
&$\Omega_{cc}^{+}\to\Lambda_{c}^{+}\pi^{0}$  &$E_{1},E_{2}$\\
&  
&$\Omega_{cc}^{+}\to\Sigma_{c}^{++}K^{-}$  &$E_{2}$  
&$\Omega_{cc}^{+}\to\Sigma_{c}^{0}\pi^{+}$  &$E_{1}$\\
&  
&$\Omega_{cc}^{+}\to\Sigma_{c}^{+}\bar{K}^{0}$  &$C^{\prime},E_{2},E^{\prime}$  
&$\Omega_{cc}^{+}\to\Sigma_{c}^{++}\pi^{-}$  &$E_{2}$\\
&  
&$\Omega_{cc}^{+}\to\Lambda_{c}^{+}\bar{K}^{0}$  &$C^{\prime},E_{2},E^{\prime}$  
&$\Omega_{cc}^{+}\to\Sigma_{c}^{+}\eta^{(\prime)}$  &$C^{\prime},E_{1},E_{2},E^{\prime}$\\
&  
&&
&$\Omega_{cc}^{+}\to\Lambda_{c}^{+}\eta^{(\prime)}$  &$C^{\prime},E_{1},E_{2},E^{\prime}$\\  
\bottomrule[0.5pt]
\bottomrule[1pt]
\end{tabular*}
\end{table*}

In theoretical study of charmed-baryon nonleptonic decay, the factorizable amplitudes $T$ and $C$ are often evaluated as the product of baryon transition matrix element and meson decay constant under the naive factorization assumption, whereas the nonfactorizable amplitudes $C^{\prime}$, $E_{1,2}$, and $E^{\prime}$ are more challenging to quantify. In previous experiments of heavy meson nonleptonic decay, the factorizable assumption works well and the nonfactorizable contributions are usually neglected. However, such experience is no longer suitable in charmed-baryon decay. The experimental measurement of pure $W$-exchange decay $\Lambda_{c}^{+}\to\Xi^{0}K^{+}$~\cite{BESIII:2023wrw} shows that $W$-exchange diagrams are important and cannot be ignored. To access the nonfactorizable contributions from $W$-exchange diagrams, pole model was widely employed in the study of charmed-baryon nonleptonic decays~\cite{Cheng:1991sn,Cheng:1992ff,Cheng:1993gf,Uppal:1994pt,Cheng:2018hwl,Zou:2019kzq,Hu:2020nkg,Meng:2020euv,Cheng:2022kea,Cheng:2022jbr,Ivanov:2023wir,Li:2025alu,Niu:2020gjw,Niu:2021qcc,Niu:2025lgt}. Within the pole model, parity-violating (PV) amplitudes mainly arise from low-lying $J^{P}=1/2^{-}$ poles, while parity-conserving (PC) amplitudes are dominated by $J^{P}=1/2^{+}$ ground-state poles. To extend the framework to double-charmed baryon decay, the theoretical predictions of $\Omega_{cc}^{+}$ nonleptonic decays have been performed in Refs.~\cite{Dhir:2018twm,Cheng:2020wmk,Liu:2023dvg}, in which the nonperturbative parameters are evaluated by bag model~\cite{Cheng:2020wmk} or flavor symmetry~\cite{Dhir:2018twm,Liu:2023dvg}. The covariant confined quark model (CCQM) has also been adopted to estimate the $W$-exchange diagram~\cite{Gutsche:2018msz,Gutsche:2019iac}. Other than these, the irreducible SU(3) approach~\cite{Wang:2017azm,Shi:2017dto} which is based on flavor symmetry while not addressed in dynamics, was also used to analyzed the double-charmed baryon two-body~\cite{Wang:2017azm} and multi-body nonleptonic decays~\cite{Shi:2017dto}.

In this work, we investigate the two-body nonleptonic decays $\Omega_{cc}^{+}\to\mathcal{B}_{c}+P$ within the framework of nonrelativistic quark model (NRQM), where $P$ denotes light-flavor pseudoscalar meson such as $\pi$, $K$ and $\eta^{(\prime)}$. The nonfactorizable contributions from $W$-exchange diagrams are evaluated within the pole model framework. It is well known that the dominate source of uncertainty in theoretical calculations of nonleptonic decays lies in hadron transition matrix element. Addressing this challenge, we propose using exact baryon wave functions obtained by solving the Schr\"{o}dinger equation with a nonrelativistic potential model~\cite{Li:2025alu,Li:2025mxz}, rather than relying on an oversimplified Gaussian-typed one. With the support of charmed baryon spectrum, our study reduces the model dependence arising from baryon wave functions in the decay amplitudes.

This paper is organized as follows: After the Introduction, we derive the weak decay amplitudes within the NRQM in Sec.~\ref{sec2}. To obtain the involved baryon spatial wave functions, a nonrelativistic potential is also introduced. And then in Sec.~\ref{sec3}, we present our numerical results of the decay amplitudes and future investigate the branching fraction and asymmetry parameter. Finally, this paper ends with a short summary in Sec.~\ref{sec4}.

\section{The theoretical framework}
\label{sec2}

\subsection{The decay amplitudes of $W$-emission diagrams}
\label{sec2.1}

\begin{figure}[htbp]\centering
  \includegraphics[width=50mm]{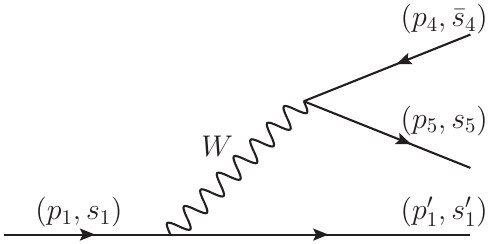}\\
  \caption{The kinematic conventions of $W$-emission process.}
  \label{fig:13}
\end{figure}

According to the conventions in Fig.~\ref{fig:13}, the effective Hamiltonian of $W$-emission process can be written as
\begin{equation}
\begin{split}
\mathcal{H}_{1\to3}=&\frac{G_{F}}{\sqrt{2}}V_{\text{CKM}}V_{\text{CKM}}^{\prime}\frac{\beta}{(2\pi)^{3}}\delta^{3}(\pmb{p}_{1}-\pmb{p}_{1}^{\prime}-\pmb{p}_{4}-\pmb{p}_{5})\\
&\times\hat{\alpha}_{1}^{(-)}\hat{I}_{P}\Big{[}c_{1}(\mu)\hat{O}_{1}+c_{2}(\mu)\hat{O}_{2}\Big{]},
\end{split}
\end{equation}
where $G_{F}=1.16637\times10^{-5}~\text{GeV}^{2}$ is the Fermi coupling constant~\cite{ParticleDataGroup:2024cfk}, and $V_{\text{CKM}}^{(\prime)}$ are Cabibbo-Kobayashi-Maskawa (CKM) matrix elements. $\pmb{p}_{i}$ is three-momentum of $i$th quark. $\hat{\alpha}_{1}^{(-)}c=s~(\text{or}~d)$ is flavor-changing operator which acts on the first quark, and the factor $\beta$ is assigned a value of 2. The $\hat{O}_{1,2}$ are abbreviations of four-fermion interaction operators $\hat{O}_{1}\equiv(\bar{q}_{5}q_{4})(\bar{q}_{1}^{\prime}q_{1})$ and $\hat{O}_{2}\equiv(\bar{q}_{1}^{\prime}q_{4})(\bar{q}_{5}q_{1})$, and $c_{1,2}$ are corresponding Wilson coefficients. We adopt the Wilson coefficients as $c_{1}=1.346$ and $c_{2}=-0.636$ at the energy scale $\mu=1.25~\text{GeV}$~\cite{Cheng:2020wmk}.

In the nonrelativistic limit, the effective Hamiltonian associated with $c_{1}(\mu)$ can be expanded as~\cite{Niu:2020gjw,Niu:2021qcc,Niu:2025lgt}
\begin{widetext}
\begin{equation}
\begin{split}
\mathcal{H}_{1,1\to3}^{\text{PC}}=&\frac{G_{F}}{\sqrt{2}}V_{\text{CKM}}V_{\text{CKM}}^{\prime}\frac{\beta}{(2\pi)^{3}}\delta^{3}(\pmb{p}_{1}-\pmb{p}_{1}^{\prime}-\pmb{p}_{4}-\pmb{p}_{5})
\Bigg{\{}\langle{s_{z,1}^{\prime}}\vert{I}\vert{s_{z,1}}\rangle\langle{s_{z,5}\bar{s}_{z,4}}\vert{\pmb{\sigma}}\vert{0}\rangle\Bigg{(}\frac{\pmb{p}_{5}}{2m_{5}}+\frac{\pmb{p}_{4}}{2m_{4}}\Bigg{)}\\
&-\Bigg{[}\Bigg{(}\frac{\pmb{p}_{1}^{\prime}}{2m_{1}^{\prime}}+\frac{\pmb{p}_{1}}{2m_{1}}\Bigg{)}\langle{s_{z,1}^{\prime}}\vert{I}\vert{s_{z,1}}\rangle-i\langle{s_{z,1}^{\prime}}\vert{\pmb{\sigma}}\vert{s_{z,1}}\rangle\times\Bigg{(}\frac{\pmb{p}_{1}}{2m_{1}}-\frac{\pmb{p}_{1}^{\prime}}{2m_{1}^{\prime}}\Bigg{)}\Bigg{]}\langle{s_{z,5}\bar{s}_{z,4}}\vert{\pmb{\sigma}}\vert{0}\rangle\\
&-\langle{s_{z,1}^{\prime}}\vert{\pmb{\sigma}}\vert{s_{z,1}}\rangle\Bigg{[}\Bigg{(}\frac{\pmb{p}_{5}}{2m_{5}}+\frac{\pmb{p}_{4}}{2m_{4}}\Bigg{)}\langle{s_{z,5}\bar{s}_{z,4}}\vert{I}\vert{0}\rangle-i\langle{s_{z,5}\bar{s}_{z,4}}\vert{\pmb{\sigma}}\vert{0}\rangle\times\Bigg{(}\frac{\pmb{p}_{4}}{2m_{4}}-\frac{\pmb{p}_{5}}{2m_{5}}\Bigg{)}\Bigg{]}\\
&+\langle{s_{z,1}^{\prime}}\vert{\pmb{\sigma}}\vert{s_{z,1}}\rangle\Bigg{(}\frac{\pmb{p}_{1}^{\prime}}{2m_{1}^{\prime}}+\frac{\pmb{p}_{1}}{2m_{1}}\Bigg{)}\langle{s_{z,5}\bar{s}_{z,4}}\vert{I}\vert{0}\rangle\Bigg{\}}\hat{\alpha}_{1}^{(-)}\hat{I}_{P},\\
\mathcal{H}_{1,1\to3}^{\text{PV}}=&\frac{G_{F}}{\sqrt{2}}V_{\text{CKM}}V_{\text{CKM}}^{\prime}\frac{\beta}{(2\pi)^{3}}\delta^{3}(\pmb{p}_{1}-\pmb{p}_{1}^{\prime}-\pmb{p}_{4}-\pmb{p}_{5})\\
&\times\Bigg{(}-\langle{s_{z,1}^{\prime}}\vert{I}\vert{s_{z,1}}\rangle\langle{s_{z,5}\bar{s}_{z,4}}\vert{I}\vert{0}\rangle+\langle{s_{z,1}^{\prime}}\vert{\pmb{\sigma}}\vert{s_{z,1}}\rangle\langle{s_{z,5}\bar{s}_{z,4}}\vert{\pmb{\sigma}}\vert{0}\rangle\Bigg{)}\hat{\alpha}_{1}^{(-)}\hat{I}_{P},
\label{eq:H13}
\end{split}
\end{equation}
in which, the PC Hamiltonian and PV one are distinguished by odd or even number of $\gamma_{5}$, respectively. $I$ is the two-rank identity matrix and $\pmb{\sigma}=(\sigma^{1},\sigma^{2},\sigma^{3})$ are the Pauli matrices. In addition, $\vert{s_{z,i}}\rangle$ is an abbreviation of $\vert{s_{i},s_{z,i}}\rangle$ and represents the spin of $i$th quark. The effective Hamiltonian $\mathcal{H}_{2,1\to3}^{\text{PC},\text{PV}}$ which is associated with $c_{2}(\mu)$, can be obtained by taking the replacements $(p_{1}^{\prime},m_{1}^{\prime})\leftrightarrows(p_{5},m_{5})$ and $s_{z,1}^{\prime}{\leftrightarrows}s_{z,5}$ in Eq~\eqref{eq:H13}.

Generally, the decay amplitudes of $W$-emission diagrams of $\mathcal{B}_{i}(\pmb{P}_{i};J^{i},J_{z}^{i})\to\mathcal{B}_{f}(\pmb{P}_{f};J^{f},J_{z}^{f})+P(\pmb{k})~(P\equiv\text{pseudoscalar~meson})$ process can be evaluated by
\begin{equation}
\begin{split}
\mathcal{M}_{T}^{J_{z}^{f},J_{z}^{i}}=\langle{\mathcal{B}_{f}(\pmb{P}_{f};J^{f},J_{z}^{f})P(\pmb{k})}{\vert}\Big{[}c_{1}(\mu)\mathcal{H}_{1,1\to3}+\frac{c_{2}(\mu)}{N_{c}}\mathcal{H}_{2,1\to3}\Big{]}{\vert}{\mathcal{B}_{i}(\pmb{P}_{i};J^{i},J_{z}^{i})}\rangle,\\
\mathcal{M}_{C}^{J_{z}^{f},J_{z}^{i}}=\langle{\mathcal{B}_{f}(\pmb{P}_{f};J^{f},J_{z}^{f})P(\pmb{k})}{\vert}\Big{[}\frac{c_{1}(\mu)}{N_{c}}\mathcal{H}_{1,1\to3}+c_{2}(\mu)\mathcal{H}_{2,1\to3}\Big{]}{\vert}{\mathcal{B}_{i}(\pmb{P}_{i};J^{i},J_{z}^{i})}\rangle,\\
\mathcal{M}_{C^{\prime}}^{J_{z}^{f},J_{z}^{i}}=\langle{\mathcal{B}_{f}(\pmb{P}_{f};J^{f},J_{z}^{f})P(\pmb{k})}{\vert}\Big{[}-\frac{c_{1}(\mu)}{N_{c}}\mathcal{H}_{1,1\to3}+\frac{c_{2}(\mu)}{N_{c}}\mathcal{H}_{2,1\to3}\Big{]}{\vert}{\mathcal{B}_{i}(\pmb{P}_{i};J^{i},J_{z}^{i})}\rangle,
\end{split}
\end{equation}
where $J^{i}=J^{f}=1/2$ are total spins of initial and final baryons and $N_{c}=3$ is a color suppressed factor. Besides, $\pmb{P}_{i}=\pmb{p}_{1}+\pmb{p}_{2}+\pmb{p}_{3}$ is the three-momentum of initial double-charmed baryon, and $\pmb{P}_{f}=\pmb{p}_{1}^{\prime}+\pmb{p}_{2}+\pmb{p}_{3}~(\pmb{P}_{f}=\pmb{p}_{2}+\pmb{p}_{3}+\pmb{p}_{5},~\pmb{P}_{f}=\pmb{p}_{1}^{\prime}+\pmb{p}_{2}+\pmb{p}_{5})$ and $\pmb{k}=\pmb{p}_{4}+\pmb{p}_{5}~(\pmb{k}=\pmb{p}_{1}^{\prime}+\pmb{p}_{4},~\pmb{k}=\pmb{p}_{3}+\pmb{p}_{4})$ are the three-momenta of final charmed baryon and pseudoscalar meson, respectively, for $T~(C,~C^{\prime})$ diagrams. In the rest frame of initial state, we have $\pmb{P}_{i}=0$ and $\pmb{P}_{f}=-\pmb{k}$. 

Taking the $T$ diagram as an example, we can separate the decay amplitude as
\begin{equation}
\begin{split}
\mathcal{M}_{T}^{J_{z}^{f},J_{z}^{i}}=&\langle{\mathcal{B}_{f}(\pmb{P}_{f};J^{f},J_{z}^{f})P(\pmb{k})}{\vert}\Big{[}c_{1}(\mu)\mathcal{H}_{1,1\to3}+\frac{c_{2}(\mu)}{N_{c}}\mathcal{H}_{2,1\to3}\Big{]}{\vert}{\mathcal{B}_{i}(\pmb{P}_{i};J^{i},J_{z}^{i})}\rangle,\\
=&\frac{G_{F}}{\sqrt{2}}V_{\text{CKM}}V_{\text{CKM}}^{\prime}\frac{\beta}{(2\pi)^{3}}\delta^{3}(\pmb{p}_{1}-\pmb{p}_{1}^{\prime}-\pmb{p}_{4}-\pmb{p}_{5})\\
&\times\sum\langle{S^{f},S_{z}^{f};L^{f},L_{z}^{f}}\vert{J^{f},J_{z}^{f}}\rangle\langle{S^{i},S_{z}^{i};L^{i},L_{z}^{i}}\vert{J^{i},J_{z}^{i}}\rangle\\
&\times\Big{[}c_{1}(\mu)\big{\langle}\phi_{f}\chi_{f}^{S^{f},S_{z}^{f}};\phi_{P}\chi_{P}\vert\hat{\mathcal{O}}_{1,1\to3}^{\text{spin}}\hat{\mathcal{O}}_{1,1\to3}^{\text{flavor}}\vert\phi_{i}\chi_{i}^{S^{i},S_{z}^{i}}\big{\rangle}
\big{\langle}\psi_{P}(\pmb{k})\psi_{L^{f},L_{z}^{f}}(\pmb{P}_{f})\vert{\hat{\mathcal{O}}_{1,1\to3}^{\text{spatial}}}\vert\psi_{L^{i},L_{z}^{i}}(\pmb{P}_{i})\big{\rangle}\\
&+\frac{c_{2}(\mu)}{N_{c}}\big{\langle}\phi_{f}\chi_{f}^{S^{f},S_{z}^{f}};\phi_{P}\chi_{P}\vert\hat{\mathcal{O}}_{2,1\to3}^{\text{spin}}\hat{\mathcal{O}}_{2,1\to3}^{\text{flavor}}\vert\phi_{i}\chi_{i}^{S^{i},S_{z}^{i}}\big{\rangle}
\big{\langle}\psi_{P}(\pmb{k})\psi_{L^{f},L_{z}^{f}}(\pmb{P}_{f})\vert{\hat{\mathcal{O}}_{2,1\to3}^{\text{spatial}}}\vert\psi_{L^{i},L_{z}^{i}}(\pmb{P}_{i})\big{\rangle}\Big{]},
\label{eq:amplitudes1}
\end{split}
\end{equation}
where $\langle{S,S_{z};L,L_{z}}\vert{J,J_{z}}\rangle$ is the Clebsch-Gordan coefficient, while $\phi$, $\chi$ and $\psi$ are the flavor, spin and spatial wave functions of corresponding hadrons, respectively. The overlap of color wave functions have been reflected in factor $1/N_{c}$. Similarly, the concrete expressions for decay amplitudes of $C$ and $C^{\prime}$ diagrams can also be written down.

In Appendix~\ref{app01}, we present the spin-flavor matrix elements $\langle{\mathcal{B}_{f}}{\downarrow}\vert\hat{\mathcal{O}}\vert{\mathcal{B}_{i}}{\downarrow}\rangle$ in Tables~\ref{tab:Tfactor}-\ref{tab:Cprimfactor}, which contribute to the PC and PV amplitudes in the $T$, $C$ and $C^{\prime}$ diagrams, respectively. We emphasize that, particularly for the $T$ diagram, the spin matrix elements associated with $\langle{s_{z,5}\bar{s}_{z,4}}\vert{\sigma_{x,y,z}}\vert{0}\rangle$ vanish~\cite{Li:2025mxz}. Likewise, those associated with $\langle{s_{z,1}^{\prime}\bar{s}_{z,4}}\vert{\sigma_{x,y,z}}\vert{0}\rangle$ vanish for the $C$ diagram. This is a consequence of the pseudoscalar meson spin wave function. Similarly, the spin-flavor matrix elements $\langle{\mathcal{B}_{f}}{\uparrow}\vert\hat{\mathcal{O}}\vert{\mathcal{B}_{i}}{\uparrow}\rangle$ can also be obtained, and satisfy
\begin{equation}
\begin{split}
\langle{\mathcal{B}_{f}}{\uparrow}\vert\hat{\mathcal{O}}\vert{\mathcal{B}_{i}}{\uparrow}\rangle
=\left\{
\begin{array}{ll}
-\langle{\mathcal{B}_{f}}{\downarrow}\vert\hat{\mathcal{O}}\vert{\mathcal{B}_{i}}{\downarrow}\rangle, &\text{for PC amplitude},\\
\langle{\mathcal{B}_{f}}{\downarrow}\vert\hat{\mathcal{O}}\vert{\mathcal{B}_{i}}{\downarrow}\rangle, &\text{for PV amplitude}.\\
\end{array}
\right.
\label{eq:relation11}
\end{split}
\end{equation}
In addition, we also have $\langle{\mathcal{B}_{f}}{\uparrow}\vert\hat{\mathcal{O}}\vert{\mathcal{B}_{i}}{\downarrow}\rangle=\langle{\mathcal{B}_{f}}{\downarrow}\vert\hat{\mathcal{O}}\vert{\mathcal{B}_{i}}{\uparrow}\rangle=0$.

After separating out the spin-flavor parts, the remaining spatial wave function overlaps can be expressed as
\begin{equation}
\begin{split}
\mathcal{I}_{i,T}^{L^{i},L_{z}^{i};L^{f},L_{z}^{f}}=&\langle{\psi_{P}(\pmb{k})\psi_{L^{f},L_{z}^{f}}(\pmb{P}_{f})}\vert\hat{\mathcal{O}}_{i,1\to3}^{\text{spatial}}(\pmb{p}_{i})\vert{\psi_{L^{i},L_{z}^{i}}(\pmb{P}_{i})}\rangle\\
=&\int{d\pmb{p}_{1}d\pmb{p}_{2}d\pmb{p}_{3}d\pmb{p}_{1}^{\prime}d\pmb{p}_{4}d\pmb{p}_{5}}\delta^{3}(\pmb{P}_{f}-\pmb{p}_{1}^{\prime}-\pmb{p}_{2}-\pmb{p}_{3})\delta^{3}(\pmb{P}_{i}-\pmb{p}_{1}-\pmb{p}_{2}-\pmb{p}_{3})\delta^{3}(\pmb{k}-\pmb{p}_{5}-\pmb{p}_{4})\\
&\times\delta^{3}(\pmb{p}_{1}-\pmb{p}_{4}-\pmb{p}_{5}-\pmb{p}_{1}^{\prime})\mathcal{O}_{i,1\to3}^{\text{spatial}}(\pmb{p}_{i})\psi^{*}_{P}(\pmb{p}_{5},\pmb{p}_{4})\psi_{L^{f},L_{z}^{f}}^{*}(\pmb{p}_{1}^{\prime},\pmb{p}_{3},\pmb{p}_{2})\psi_{L^{i},L_{z}^{i}}(\pmb{p}_{1},\pmb{p}_{2},\pmb{p}_{3}),
\end{split}
\end{equation}

\begin{equation}
\begin{split}
\mathcal{I}_{i,C}^{L^{i},L_{z}^{i};L^{f},L_{z}^{f}}=&\langle{\psi_{P}(\pmb{k})\psi_{L^{f},L_{z}^{f}}(\pmb{P}_{f})}\vert\hat{\mathcal{O}}_{i,1\to3}^{\text{spatial}}(\pmb{p}_{i})\vert{\psi_{L^{i},L_{z}^{i}}(\pmb{P}_{i})}\rangle\\
=&\int{d\pmb{p}_{1}d\pmb{p}_{2}d\pmb{p}_{3}d\pmb{p}_{1}^{\prime}d\pmb{p}_{4}d\pmb{p}_{5}}\delta^{3}(\pmb{P}_{f}-\pmb{p}_{5}-\pmb{p}_{2}-\pmb{p}_{3})\delta^{3}(\pmb{P}_{i}-\pmb{p}_{1}-\pmb{p}_{2}-\pmb{p}_{3})\delta^{3}(\pmb{k}-\pmb{p}_{1}^{\prime}-\pmb{p}_{4})\\
&\times\delta^{3}(\pmb{p}_{1}-\pmb{p}_{4}-\pmb{p}_{5}-\pmb{p}_{1}^{\prime})\mathcal{O}_{i,1\to3}^{\text{spatial}}(\pmb{p}_{i})\psi^{*}_{P}(\pmb{p}_{1}^{\prime},\pmb{p}_{4})\psi_{L^{f},L_{z}^{f}}^{*}(\pmb{p}_{3},\pmb{p}_{5},\pmb{p}_{2})\psi_{L^{i},L_{z}^{i}}(\pmb{p}_{1},\pmb{p}_{2},\pmb{p}_{3}),
\end{split}
\end{equation}

\begin{equation}
\begin{split}
\mathcal{I}_{i,C^{\prime}}^{L^{i},L_{z}^{i};L^{f},L_{z}^{f}}=&\langle{\psi_{P}(\pmb{k})\psi_{L^{f},L_{z}^{f}}(\pmb{P}_{f})}\vert\hat{\mathcal{O}}_{i,1\to3}^{\text{spatial}}(\pmb{p}_{i})\vert{\psi_{L^{i},L_{z}^{i}}(\pmb{P}_{i})}\rangle\\
=&\int{d\pmb{p}_{1}d\pmb{p}_{2}d\pmb{p}_{3}d\pmb{p}_{1}^{\prime}d\pmb{p}_{4}d\pmb{p}_{5}}\delta^{3}(\pmb{P}_{f}-\pmb{p}_{1}^{\prime}-\pmb{p}_{5}-\pmb{p}_{2})\delta^{3}(\pmb{P}_{i}-\pmb{p}_{1}-\pmb{p}_{2}-\pmb{p}_{3})\delta^{3}(\pmb{k}-\pmb{p}_{3}-\pmb{p}_{4})\\
&\times\delta^{3}(\pmb{p}_{1}-\pmb{p}_{4}-\pmb{p}_{5}-\pmb{p}_{1}^{\prime})\mathcal{O}_{i,1\to3}^{\text{spatial}}(\pmb{p}_{i})\psi^{*}_{P}(\pmb{p}_{3},\pmb{p}_{4})\psi_{L^{f},L_{z}^{f}}^{*}(\pmb{p}_{1}^{\prime},\pmb{p}_{5},\pmb{p}_{2})\psi_{L^{i},L_{z}^{i}}(\pmb{p}_{1},\pmb{p}_{2},\pmb{p}_{3}),
\end{split}
\end{equation}
\end{widetext}
where $\mathcal{O}_{i,1\to3}^{\text{spatial}}(\pmb{p}_{i})$ is the function of $\pmb{p}_{i}/(2m_{i})$ for $\text{PC}$ amplitude and $\mathcal{O}_{i,1\to3}^{\text{spatial}}(\pmb{p}_{i})\equiv1$ for $\text{PV}$ amplitude.

After synthesizing the above discussion, especially considering the Eq.~\eqref{eq:relation11}, we have the following relations:
\begin{equation}
\begin{split}
\mathcal{M}_{\text{PC}}^{-1/2,-1/2}=&-\mathcal{M}_{\text{PC}}^{1/2,1/2},~\mathcal{M}_{\text{PV}}^{-1/2,-1/2}=\mathcal{M}_{\text{PV}}^{1/2,1/2},\\
\mathcal{M}_{\text{PV},\text{PC}}^{-1/2,1/2}=&\mathcal{M}_{\text{PV},\text{PC}}^{1/2,-1/2}=0,
\label{eq:relation1}
\end{split}
\end{equation}
for $T$, $C$ and $C^{\prime}$ diagrams.

To describe the spatial wave function of pseudoscalar meson, we employ the simple harmonic oscillator wave function~\cite{Niu:2020gjw,Niu:2021qcc,Niu:2025lgt}
\begin{equation}
\psi_{P}(\pmb{k}_{1},\pmb{k}_{2})=\frac{1}{\pi^{3/4}R_{P}^{3/2}}\exp\Big{[}-\frac{(\pmb{k}_{1}-\pmb{k}_{2})^{2}}{8R_{P}^{2}}\Big{]},
\end{equation}
where $R_{P}$ is a phenomenal parameter. In this work, we adopt $R_{\pi}=0.28~\text{GeV}$, $R_{K}=0.50~\text{GeV}$, $R_{\eta}=0.40~\text{GeV}$ and $R_{\eta^{\prime}}=0.90~\text{GeV}$~\cite{Niu:2025lgt}. Besides, in order to describe the spatial wave function of charmed baryon, we adopt the expression as
\begin{equation}
\begin{split}
\psi_{L,L_{z}}(\pmb{k}_{1},\pmb{k}_{2},\pmb{k}_{3})=&\sum_{n_{\rho},n_{\lambda}}\sum_{m_{\rho},m_{\lambda}}\langle{l_{\rho},m_{\rho};l_{\lambda},m_{\lambda}}\vert{L,L_{z}}\rangle\\
&\times{C^{(n_{\rho},n_{\lambda})}}\psi_{n_{\rho},l_{\rho},m_{\rho}}(\pmb{\rho})\psi_{n_{\lambda},l_{\lambda},m_{\lambda}}(\pmb{\lambda}),
\end{split}
\end{equation}
where the three-momentums of $\rho$-mode and $\lambda$-mode are defined as
\begin{equation}
\begin{split}
\pmb{\rho}=&\frac{m_{1}\pmb{k}_{2}-m_{2}\pmb{k}_{1}}{m_{1}+m_{2}},\\
\pmb{\lambda}=&\frac{(m_{1}+m_{2})\pmb{k}_{3}-m_{3}(\pmb{k}_{1}+\pmb{k}_{2})}{m_{1}+m_{2}+m_{3}},
\end{split}
\end{equation}
respectively. $C^{(n_{\rho},n_{\lambda})}$ is a set of coefficients and $\psi_{n,l,m}(\pmb{p})$ is the Gaussian basis function~\cite{Hiyama:2003cu}
\begin{equation}
\begin{split}
\psi_{n,l,m}(\pmb{p})=&N_{nl}~k^{l}~e^{-\nu_{n}p^{2}}Y_{lm}(\hat{\pmb{p}}),\\
N_{nl}=&\Bigg{(}\frac{2^{l+2}(2\nu_{n})^{l+3/2}}{\sqrt{\pi}(2l+1)!!}\Bigg{)}^{1/2},
\end{split}
\end{equation}
where $\hat{\pmb{p}}=\pmb{p}/p$ is a unit vector and $\nu_{n}$ is the Gaussian size parameter. The details of wave functions of charmed baryons will be presented in Sec.~\ref{sec2.3}.

\begin{figure}[htbp]\centering
  \includegraphics[width=50mm]{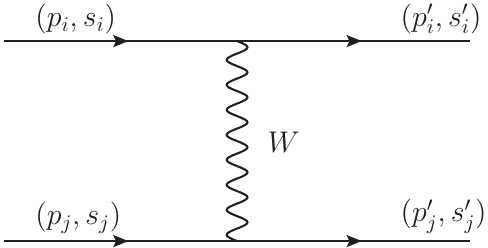}\\
  \caption{The kinematic conventions of $W$-exchange process.}
  \label{fig:22}
\end{figure}

\begin{widetext}
\subsection{The decay amplitudes of $W$-exchange diagrams}
\label{sec2.2}

Next, we compute the decay amplitudes for the $W$-exchange diagrams. Adopting the conventions of Fig.~\ref{fig:22}, the effective Hamiltonian associated with $c_{1}(\mu)$ can be written as~\cite{Niu:2020gjw,Niu:2021qcc,Niu:2025lgt}
\begin{equation}
\begin{split}
\mathcal{H}_{1,2\to2}^{\text{PC}}=&\frac{G_{F}}{\sqrt{2}}V_{\text{CKM}}V_{\text{CKM}}^{\prime}\frac{1}{(2\pi)^{3}}\sum_{i{\neq}j}\hat{\alpha}_{i}^{(-)}\hat{\beta}_{j}^{(+)}\delta^{3}(\pmb{p}_{i}^{\prime}+\pmb{p}_{j}^{\prime}-\pmb{p}_{i}-\pmb{p}_{j})(1-\langle{s_{z,i}^{\prime}}\vert{\pmb{\sigma}_{i}}\vert{s_{z,i}}\rangle\langle{s_{z,j}^{\prime}}\vert{\pmb{\sigma}_{j}}\vert{s_{z,j}}\rangle),\\
\mathcal{H}_{1,2\to2}^{\text{PV}}=&\frac{G_{F}}{\sqrt{2}}V_{\text{CKM}}V_{\text{CKM}}^{\prime}\frac{1}{(2\pi)^{3}}\sum_{i{\neq}j}\hat{\alpha}_{i}^{(-)}\hat{\beta}_{j}^{(+)}\delta^{3}(\pmb{p}_{i}^{\prime}+\pmb{p}_{j}^{\prime}-\pmb{p}_{i}-\pmb{p}_{j})\\
&\times\Bigg{\{}-\big{(}\langle{s_{z,i}^{\prime}}\vert{\pmb{\sigma}_{i}}\vert{s_{z,i}}\rangle-\langle{s_{z,j}^{\prime}}\vert{\pmb{\sigma}_{j}}\vert{s_{z,j}}\rangle\big{)}\Bigg{[}\Bigg{(}\frac{\pmb{p}_{i}}{2m_{i}}-\frac{\pmb{p}_{j}}{2m_{j}}\Bigg{)}+\Bigg{(}\frac{\pmb{p}_{i}^{\prime}}{2m_{i}^{\prime}}-\frac{\pmb{p}_{j}^{\prime}}{2m_{j}^{\prime}}\Bigg{)}\Bigg{]}\\
&+i\big{(}\langle{s_{z,i}^{\prime}}\vert{\pmb{\sigma}_{i}}\vert{s_{z,i}}\rangle\times\langle{s_{z,j}^{\prime}}\vert{\pmb{\sigma}_{j}}\vert{s_{z,j}}\rangle\big{)}\Bigg{[}\Bigg{(}\frac{\pmb{p}_{i}}{2m_{i}}-\frac{\pmb{p}_{j}}{2m_{j}}\Bigg{)}-\Bigg{(}\frac{\pmb{p}_{i}^{\prime}}{2m_{i}^{\prime}}-\frac{\pmb{p}_{j}^{\prime}}{2m_{j}^{\prime}}\Bigg{)}\Bigg{]}\Bigg{\}},
\label{eq:Hamilton22}
\end{split}
\end{equation}
\end{widetext}
where, $\hat{\alpha}_{i}^{(-)}$ and $\hat{\beta}_{j}^{(+)}$ are flavor-changing operators acting on the $i$th and $j$th quarks, respectively. In particular, for the indicated transitions we take $\hat{\alpha}_{i}^{(-)}c=s$ and $\hat{\beta}_{j}^{(+)}d=u$ for $cd\to{su}$ transition, $\hat{\alpha}_{i}^{(-)}c=d$ and $\hat{\beta}_{j}^{(+)}d=u$ for $cd\to{du}$ transition, $\hat{\alpha}_{i}^{(-)}c=s$ and $\hat{\beta}_{j}^{(+)}s=u$ for $cs\to{su}$ transition, and $\hat{\alpha}_{i}^{(-)}c=d$ and $\hat{\beta}_{j}^{(+)}s=u$ for $cs\to{du}$ transition. Furthermore, the effective Hamiltonian $\mathcal{H}_{2,2\to2}^{\text{PC},\text{PV}}$ associated with $c_{2}(\mu)$ can be obtained from Eq.~\eqref{eq:Hamilton22} by the replacements $(p_{i}^{\prime},m_{i}^{\prime})\leftrightarrows(p_{j}^{\prime},m_{j}^{\prime})$ and $s_{z,i}^{\prime}{\leftrightarrows}s_{z,j}^{\prime}$.

\begin{figure}[htbp]\centering
  \includegraphics[width=60mm]{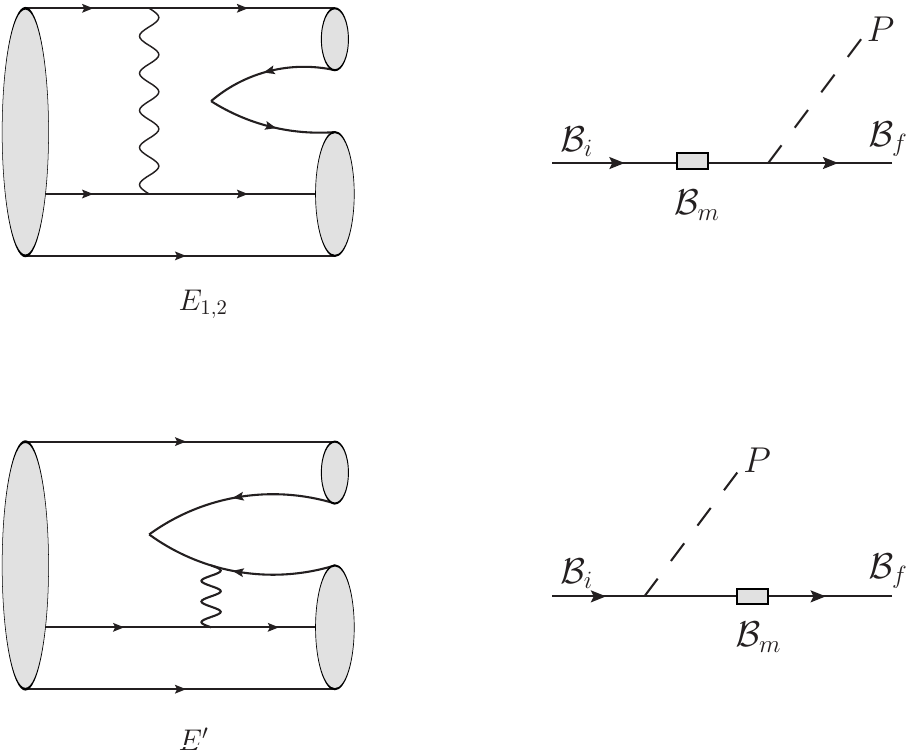}\\
  \caption{The corresponding pole diagrams of $W$-exchange processes.}
  \label{fig:W}
\end{figure}

In this work, we adopt the pole model to estimate the decay amplitudes for $W$-exchange diagrams. The pole model has been widely and successful applied to evaluate the nonfactorizable amplitudes of charmed baryons nonleptonic decays~\cite{Cheng:1991sn,Cheng:1992ff,Cheng:1993gf,Uppal:1994pt,Cheng:2018hwl,Zou:2019kzq,Hu:2020nkg,Meng:2020euv,Niu:2021qcc,Cheng:2022kea,Cheng:2022jbr,Ivanov:2023wir,Li:2025alu,Cheng:2020wmk,Liu:2023dvg}. In the pole model, the general expressions for the decay amplitudes can be written as~\cite{Niu:2020gjw,Niu:2021qcc,Li:2025mxz,Niu:2025lgt}
\begin{widetext}
\begin{equation}
\begin{split}
i\mathcal{M}_{\mathcal{B}_{m}-pole}^{J_{z}^{f},J_{z}^{i}}=\langle{\mathcal{B}_{f}(\pmb{P}_{f};J^{f},J_{z}^{f})}\vert{\mathcal{H}_{P}}\vert{\mathcal{B}_{m}(\pmb{P}_{m};J^{m},J_{z}^{m})}\rangle
\frac{i}{\slashed{p}_{m}-M_{m}+i\Gamma_{m}/2}
\langle{\mathcal{B}_{m}(\pmb{P}_{m};J^{m},J_{z}^{m})}\vert{\mathcal{H}_{2\to2}}\vert{\mathcal{B}_{i}(\pmb{P}_{i};J^{i},J_{z}^{i})}\rangle,
\label{eq:amplitudes21}
\end{split}
\end{equation}
\begin{equation}
\begin{split}
i\mathcal{M}_{\mathcal{B}_{m}-pole}^{J_{z}^{f},J_{z}^{i}}=\langle{\mathcal{B}_{f}(\pmb{P}_{f};J^{f},J_{z}^{f})}\vert{\mathcal{H}_{2\to2}}\vert{\mathcal{B}_{m}(\pmb{P}_{m};J^{m},J_{z}^{m})}\rangle
\frac{i}{\slashed{p}_{m}-M_{m}+i\Gamma_{m}/2}
\langle{\mathcal{B}_{m}(\pmb{P}_{m};J^{m},J_{z}^{m})}\vert{\mathcal{H}_{P}}\vert{\mathcal{B}_{i}(\pmb{P}_{i};J^{i},J_{z}^{i})}\rangle,
\label{eq:amplitudes22}
\end{split}
\end{equation}
\end{widetext}
for $E_{1,2}$ and $E^{\prime}$ diagrams, respectively, according to Fig.~\ref{fig:W}. The state $\mathcal{B}_{m}$ denotes intermediate baryon with $J^{P}=1/2^{+}$ or $1/2^{-}$. Particularly in this work, we take $\mathcal{B}_{m}=(\Xi_{c}^{(\prime)+},\Lambda_{c}^{+},\Sigma_{c}^{+})$ for the $E_{1,2}$ diagrams, and $\mathcal{B}_{m}=(\Xi_{cc}^{+},\Omega_{cc}^{+})$ for the $E^{\prime}$ diagrams. The $p_{m}$ is four-momentum, and $M_{m}$ and $\Gamma_{m}$ are mass and total width, respectively, of the intermediate state $\mathcal{B}_{m}$. Besides, it is worthy to mention that we have $\pmb{P}_{m}=\pmb{P}_{i}$ and $(J^{m},J_{z}^{m})=(J^{i},J_{z}^{i})$ in Eq.~\eqref{eq:amplitudes21}, while $\pmb{P}_{m}=\pmb{P}_{f}$ and $(J^{m},J_{z}^{m})=(J^{f},J_{z}^{f})$ in Eq.~\eqref{eq:amplitudes22}. In the numerical evaluation, the approximation of propagator
\begin{equation}
\frac{1}{\slashed{p}-m+i\Gamma/2}\approx\frac{2m}{p^{2}-m^{2}+i\Gamma m},
\end{equation}
is employed. Moreover, considering the narrow widths of involved charmed baryons and large masses differences between double-charmed baryons and single-charmed baryons, we further neglect the $i\Gamma{m}$ term in the numerical calculation.

The Hamiltonian $\mathcal{H}_{P}$ is adopted as quark-pseudoscalar meson coupling as
\begin{equation}
\mathcal{H}_{P}=\sum_{j}{\int}d{\pmb{x}}\frac{1}{f_{P}}\bar{q}_{j}(\pmb{x})\gamma_{\mu}\gamma_{5}q_{j}(\pmb{x})\partial^{\mu}\phi_{P}(\pmb{x}),
\end{equation}
where $f_{P}$ is the pseudoscalar meson decay constant, and $q(\pmb{x})$ and $\phi_{P}$ denote the quark and pseudoscalar meson fields, respectively. In our calculation, we adopted the experimental values $f_{\pi}=130~\text{MeV}$ and $f_{K}=156~\text{MeV}$~\cite{ParticleDataGroup:2024cfk}, and the LQCD's results $f_{q}=139~\text{MeV}$ and $f_{s}=171~\text{MeV}$ for the $\eta^{(\prime)}$ mesons~\cite{Ottnad:2025zxq}. In the nonrelativistic limit, the effective Hamiltonian can be expanded as~\cite{Niu:2020gjw,Niu:2021qcc,Niu:2025lgt}
\begin{equation}
\begin{split}
\mathcal{H}_{P}=&\frac{1}{\sqrt{(2\pi)^{3}2\omega_{P}}}\sum_{j}\frac{1}{f_{P}}\Bigg{[}\omega_{P}\bigg{(}\frac{\pmb{\sigma}\cdot\pmb{p}_{f}^{j}}{2m_{f}}+\frac{\pmb{\sigma}\cdot\pmb{p}_{i}^{j}}{2m_{i}}\bigg{)}-\pmb{\sigma}\cdot\pmb{k}\Bigg{]}\\
&\times\hat{I}_{P}^{j}\delta^{3}(\pmb{p}_{f}^{j}+\pmb{k}-\pmb{p}_{i}^{j}),
\end{split}
\end{equation}
where $\omega_{P}$ and $\pmb{k}$ are the energy and three-momentum of the pseudoscalar meson in the rest frame of the initial state, and $\pmb{p}_{i}^{j}$ and $\pmb{p}_{f}^{j}$ are the initial and final three-momenta of the $j$th quark, respectively. Besides $\hat{I}_{P}^{j}$ represents the isospin operator acting on the $j$th quark, with the explicit form being~\cite{Niu:2025lgt}
\begin{equation}
\begin{split}
\hat{I}_{P}
=\left\{
\begin{array}{ll}
b_{d}^{\dagger}b_{u}, &\text{for}~\pi^{+},\\
\frac{1}{\sqrt{2}}[b_{u}^{\dagger}b_{u}-b_{d}^{\dagger}b_{d}], &\text{for}~\pi^{0},\\
b_{s}^{\dagger}b_{u}, &\text{for}~K^{+},\\
b_{u(d)}^{\dagger}b_{s}, &\text{for}~K^{-}(\bar{K}^{0}),\\
\frac{\cos\zeta}{\sqrt{2}}[b_{u}^{\dagger}b_{u}+b_{d}^{\dagger}b_{d}]-\sin\zeta[b_{s}^{\dagger}b_{s}], &\text{for}~\eta,\\
\frac{\sin\zeta}{\sqrt{2}}[b_{u}^{\dagger}b_{u}+b_{d}^{\dagger}b_{d}]+\cos\zeta[b_{s}^{\dagger}b_{s}], &\text{for}~\eta^{\prime}.\\
\end{array}
\right.
\end{split}
\end{equation}
Here $\zeta=39.3^{\circ}$~\cite{Feldmann:1998vh} is the mixing angle of $\eta$ and $\eta^{\prime}$, which is introduced by
\begin{equation}
\left(\begin{array}{c}
\eta\\
\eta^{\prime}
\end{array}\right)
=\left(\begin{array}{cc}
\cos{\zeta} & -\sin{\zeta}\\
\sin{\zeta} & \cos{\zeta}
\end{array}\right)
\left(\begin{array}{c}
\eta_{q}\\
\eta_{s}
\end{array}\right).
\end{equation}

Taking the decay amplitude in Eq.~\eqref{eq:amplitudes21} as an example, the concrete expressions of the baryon weak transition matrix elements for the PC and PV amplitudes can be obtained by performing the integrations
\begin{widetext}
\begin{equation}
\begin{split}
\langle{\mathcal{B}_{m}(\pmb{P}_{i};J^{i},J_{z}^{i})}\vert{\mathcal{H}_{1(2),2\to2}^{\text{PC}}}\vert&{\mathcal{B}_{i}(\pmb{P}_{i};J^{i},J_{z}^{i})}\rangle=
(\pm1)\times\frac{G_{F}}{\sqrt{2}}V_{\text{CKM}}V_{\text{CKM}}^{\prime}\frac{6}{(2\pi)^{3}}\int{d\pmb{p}_{1}}{d\pmb{p}_{2}}{d\pmb{p}_{3}}{d\pmb{p}_{1}^{\prime}}{d\pmb{p}_{2}^{\prime}}{d\pmb{p}_{3}^{\prime}}\delta^{3}(\pmb{p}_{3}^{\prime}-\pmb{p}_{3})\\
&\times\sum_{S_{z}^{i},S_{z}^{m}}\sum_{L_{z}^{i},L_{z}^{m}}\langle{S^{m},S_{z}^{m};L^{m},L_{z}^{m}}\vert{J^{i},J_{z}^{i}}\rangle\langle{S^{i},S_{z}^{i};L^{i},L_{z}^{i}}\vert{J^{i},J_{z}^{i}}\rangle\\
&\times\langle{\mathcal{B}_{m}(S^{m},S_{z}^{m})}{\vert}\hat{\alpha}_{1}^{(-)}\hat{\beta}_{2}^{(+)}\hat{O}_{1(2),2\to2}^{\text{spin}}{\vert}{\mathcal{B}_{i}(S^{i},S_{z}^{i})}\rangle\delta^{3}(\pmb{p}_{1}^{\prime}+\pmb{p}_{2}^{\prime}-\pmb{p}_{1}-\pmb{p}_{2})\\
&\times\psi_{L^{m},L_{z}^{m}}^{*}(\pmb{p}_{1}^{\prime},\pmb{p}_{2}^{\prime},\pmb{p}_{3}^{\prime})\psi_{L^{i},L_{z}^{i}}(\pmb{p}_{1},\pmb{p}_{2},\pmb{p}_{3})
\delta^{3}(\pmb{P}_{i}-\pmb{p}_{1}-\pmb{p}_{2}-\pmb{p}_{3})\delta^{3}(\pmb{P}_{i}-\pmb{p}_{1}^{\prime}-\pmb{p}_{2}^{\prime}-\pmb{p}_{3}^{\prime}),
\end{split}
\end{equation}
\begin{equation}
\begin{split}
\langle{\mathcal{B}_{m}(\pmb{P}_{i};J^{i},J_{z}^{i})}\vert{\mathcal{H}_{1(2),2\to2}^{\text{PV}}}\vert&{\mathcal{B}_{i}(\pmb{P}_{i};J^{i},J_{z}^{i})}\rangle=
(\pm1)\times\frac{G_{F}}{\sqrt{2}}V_{\text{CKM}}V_{\text{CKM}}^{\prime}\frac{6}{(2\pi)^{3}}\int{d\pmb{p}_{1}}{d\pmb{p}_{2}}{d\pmb{p}_{3}}{d\pmb{p}_{1}^{\prime}}{d\pmb{p}_{2}^{\prime}}{d\pmb{p}_{3}^{\prime}}\delta^{3}(\pmb{p}_{3}^{\prime}-\pmb{p}_{3})\\
&\times\sum_{S_{z}^{i},S_{z}^{m}}\sum_{L_{z}^{i},L_{z}^{m}}\langle{S^{m},S_{z}^{m};L^{m},L_{z}^{m}}\vert{J^{i},J_{z}^{i}}\rangle\langle{S^{i},S_{z}^{i};L^{i},L_{z}^{i}}\vert{J^{i},J_{z}^{i}}\rangle\\
&\times\langle{\mathcal{B}_{m}(S^{m},S_{z}^{m})}\vert\hat{\alpha}_{1}^{(-)}\hat{\beta}_{2}^{(+)}\hat{\mathcal{O}}_{1(2),2\to2}^{\text{spin}}\vert{\mathcal{B}_{i}(S^{i},S_{z}^{i})}\rangle\delta^{3}(\pmb{p}_{1}^{\prime}+\pmb{p}_{2}^{\prime}-\pmb{p}_{1}-\pmb{p}_{2})\\
&\times\psi_{L^{m},L_{z}^{m}}^{*}(\pmb{p}_{1}^{\prime},\pmb{p}_{2}^{\prime},\pmb{p}_{3}^{\prime})\mathcal{O}_{1(2),2\to2}^{\text{spatial}}(\pmb{p}_{i})\psi_{L^{i},L_{z}^{i}}(\pmb{p}_{1},\pmb{p}_{2},\pmb{p}_{3})
\delta^{3}(\pmb{P}_{i}-\pmb{p}_{1}-\pmb{p}_{2}-\pmb{p}_{3})\delta^{3}(\pmb{P}_{i}-\pmb{p}_{1}^{\prime}-\pmb{p}_{2}^{\prime}-\pmb{p}_{3}^{\prime}),
\end{split}
\end{equation}
where the factor $\pm1$ is color wave function overlap and $6$ is flavor symmetry factor. Besides, the concerned spin-flavor matrix elements for PC and PV amplitudes of $W$-exchange diagrams are summarized in Tables~\ref{tab:polefactorPC} and \ref{tab:polefactorPV} in Appendix~\ref{app02}, respectively.

As for the transition matrix element $\langle{\mathcal{B}_{f}(\pmb{P}_{f})}\vert{\mathcal{H}_{P}}\vert{\mathcal{B}_{m}(\pmb{P}_{i})}\rangle$, we can evaluate by
\begin{equation}
\begin{split}
\langle{\mathcal{B}_{f}(\pmb{P}_{f};J^{f},J_{z}^{f})}\vert{\mathcal{H}_{P}}\vert&{\mathcal{B}_{m}(\pmb{P}_{i};J^{i},J_{z}^{i})}\rangle=\frac{1}{\sqrt{(2\pi)^{3}2\omega_{P}}}\frac{1}{f_{P}}
\int{d\pmb{p}_{1}}{d\pmb{p}_{2}}{d\pmb{p}_{3}}{d\pmb{p}_{1}^{\prime}}{d\pmb{p}_{2}^{\prime}}{d\pmb{p}_{3}^{\prime}}\delta^{3}(\pmb{p}_{1}^{\prime}-\pmb{p}_{1})\delta^{3}(\pmb{p}_{2}^{\prime}-\pmb{p}_{2})\\
&\times\sum_{S_{z}^{f},S_{z}^{m}}\sum_{L_{z}^{f},L_{z}^{m}}\langle{S^{f},S_{z}^{f};L^{f},L_{z}^{f}}\vert{J^{f},J_{z}^{f}}\rangle\langle{S^{m},S_{z}^{m};L^{m},L_{z}^{m}}\vert{J^{i},J_{z}^{i}}\rangle\\
&\times\langle{\mathcal{B}_{f}(S^{f},S_{z}^{f})}\vert\hat{I}_{P}(-\pmb{\sigma}\cdot\pmb{k})\vert{\mathcal{B}_{m}(S^{m},S_{z}^{m})}\rangle{\frac{\omega_{P}+2(m_{1}+m_{2}+m_{3})}{2(m_{1}+m_{2}+m_{3})}}\delta^{3}(\pmb{p}_{3}^{\prime}-\pmb{p}_{3}+\pmb{k})\\
&\times\psi_{L^{f},L_{z}^{f}}^{*}(\pmb{p}_{1}^{\prime},\pmb{p}_{2}^{\prime},\pmb{p}_{3}^{\prime})\psi_{L^{m},L_{z}^{m}}(\pmb{p}_{1},\pmb{p}_{2},\pmb{p}_{3})
\delta^{3}(\pmb{P}_{i}-\pmb{p}_{1}-\pmb{p}_{2}-\pmb{p}_{3})\delta^{3}(\pmb{P}_{f}-\pmb{p}_{1}^{\prime}-\pmb{p}_{2}^{\prime}-\pmb{p}_{3}^{\prime}),
\end{split}
\end{equation}
in which the spin-flavor matrix element can be divided as
\begin{equation}
\langle{\mathcal{B}^{\prime}}\vert\hat{I}_{P}(-\pmb{\sigma}\cdot\pmb{k})\vert{\mathcal{B}}\rangle={-k}\times\langle{\mathcal{B}^{\prime}}\vert\hat{I}_{P}\sigma_{z}\vert{\mathcal{B}}\rangle,
\label{eq:gA}
\end{equation}
where $k\equiv\vert{\pmb{k}}\vert$. We summarize the values of matrix elements $\langle{\mathcal{B}^{\prime}{\downarrow}}\vert\hat{I}_{P}\sigma_{z}\vert{\mathcal{B}{\downarrow}}\rangle$ in Tables~\ref{tab:gA} and \ref{tab:gA2} in Appendix~\ref{app02}. Obviously, we have $\langle{\mathcal{B}_{\bar{3}}^{\prime}}\vert\hat{I}_{P}\sigma_{z}\vert{\mathcal{B}_{\bar{3}}}\rangle=0$ due to the heavy quark symmetry.

Analogously, the decay amplitude in Eq.~\eqref{eq:amplitudes22} can also be obtained. Besides, based on the above discussion we have the following relations:
\begin{equation}
\begin{split}
\mathcal{M}_{\text{PC}}^{-1/2,-1/2}=&-\mathcal{M}_{\text{PC}}^{1/2,1/2},~\mathcal{M}_{\text{PV}}^{-1/2,-1/2}=\mathcal{M}_{\text{PV}}^{1/2,1/2},\\
\mathcal{M}_{\text{PV},\text{PC}}^{-1/2,1/2}=&\mathcal{M}_{\text{PV},\text{PC}}^{1/2,-1/2}=0,
\label{eq:relation2}
\end{split}
\end{equation}
for $W$-exchange diagram.
\end{widetext}

\subsection{The nonrelativistic potential and baryon wave functions}
\label{sec2.3}

In this section, we illustrate how to obtain the wave functions of baryons. For this purpose, we employ a nonrelativistic Hamiltonian~\cite{Copley:1979wj,Pervin:2007wa,Roberts:2007ni}
\begin{equation}
\mathcal{H}=\sum_{i=1,2,3}\Big{(}\frac{p_{i}^{2}}{2m_{i}}+m_{i}\Big{)}+\sum_{i<j}V_{ij},
\end{equation}
to describe a baryon system, where $m_{i}$ and $p_{i}$ are mass and momentum of $i$th quark, respectively. The nonrelativistic quark-quark interaction is given by $V_{ij}=V_{ij}^{\text{con}}+V_{ij}^{\text{hyp}}+V_{ij}^{\text{so(cm)}}+V_{ij}^{\text{so(tp)}}$, where the confinement potential $V_{ij}^{\text{con}}$, hyperfine potential $V_{ij}^{\text{hyp}}$, color-magnetic spin-orbit potential $V_{ij}^{\text{so(cm)}}$, and Thomas-procession spin-orbit potential $V_{ij}^{\text{so(tp)}}$ show the explicit expressions as~\cite{Luo:2023sra,Luo:2023sne,Peng:2024pyl,Zhou:2025fpp}
\begin{equation}
\begin{split}
V_{ij}^{\text{conf}}=-\frac{2}{3}\frac{\alpha_{s}}{r_{ij}}+\frac{b}{2}r_{ij}+\frac{C}{2},
\label{eq:potential1}
\end{split}
\end{equation}
\begin{equation}
\begin{split}
V_{ij}^{\text{hyp}}=&\frac{2\alpha_{s}}{3m_{i}m_{j}}\bigg{[}\frac{8\pi}{3}\tilde{\delta}(r_{ij})\pmb{S}_{i}\cdot\pmb{S}_{j}+\frac{1}{r_{ij}^{3}}\Big{(}\frac{3\pmb{s}_{i}\cdot\pmb{r}_{ij}\pmb{s}_{j}\cdot\pmb{r}_{ij}}{r_{ij}^{2}}\\
&-\pmb{S}_{i}\cdot\pmb{S}_{j}\Big{)}\bigg{]},\\
\label{eq:potential2}
\end{split}
\end{equation}
\begin{equation}
\begin{split}
V_{ij}^{\text{so(cm)}}=&\frac{2\alpha_{s}}{3r_{ij}^{3}}\Big{(}\frac{\pmb{r}_{ij}\times\pmb{p}_{i}\cdot\pmb{s}_{i}}{m_{i}^{2}}-\frac{\pmb{r}_{ij}\times\pmb{p}_{j}\cdot\pmb{s}_{j}}{m_{j}^{2}}\\
&-\frac{\pmb{r}_{ij}\times\pmb{p}_{j}\cdot\pmb{s}_{i}-\pmb{r}_{ij}\times\pmb{p}_{i}\cdot\pmb{s}_{j}}{m_{i}m_{j}}\Big{)},\\
\label{eq:potential3}
\end{split}
\end{equation}
\begin{equation}
\begin{split}
V_{ij}^{\text{so(tp)}}=-\frac{1}{2r_{ij}}\frac{\partial{V_{ij}^{\text{conf}}}}{\partial{r_{ij}}}\Big{(}\frac{\pmb{r_{ij}}\times\pmb{p}_{i}\cdot\pmb{s}_{i}}{m_{i}^{2}}-\frac{\pmb{r_{ij}}\times\pmb{p}_{j}\cdot\pmb{s}_{j}}{m_{j}^{2}}\Big{)}.
\label{eq:potential4}
\end{split}
\end{equation}
In the nonrelativistic quark potential, there are four phenomenological parameters:
\begin{enumerate}
\item[{$\bullet$}] the one-gluon exchange coupling constant $\alpha_{s}$,
\item[{$\bullet$}] the linear-confinement strength $b$,
\item[{$\bullet$}] the constant $C$,
\item[{$\bullet$}] the smearing parameter $\sigma$,
which is introduced by
\begin{equation}
\tilde{\delta}(r_{ij})=\frac{\sigma^{3}}{\pi^{3/2}}\exp(-\sigma^{2}r_{ij}^{2}).
\end{equation}
\end{enumerate}
The phenomenological parameters of nonrelativistic potential model are shown in Table~\ref{tab:PotentialParameters}, which can be constrained by the well-established low-lying charmed baryon spectrum.

\begin{table}[htbp]
\centering
\caption{The parameters used in nonrelativistic potential model. Besides, the masses of constituent quarks are taken as $m_{u,d}=370~\text{MeV}$, $m_{s}=600~\text{MeV}$ and $m_{c}=1880~\text{MeV}$~\cite{Luo:2023sra,Luo:2023sne,Peng:2024pyl}.}
\label{tab:PotentialParameters}
\renewcommand\arraystretch{1.05}
\begin{tabular*}{76mm}{c@{\extracolsep{\fill}}ccc}
\toprule[1pt]
\toprule[0.5pt]
Parameters   &Values   &Parameters   &Values\\
\midrule[0.5pt]
$\alpha_{s}$            &$0.560$   &$b~(\text{GeV}^{2})$   &$0.125$\\
$\sigma~(\text{GeV})$   &$1.600$   &$C~(\text{GeV})$       &$-0.644$\\
\bottomrule[0.5pt]
\bottomrule[1pt]
\end{tabular*}
\end{table}

\begin{figure}[htbp]\centering
  \includegraphics[width=50mm]{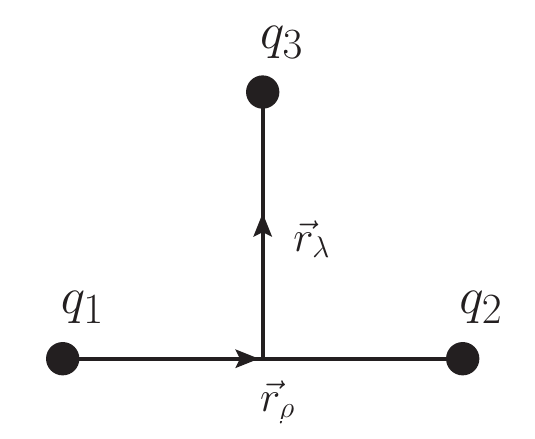}\\
  \caption{The definition of Jacobi coordinates of $\rho$-mode and $\lambda$-mode of charmed baryon. We set $q_{1}q_{2}q_{3}=(qqc)$ for single-charmed baryons and $q_{1}q_{2}q_{3}=(ccq)$ for double charmed baryons.}
  \label{fig:Jacobi}
\end{figure}

In the study of baryon spectrum, the masses and wave functions can be obtained by the solving Schr\"{o}dinger equation with the nonrelativistic Hamiltonian:
\begin{equation}
\mathcal{H}\vert\Psi_{J,M_{J}}\rangle=E\vert\Psi_{J,M_{J}}\rangle,
\label{eq:Schrodinger}
\end{equation}
in which the baryon wave function $\Psi_{J,M_{J}}$ can be constructed as $L-S$ coupling:
\begin{equation}
\begin{split}
\Psi_{J,M_{J}}=&\sum_{\alpha}C^{(\alpha)}\Psi_{J,M_{J}}^{(\alpha)},\\
\Psi_{J,M_{J}}^{(\alpha)}=&\chi^{\text{color}}
\Big{[}\big{[}[s_{q_{1}}s_{q_{2}}]_{S_{\rho}}s_{q_{3}}\big{]}_{S}\psi_{l_{\rho},l_{\lambda},L}^{\text{spatial}}(\pmb{r}_{\rho},\pmb{r}_{\lambda})\Big{]}_{J,M_{J}}\psi^{\text{flavor}}.
\end{split}
\end{equation}
Here, $C^{(\alpha)}$ is a coefficient, and $\alpha$ represent all possible quantum numbers. Besides, the spatial wave function $\psi_{l_{\rho},l_{\lambda},L}^{\text{spatial}}$ consists of $\rho$-mode and $\lambda$-mode excitations as
\begin{equation}
\begin{split}
\psi_{l_{\rho},l_{\lambda},L}^{\text{spatial}}(\pmb{r}_{\rho},\pmb{r}_{\lambda})=&\sum_{m_{\rho},m_{\lambda}}\langle{l_{\rho},m_{\rho};l_{\lambda},m_{\lambda}}\vert{L,m_{\rho}+m_{\lambda}}\rangle\\
&\times\psi_{l_{\rho},m_{\rho}}(\pmb{r}_{\rho})\psi_{l_{\lambda},m_{\lambda}}(\pmb{r}_{\lambda}),
\label{eq:wavefunction2}
\end{split}
\end{equation}
with the corresponding Jacobi coordinates being defined as
\begin{equation}
\begin{split}
\pmb{r}_{\rho}=&\pmb{r}_{2}-\pmb{r}_{1},\\
\pmb{r}_{\lambda}=&\pmb{r}_{3}-\frac{m_{1}\pmb{r}_{1}+m_{2}\pmb{r}_{2}}{m_{1}+m_{2}},
\end{split}
\end{equation}
respectively. This is consistent with the picture, in which a single-charmed baryon is treated as a bound state of a light-quark cluster and a charm quark, whereas a double-charmed baryon is regarded as a bound state of a $(cc)$ cluster and a light quark, as shown in Fig.~\ref{fig:Jacobi}.

In the Gaussian expansion method (GEM), the infinitesimally-shift Gaussian basis~\cite{Hiyama:2003cu,Hiyama:2018ivm}
\begin{equation}
\begin{split}
\phi_{nlm}^{G}(\pmb{r})=&\phi^{G}_{nl}(r)~Y_{lm}(\pmb{\hat{r}})\\
=&\sqrt{\frac{2^{l+2}(2\nu_{n})^{l+3/2}}{\sqrt{\pi}(2l+1)!!}}\lim_{\varepsilon\rightarrow0}
\frac{1}{(\nu_{n}\varepsilon)^l}\sum_{k=1}^{k_{\text{max}}}C_{lm,k}e^{-\nu_{n}\big{(}\pmb{r}-\varepsilon\pmb{D}_{lm,k}\big{)}^2},
\label{eq:Gaussianbasis}
\end{split}
\end{equation}
is adopted to expand the spatial wave functions $\psi_{l_{\rho},m_{\rho}}$ and $\psi_{l_{\lambda},m_{\lambda}}$, in which the Gaussian size parameter $\nu_{n}$ in our computations follows a geometric progression sequence:~\cite{Li:2021qod,Luo:2022cun,Li:2025alu}
\begin{equation}
\begin{split}
\nu_{n}=&1/r^2_{n},~~~r_{n}=r_{min}~a^{n-1},\\
a=&\Big{(}r_{max}/r_{min}\Big{)}^{\frac{1}{n_{max}-1}}
\end{split}
\end{equation}
with $(r_{\rho_{min}},r_{\rho_{max}},n_{\rho_{max}})=(0.1~\text{fm},3.0~\text{fm},10)$ and $(r_{\lambda_{min}},r_{\lambda_{max}},n_{\lambda_{max}})=(0.1~\text{fm},3.0~\text{fm},10)$. The Gaussian basis in momentum space will be obtained by the replacements $r\to{p}$ and $\nu_{n}\to1/(4\nu_{n})$.

Finally, with the nonrelativistic potential and baryon wave function prepared, we further solve the Schr\"{o}dinger equation assisted by the GEM~\cite{Hiyama:2003cu,Hiyama:2018ivm}:
\begin{equation}
\Big{(}T^{\alpha^{\prime},\alpha}+V^{\alpha^{\prime},\alpha}\Big{)}C^{(\alpha)}=EN^{\alpha^{\prime},\alpha}C^{(\alpha)},
\end{equation}
where the matrix elements can be evaluated by
\begin{equation}
\begin{split}
T^{\alpha^{\prime},\alpha}=&\Big{\langle}\Psi_{{J},{M_J}}^{(\alpha^{\prime})}\big{\vert}
\Bigg{[}\sum_{i=1,2,3}\Big{(}\frac{p_{i}^{2}}{2m_{i}}+m_{i}\Big{)}\Bigg{]}
\big{\vert}\Psi_{{J},{M_J}}^{(\alpha)}\Big{\rangle},\\
V^{\alpha^{\prime},\alpha}=&\Big{\langle}\Psi_{{J},{M_J}}^{(\alpha^{\prime})}\big{\vert}
\sum_{i<j}V_{ij}
\big{\vert}\Psi_{{J},{M_J}}^{(\alpha)}\Big{\rangle},\\
N^{\alpha^{\prime},\alpha}=&\Big{\langle}\Psi_{{J},{M_J}}^{(\alpha^{\prime})}
\big{\vert}\Psi_{{J},{M_J}}^{(\alpha)}\Big{\rangle}.
\label{eq:matrixelement}
\end{split}
\end{equation}

\section{Numerical results}
\label{sec3}

Before calculating the weak decay amplitudes, we first discuss the masses and spatial wave functions of the relevant low-lying charmed baryons. Adopting the nonrelativistic potential in Eqs.~(\ref{eq:potential1}-\ref{eq:potential4}) with the phenomenological parameters in Table~\ref{tab:PotentialParameters}, we are able to get the masses of corresponding charmed baryons along with their spatial wave functions, after solving the three-body Schr\"{o}dinger equation. The computed results of masses are summarized in Table~\ref{tab:mass1}. Overall, our results of masses of low-lying charmed baryons are comparable to the experimental values.

\begin{table*}[htbp]\centering
\caption{The comparison of theoretical and experimental values of masses of concerned charmed baryons.}
\label{tab:mass1}
\renewcommand\arraystretch{1.05}
\begin{tabular*}{110mm}{c@{\extracolsep{\fill}}cccc}
\toprule[1pt]
\toprule[0.5pt]
\multirow{2}*{States}   &Quantum numbers   &\multirow{2}*{PDG name}   &This work   &Expt.~\cite{ParticleDataGroup:2024cfk}\\
                        &$(l_{\rho},l_{\lambda},L,S_{\rho},S,J^{P})$   &   &$[\text{MeV}]$   &$[\text{MeV}]$\\
\midrule[0.5pt]
$\Lambda_{c}(1S)$        &$(0,0,0,0,\frac{1}{2},\frac{1}{2}^{+})$   &$\Lambda_{c}^{+}$         &$2280$          &$2286$\\
$\Lambda_{c}(2S)$        &$(0,0,0,0,\frac{1}{2},\frac{1}{2}^{+})$   &$\Lambda_{c}(2765)^{+}$   &$2786$          &$2765$\\
$\Lambda_{c}(1^{2}P_{\lambda})$        &$(0,1,1,0,\frac{1}{2},\frac{1}{2}^{-})$   &$\Lambda_{c}(2595)^{+}$   &$2593$          &$2595$\\
\specialrule{0em}{1pt}{1pt}
$\Sigma_{c}(1S)$         &$(0,0,0,1,\frac{1}{2},\frac{1}{2}^{+})$   &$\Sigma_{c}(2455)^{++,+,0}$   &$2470$         &$2455$\\
$\Sigma_{c}(2S)$         &$(0,0,0,1,\frac{1}{2},\frac{1}{2}^{+})$   &                          &$2953$          &$$\\
$\Sigma_{c}(1^{2}P_{\lambda})$         &$(0,1,1,1,\frac{1}{2},\frac{1}{2}^{-})$   &                          &$2798$          &$$\\
$\Sigma_{c}(1^{4}P_{\lambda})$         &$(0,1,1,1,\frac{3}{2},\frac{1}{2}^{-})$   &                          &$2787$          &$$\\
\specialrule{0em}{1pt}{1pt}
$\Xi_{c}(1S)$            &$(0,0,0,0,\frac{1}{2},\frac{1}{2}^{+})$   &$\Xi_{c}^{+,0}$           &$2481$          &$2470$\\
$\Xi_{c}(2S)$            &$(0,0,0,0,\frac{1}{2},\frac{1}{2}^{+})$   &$\Xi_{c}(2970)^{+,0}$     &$2961$          &$2970$\\
$\Xi_{c}(1^{2}P_{\lambda})$            &$(0,1,1,0,\frac{1}{2},\frac{1}{2}^{-})$   &$\Xi_{c}(2790)^{+,0}$     &$2780$          &$2790$\\
\specialrule{0em}{1pt}{1pt}
$\Xi_{c}^{\prime}(1S)$   &$(0,0,0,1,\frac{1}{2},\frac{1}{2}^{+})$   &$\Xi_{c}^{\prime+,0}$     &$2589$          &$2580$\\
$\Xi_{c}^{\prime}(2S)$   &$(0,0,0,1,\frac{1}{2},\frac{1}{2}^{+})$   &                          &$3058$          &$$\\
$\Xi_{c}^{\prime}(1^{2}P_{\lambda})$   &$(0,1,1,1,\frac{1}{2},\frac{1}{2}^{-})$   &                          &$2908$          &$$\\
$\Xi_{c}^{\prime}(1^{4}P_{\lambda})$   &$(0,1,1,1,\frac{3}{2},\frac{1}{2}^{-})$   &                          &$2894$          &$$\\
\specialrule{0em}{1pt}{1pt}
$\Omega_{c}(1S)$         &$(0,0,0,1,\frac{1}{2},\frac{1}{2}^{+})$   &$\Omega_{c}^{0}$          &$2701$          &$2695$\\
$\Omega_{c}(2S)$         &$(0,0,0,1,\frac{1}{2},\frac{1}{2}^{+})$   &                          &$3163$          &$$\\
$\Omega_{c}(1^{2}P_{\lambda})$         &$(0,1,1,1,\frac{1}{2},\frac{1}{2}^{-})$   &                          &$3018$          &$$\\
$\Omega_{c}(1^{4}P_{\lambda})$         &$(0,1,1,1,\frac{3}{2},\frac{1}{2}^{-})$   &                          &$3000$          &$$\\
\specialrule{0em}{1pt}{1pt}
$\Xi_{cc}(1S)$                &$(0,0,0,1,\frac{1}{2},\frac{1}{2}^{+})$   &$\Xi_{cc}^{++}$      &$3649$          &$3621$\\
$\Xi_{cc}(2S)$                &$(0,0,0,1,\frac{1}{2},\frac{1}{2}^{+})$   &                     &$4064$          &$$\\
$\Xi_{cc}(1^{2}P_{\rho})$         &$(1,0,1,0,\frac{1}{2},\frac{1}{2}^{-})$   &                     &$3965$          &$$\\
$\Xi_{cc}(1^{2}P_{\lambda})$      &$(0,1,1,1,\frac{1}{2},\frac{1}{2}^{-})$   &                     &$4064$          &$$\\
$\Xi_{cc}(1^{4}P_{\lambda})$      &$(0,1,1,1,\frac{3}{2},\frac{1}{2}^{-})$   &                     &$4011$          &$$\\
\specialrule{0em}{1pt}{1pt}
$\Omega_{cc}(1S)$             &$(0,0,0,1,\frac{1}{2},\frac{1}{2}^{+})$   &                     &$3751$          &$$\\
$\Omega_{cc}(2S)$             &$(0,0,0,1,\frac{1}{2},\frac{1}{2}^{+})$   &                     &$4185$          &$$\\
$\Omega_{cc}(1^{2}P_{\rho})$      &$(1,0,1,0,\frac{1}{2},\frac{1}{2}^{-})$   &                     &$4077$          &$$\\
$\Omega_{cc}(1^{2}P_{\lambda})$   &$(0,1,1,1,\frac{1}{2},\frac{1}{2}^{-})$   &                     &$4149$          &$$\\
$\Omega_{cc}(1^{4}P_{\lambda})$   &$(0,1,1,1,\frac{3}{2},\frac{1}{2}^{-})$   &                     &$4093$          &$$\\
\bottomrule[0.5pt]
\bottomrule[1pt]
\end{tabular*}
\end{table*}

Having prepared the charmed-baryon wave functions, we are subsequently able to calculate the weak decay amplitudes numerically. The CKM matrix elements are adopted in the Wolfenstein parameterization as
\begin{equation}
\begin{split}
V_{ud}=&1-\lambda^{2}/2,~~V_{us}=\lambda,\\
V_{cd}=&-\lambda,~~V_{cs}=1-\lambda^{2}/2,
\end{split}
\end{equation}
where $\lambda=0.22501$~\cite{ParticleDataGroup:2024cfk}. The decay amplitudes for CF, SCS and DCS processes are presented in Tables~\ref{tab:CFamplitudes}, \ref{tab:SCSamplitudes} and \ref{tab:DCSamplitudes}, respectively. Here, we use the symbols $\times(\text{w})$ and $\times(\text{s})$ to mark that the corresponding decay amplitude vanishes due to the relevant weak or strong interaction vertex, respectively. According to the pole model, the nonfactorizable amplitudes arising from $W$-exchange diagrams are evaluated by considering intermediate low-lying poles, specifically $\Lambda_{c}^{+}$, $\Sigma_{c}^{+}$, $\Xi_{c}^{(\prime)+}$, $\Xi_{cc}^{+}$ and $\Omega_{cc}^{+}$ in this work. In our calculation, for the PC amplitudes induced by double-charmed baryon poles, the contribution from $2S$ pole is significantly suppressed compared to that of $1S$ pole. However, for single-charmed baryon poles, the $2S$ pole contribution remains comparable to $1S$ pole. For the PV amplitudes, the main contribution arises from double-charmed baryon poles, who have quantum numbers $1^{4}P_{\lambda}$. In comparison, the contributions from either single-charmed or double-charmed baryon poles with quantum numbers $1^{2}P_{\lambda}$, are small.

\begin{table*}[htbp]\centering
\caption{The decay amplitudes $\mathcal{M}_{\text{PC},\text{PV}}^{-1/2,-1/2}$ (in the units of $10^{-4}G_{F}~\text{GeV}^{3/2}$) of CF decays.}
\label{tab:CFamplitudes}
\renewcommand\arraystretch{1.05}
\begin{tabular*}{120mm}{l@{\extracolsep{\fill}}llll}
\toprule[1pt]
\toprule[0.5pt]
&\multicolumn{2}{c}{{\normalsize $\mathcal{M}_{\text{PC}}^{-1/2,-1/2}$}}  &\multicolumn{2}{c}{{\normalsize $\mathcal{M}_{\text{PV}}^{-1/2,-1/2}$}}\\
\midrule[0.5pt]
\multirow{1}*{\shortstack{$\Omega_{cc}^{+}\to\Omega_{c}^{0}\pi^{+}$}}   &$T-diagram$   &$-53.96$  &$T-diagram$   &$22.21$\\
\specialrule{0em}{2pt}{2pt}
\multirow{1}*{\shortstack{$\Omega_{cc}^{+}\to\Xi_{c}^{+}\bar{K}^{0}$}}   &$C-diagram$   &$17.69$   &$C-diagram$   &$-35.61$  \\
                                                                     &$C^{\prime}-diagram$   &$-3.61$   &$C^{\prime}-diagram$   &$12.14$  \\
                                                                     &$\Xi_{cc}^{+}(1S)-pole$   &$-54.75$   &$\Xi_{cc}^{+}(1^{2}P_{\rho})-pole$   &$\times~(\text{s})$ \\
                                                                     &$\Xi_{cc}^{+}(2S)-pole$   &$-2.57$   &$\Xi_{cc}^{+}(1^{2}P_{\lambda})-pole$   &$-1.90$ \\
                                                                     &   &   &$\Xi_{cc}^{+}(1^{4}P_{\lambda})-pole$   &$70.04$\\
                                                                     &$\text{total}$   &$-43.24$   &$\text{total}$   &$44.67$  \\
\specialrule{0em}{2pt}{2pt}
\multirow{1}*{\shortstack{$\Omega_{cc}^{+}\to\Xi_{c}^{\prime+}\bar{K}^{0}$}}   &$C-diagram$   &$-43.65$   &$C-diagram$   &$20.32$  \\
                                                                     &$C^{\prime}-diagram$   &$0$   &$C^{\prime}-diagram$   &$0$  \\
                                                    &$\Xi_{cc}^{+}(1S)-pole$   &$\times~(\text{w})$   &$\Xi_{cc}^{+}(1^{2}P_{\rho})-pole$   &$\times~(\text{s})$ \\
                                                    &$\Xi_{cc}^{+}(2S)-pole$   &$\times~(\text{w})$   &$\Xi_{cc}^{+}(1^{2}P_{\lambda})-pole$   &$0$ \\
                                                    &   &   &$\Xi_{cc}^{+}(1^{4}P_{\lambda})-pole$   &$\times~(\text{w})$\\
                                                    &$\text{total}$   &$-43.65$   &$\text{total}$   &$20.32$  \\
\bottomrule[0.5pt]
\bottomrule[1pt]
\end{tabular*}
\end{table*}

\begin{table*}[htbp]\centering
\caption{Same as Table~\ref{tab:CFamplitudes} but for SCS decays.}
\label{tab:SCSamplitudes}
\renewcommand\arraystretch{1.05}
\begin{tabular*}{165mm}{l@{\extracolsep{\fill}}lclc|llclc}
\toprule[1pt]
\toprule[0.5pt]
&\multicolumn{2}{c}{{\normalsize $\mathcal{M}_{\text{PC}}^{-1/2,-1/2}$}}  &\multicolumn{2}{c}{{\normalsize $\mathcal{M}_{\text{PV}}^{-1/2,-1/2}$}} &  &\multicolumn{2}{c}{{\normalsize $\mathcal{M}_{\text{PC}}^{-1/2,-1/2}$}}  &\multicolumn{2}{c}{{\normalsize $\mathcal{M}_{\text{PV}}^{-1/2,-1/2}$}}\\
\midrule[0.5pt]
\multirow{1}*{\shortstack{$\Omega_{cc}^{+}\to\Xi_{c}^{0}\pi^{+}$}}   &$T-diagram$   &$2.78$   &$T-diagram$   &$-4.43$  &\multirow{1}*{\shortstack{$\Omega_{cc}^{+}\to\Xi_{c}^{\prime0}\pi^{+}$}}   &$T-diagram$   &$-6.91$   &$T-diagram$   &$2.49$\\
            &$\Xi_{c}^{+}(1S)-pole$   &$\times~(\text{s})$   &$\Xi_{c}^{+}(1^{2}P_{\lambda})-pole$   &$\times~(\text{s})$  &  &$\Xi_{c}^{+}(1S)-pole$   &$16.20$   &$\Xi_{c}^{+}(1^{2}P_{\lambda})-pole$   &$-0.14$\\
            &$\Xi_{c}^{+}(2S)-pole$   &$\times~(\text{s})$   &$\Xi_{c}^{\prime+}(1^{2}P_{\lambda})-pole$   &$0$  &  &$\Xi_{c}^{+}(2S)-pole$   &$8.48$   &$\Xi_{c}^{\prime+}(1^{2}P_{\lambda})-pole$   &$0$\\
            &$\Xi_{c}^{\prime+}(1S)-pole$   &$\times~(\text{w})$   &$\Xi_{c}^{\prime+}(1^{4}P_{\lambda})-pole$   &$0$  &  &$\Xi_{c}^{\prime+}(1S)-pole$   &$\times~(\text{w})$   &$\Xi_{c}^{\prime+}(1^{4}P_{\lambda})-pole$   &$0$\\
            &$\Xi_{c}^{\prime+}(2S)-pole$   &$\times~(\text{w})$   &   &  &  &$\Xi_{c}^{\prime+}(2S)-pole$   &$\times~(\text{w})$   &   &\\
            &$\text{total}$   &$2.78$   &$\text{total}$   &$-4.43$  &  &$\text{total}$   &$17.77$   &$\text{total}$   &$2.35$\\
\specialrule{0em}{2pt}{2pt}
\multirow{1}*{\shortstack{$\Omega_{cc}^{+}\to\Omega_{c}^{0}K^{+}$}}   &$T-diagram$   &$-24.56$   &$T-diagram$   &$13.54$  &\multirow{1}*{\shortstack{$\Omega_{cc}^{+}\to\Sigma_{c}^{++}K^{-}$}}  &$\Xi_{c}^{+}(1S)-pole$   &$17.15$   &$\Xi_{c}^{+}(1^{2}P_{\lambda})-pole$   &$-0.17$\\
                     &$\Xi_{c}^{+}(1S)-pole$   &$-30.26$   &$\Xi_{c}^{+}(1^{2}P_{\lambda})-pole$   &$0.18$  &  &$\Xi_{c}^{+}(2S)-pole$   &$8.27$   &$\Xi_{c}^{\prime+}(1^{2}P_{\lambda})-pole$   &$0$\\
                     &$\Xi_{c}^{+}(2S)-pole$   &$-9.55$   &$\Xi_{c}^{\prime+}(1^{2}P_{\lambda})-pole$   &$0$  &  &$\Xi_{c}^{\prime+}(1S)-pole$   &$\times~(\text{w})$   &$\Xi_{c}^{\prime+}(1^{4}P_{\lambda})-pole$   &$0$\\
                     &$\Xi_{c}^{\prime+}(1S)-pole$   &$\times~(\text{w})$   &$\Xi_{c}^{\prime+}(1^{4}P_{\lambda})-pole$   &$0$  &  &$\Xi_{c}^{\prime+}(2S)-pole$   &$\times~(\text{w})$   &   &\\
                     &$\Xi_{c}^{\prime+}(2S)-pole$   &$\times~(\text{w})$   &   &  &  &$\text{total}$   &$25.42$   &$\text{total}$   &$-0.17$\\
                     &$\text{total}$   &$-64.37$   &$\text{total}$   &$13.72$  &&&&&\\
\specialrule{0em}{2pt}{2pt}
\multirow{1}*{\shortstack{$\Omega_{cc}^{+}\to\Xi_{c}^{+}\eta$}}   &$C-diagram$   &$-3.43$   &$C-diagram$   &$7.14$  &\multirow{1}*{\shortstack{$\Omega_{cc}^{+}\to\Xi_{c}^{\prime+}\eta$}}   &$C-diagram$   &$8.45$   &$C-diagram$   &$-4.09$\\
                    &$C^{\prime}-diagram$   &$0.29$   &$C^{\prime}-diagram$   &$-1.72$  &  &$C^{\prime}-diagram$   &$0$   &$C^{\prime}-diagram$   &$0$\\
                    &$\Omega_{cc}^{+}(1S)-pole$   &$10.43$   &$\Omega_{cc}^{+}(1^{2}P_\rho)-pole$   &$\times~(\text{s})$  &  &$\Omega_{cc}^{+}(1S)-pole$   &$\times~(\text{w})$   &$\Omega_{cc}^{+}(1^{2}P_\rho)-pole$   &$\times~(\text{s})$\\
                    &$\Omega_{cc}^{+}(2S)-pole$   &$0.62$   &$\Omega_{cc}^{+}(1^{2}P_\lambda)-pole$   &$0.34$  &  &$\Omega_{cc}^{+}(2S)-pole$   &$\times~(\text{w})$   &$\Omega_{cc}^{+}(1^{2}P_\lambda)-pole$   &$0$\\
                    &$\Xi_{c}^{+}(1S)-pole$   &$\times~(\text{s})$   &$\Omega_{cc}^{+}(1^{4}P_\lambda)-pole$   &$-7.20$  &  &$\Xi_{c}^{+}(1S)-pole$   &$18.09$   &$\Omega_{cc}^{+}(1^{4}P_\lambda)-pole$   &$\times~(\text{w})$\\
                    &$\Xi_{c}^{+}(2S)-pole$   &$\times~(\text{s})$   &$\Xi_{c}^{+}(1^{2}P_{\lambda})-pole$   &$\times~(\text{s})$  &  &$\Xi_{c}^{+}(2S)-pole$   &$7.76$   &$\Xi_{c}^{+}(1^{2}P_{\lambda})-pole$   &$-0.15$\\
                    &$\Xi_{c}^{\prime+}(1S)-pole$   &$\times~(\text{w})$   &$\Xi_{c}^{\prime+}(1^{2}P_{\lambda})-pole$   &$0$  &  &$\Xi_{c}^{\prime+}(1S)-pole$   &$\times~(\text{w})$   &$\Xi_{c}^{\prime+}(1^{2}P_{\lambda})-pole$   &$0$\\
                    &$\Xi_{c}^{\prime+}(2S)-pole$   &$\times~(\text{w})$   &$\Xi_{c}^{\prime+}(1^{4}P_{\lambda})-pole$   &$0$  &  &$\Xi_{c}^{\prime+}(2S)-pole$   &$\times~(\text{w})$   &$\Xi_{c}^{\prime+}(1^{4}P_{\lambda})-pole$   &$0$\\
                    &$\text{total}$   &$7.91$   &$\text{total}$   &$-1.44$  &  &$\text{total}$   &$34.3$   &$\text{total}$   &$-4.24$\\
\specialrule{0em}{2pt}{2pt}
\multirow{1}*{\shortstack{$\Omega_{cc}^{+}\to\Xi_{c}^{+}\eta^{\prime}$}}   &$C-diagram$   &$1.17$   &$C-diagram$   &$-9.04$  &\multirow{1}*{\shortstack{$\Omega_{cc}^{+}\to\Xi_{c}^{\prime+}\eta^{\prime}$}}   &$C-diagram$   &$-2.43$   &$C-diagram$   &$5.42$\\
                    &$C^{\prime}-diagram$   &$-0.30$   &$C^{\prime}-diagram$   &$2.19$  &  &$C^{\prime}-diagram$   &$0$   &$C^{\prime}-diagram$   &$0$\\
                    &$\Omega_{cc}^{+}(1S)-pole$   &$-12.64$   &$\Omega_{cc}^{+}(1^{2}P_\rho)-pole$   &$\times~(\text{s})$  &  &$\Omega_{cc}^{+}(1S)-pole$   &$\times~(\text{w})$   &$\Omega_{cc}^{+}(1^{2}P_\rho)-pole$   &$\times~(\text{s})$\\
                    &$\Omega_{cc}^{+}(2S)-pole$   &$-0.42$   &$\Omega_{cc}^{+}(1^{2}P_\lambda)-pole$   &$-0.34$  &  &$\Omega_{cc}^{+}(2S)-pole$   &$\times~(\text{w})$   &$\Omega_{cc}^{+}(1^{2}P_\lambda)-pole$   &$0$\\
                    &$\Xi_{c}^{+}(1S)-pole$   &$\times~(\text{s})$   &$\Omega_{cc}^{+}(1^{4}P_\lambda)-pole$   &$6.74$  &  &$\Xi_{c}^{+}(1S)-pole$   &$-3.44$   &$\Omega_{cc}^{+}(1^{4}P_\lambda)-pole$   &$\times~(\text{w})$\\
                    &$\Xi_{c}^{+}(2S)-pole$   &$\times~(\text{s})$   &$\Xi_{c}^{+}(1^{2}P_{\lambda})-pole$   &$\times~(\text{s})$  &  &$\Xi_{c}^{+}(2S)-pole$   &$-0.62$   &$\Xi_{c}^{+}(1^{2}P_{\lambda})-pole$   &$0.020$\\
                    &$\Xi_{c}^{\prime+}(1S)-pole$   &$\times~(\text{w})$   &$\Xi_{c}^{\prime+}(1^{2}P_{\lambda})-pole$   &$0$  &  &$\Xi_{c}^{\prime+}(1S)-pole$   &$\times~(\text{w})$   &$\Xi_{c}^{\prime+}(1^{2}P_{\lambda})-pole$   &$0$\\
                    &$\Xi_{c}^{\prime+}(2S)-pole$   &$\times~(\text{w})$   &$\Xi_{c}^{\prime+}(1^{4}P_{\lambda})-pole$   &$0$  &  &$\Xi_{c}^{\prime+}(2S)-pole$   &$\times~(\text{w})$   &$\Xi_{c}^{\prime+}(1^{4}P_{\lambda})-pole$   &$0$\\
                    &$\text{total}$   &$-12.19$   &$\text{total}$   &$-0.45$  &  &$\text{total}$   &$-6.49$   &$\text{total}$   &$5.44$\\
\specialrule{0em}{2pt}{2pt}
\multirow{1}*{\shortstack{$\Omega_{cc}^{+}\to\Xi_{c}^{+}\pi^{0}$}}   &$C-diagram$   &$1.37$   &$C-diagram$   &$-2.18$  &\multirow{1}*{\shortstack{$\Omega_{cc}^{+}\to\Xi_{c}^{\prime+}\pi^{0}$}}   &$C-diagram$   &$-3.40$   &$C-diagram$   &$1.22$\\
                    &$\Xi_{c}^{+}(1S)-pole$   &$\times~(\text{s})$   &$\Xi_{c}^{+}(1^{2}P_{\lambda})-pole$   &$\times~(\text{s})$  &  &$\Xi_{c}^{+}(1S)-pole$   &$11.45$   &$\Xi_{c}^{+}(1^{2}P_{\lambda})-pole$   &$-0.10$\\
                    &$\Xi_{c}^{+}(2S)-pole$   &$\times~(\text{s})$   &$\Xi_{c}^{\prime+}(1^{2}P_{\lambda})-pole$   &$0$  &  &$\Xi_{c}^{+}(2S)-pole$   &$5.99$   &$\Xi_{c}^{\prime+}(1^{2}P_{\lambda})-pole$   &$0$\\
                    &$\Xi_{c}^{\prime+}(1S)-pole$   &$\times~(\text{w})$   &$\Xi_{c}^{\prime+}(1^{4}P_{\lambda})-pole$   &$0$  &  &$\Xi_{c}^{\prime+}(1S)-pole$   &$\times~(\text{w})$   &$\Xi_{c}^{\prime+}(1^{4}P_{\lambda})-pole$   &$0$\\
                    &$\Xi_{c}^{\prime+}(2S)-pole$   &$\times~(\text{w})$   &   &  &  &$\Xi_{c}^{\prime+}(2S)-pole$   &$\times~(\text{w})$   &   &\\
                    &$\text{total}$   &$1.37$   &$\text{total}$   &$-2.18$  &  &$\text{total}$   &$14.04$   &$\text{total}$   &$1.12$\\
\specialrule{0em}{2pt}{2pt}
\multirow{1}*{\shortstack{$\Omega_{cc}^{+}\to\Sigma_{c}^{+}\bar{K}^{0}$}}   &$C^{\prime}-diagram$   &$0$   &$C^{\prime}-diagram$   &$0$  &\multirow{1}*{\shortstack{$\Omega_{cc}^{+}\to\Lambda_{c}^{+}\bar{K}^{0}$}}   &$C^{\prime}-diagram$   &$0.85$   &$C^{\prime}-diagram$   &$-2.60$\\
                     &$\Xi_{cc}^{+}(1S)-pole$   &$\times~(\text{w})$   &$\Xi_{cc}^{+}(1^{2}P_\rho)-pole$   &$\times~(\text{s})$  &  &$\Xi_{cc}^{+}(1S)-pole$   &$10.71$   &$\Xi_{cc}^{+}(1^{2}P_\rho)-pole$   &$\times~(\text{s})$\\
                     &$\Xi_{cc}^{+}(2S)-pole$   &$\times~(\text{w})$   &$\Xi_{cc}^{+}(1^{2}P_\lambda)-pole$   &$0$  &  &$\Xi_{cc}^{+}(2S)-pole$   &$0.57$   &$\Xi_{cc}^{+}(1^{2}P_\lambda)-pole$   &$0.40$\\
                     &$\Xi_{c}^{+}(1S)-pole$   &$12.13$   &$\Xi_{cc}^{+}(1^{4}P_\lambda)-pole$   &$\times~(\text{w})$  &  &$\Xi_{c}^{+}(1S)-pole$   &$\times~(\text{s})$   &$\Xi_{cc}^{+}(1^{4}P_\lambda)-pole$   &$-15.66$\\
                     &$\Xi_{c}^{+}(2S)-pole$   &$5.85$   &$\Xi_{c}^{+}(1^{2}P_{\lambda})-pole$   &$-0.12$  &  &$\Xi_{c}^{+}(2S)-pole$   &$\times~(\text{s})$   &$\Xi_{c}^{+}(1^{2}P_{\lambda})-pole$   &$\times~(\text{s})$\\
                     &$\Xi_{c}^{\prime+}(1S)-pole$   &$\times~(\text{w})$   &$\Xi_{c}^{\prime+}(1^{2}P_{\lambda})-pole$   &$0$  &  &$\Xi_{c}^{\prime+}(1S)-pole$   &$\times~(\text{w})$   &$\Xi_{c}^{\prime+}(1^{2}P_{\lambda})-pole$   &$0$\\
                     &$\Xi_{c}^{\prime+}(2S)-pole$   &$\times~(\text{w})$   &$\Xi_{c}^{\prime+}(1^{4}P_{\lambda})-pole$   &$0$  &  &$\Xi_{c}^{\prime+}(2S)-pole$   &$\times~(\text{w})$   &$\Xi_{c}^{\prime+}(1^{4}P_{\lambda})-pole$   &$0$\\
                     &$\text{total}$   &$17.98$   &$\text{total}$   &$-0.12$  &  &$\text{total}$   &$12.13$   &$\text{total}$   &$-17.86$\\
\bottomrule[0.5pt]
\bottomrule[1pt]
\end{tabular*}
\end{table*}

\begin{table*}[htbp]\centering
\caption{Same as Table~\ref{tab:CFamplitudes} but for DCS decays.}
\label{tab:DCSamplitudes}
\renewcommand\arraystretch{1.05}
\begin{tabular*}{165mm}{l@{\extracolsep{\fill}}lclc|llclc}
\toprule[1pt]
\toprule[0.5pt]
&\multicolumn{2}{c}{{\normalsize $\mathcal{M}_{\text{PC}}^{-1/2,-1/2}$}}  &\multicolumn{2}{c}{{\normalsize $\mathcal{M}_{\text{PV}}^{-1/2,-1/2}$}} & &\multicolumn{2}{c}{{\normalsize $\mathcal{M}_{\text{PC}}^{-1/2,-1/2}$}}  &\multicolumn{2}{c}{{\normalsize $\mathcal{M}_{\text{PV}}^{-1/2,-1/2}$}}\\
\midrule[0.5pt]
\multirow{1}*{\shortstack{$\Omega_{cc}^{+}\to\Xi_{c}^{0}K^{+}$}}   &$T-diagram$   &$1.35$   &$T-diagram$   &$-2.73$  &\multirow{1}*{\shortstack{$\Omega_{cc}^{+}\to\Xi_{c}^{\prime0}K^{+}$}}   &$T-diagram$   &$-3.34$   &$T-diagram$   &$1.56$\\
                             &$\Lambda_{c}^{+}(1S)-pole$   &$\times~(\text{s})$   &$\Lambda_{c}^{+}(1^{2}P_{\lambda})-pole$   &$\times~(\text{s})$  &  &$\Lambda_{c}^{+}(1S)-pole$   &$-3.41$   &$\Lambda_{c}^{+}(1^{2}P_{\lambda})-pole$   &$0.028$\\
                             &$\Lambda_{c}^{+}(2S)-pole$   &$\times~(\text{s})$   &$\Sigma_{c}^{+}(1^{2}P_{\lambda})-pole$    &$0$  &  &$\Lambda_{c}^{+}(2S)-pole$   &$-1.64$   &$\Sigma_{c}^{+}(1^{2}P_{\lambda})-pole$    &$0$\\
                             &$\Sigma_{c}^{+}(1S)-pole$    &$\times~(\text{w})$   &$\Sigma_{c}^{+}(1^{4}P_{\lambda})-pole$   &$0$  &  &$\Sigma_{c}^{+}(1S)-pole$    &$\times~(\text{w})$   &$\Sigma_{c}^{+}(1^{4}P_{\lambda})-pole$   &$0$\\
                             &$\Sigma_{c}^{+}(2S)-pole$    &$\times~(\text{w})$   &   &  &  &$\Sigma_{c}^{+}(2S)-pole$    &$\times~(\text{w})$   &   &\\
                             &$\text{total}$   &$1.35$   &$\text{total}$   &$-2.73$  &  &$\text{total}$   &$-8.39$   &$\text{total}$   &$1.59$\\
\specialrule{0em}{2pt}{2pt}
\multirow{1}*{\shortstack{$\Omega_{cc}^{+}\to\Xi_{c}^{+}K^{0}$}}   &$C-diagram$   &$-0.94$   &$C-diagram$   &$1.90$  &\multirow{1}*{\shortstack{$\Omega_{cc}^{+}\to\Xi_{c}^{\prime+}K^{0}$}}   &$C-diagram$   &$2.33$   &$C-diagram$   &$-1.08$\\
              &$\Lambda_{c}^{+}(1S)-pole$   &$\times~(\text{s})$   &$\Lambda_{c}^{+}(1^{2}P_{\lambda})-pole$   &$\times~(\text{s})$  &  &$\Lambda_{c}^{+}(1S)-pole$   &$3.41$   &$\Lambda_{c}^{+}(1^{2}P_{\lambda})-pole$   &$-0.028$\\
              &$\Lambda_{c}^{+}(2S)-pole$   &$\times~(\text{s})$   &$\Sigma_{c}^{+}(1^{2}P_{\lambda})-pole$    &$0$  &  &$\Lambda_{c}^{+}(2S)-pole$   &$1.64$   &$\Sigma_{c}^{+}(1^{2}P_{\lambda})-pole$    &$0$\\
              &$\Sigma_{c}^{+}(1S)-pole$    &$\times~(\text{w})$   &$\Sigma_{c}^{+}(1^{4}P_{\lambda})-pole$   &$0$  &  &$\Sigma_{c}^{+}(1S)-pole$    &$\times~(\text{w})$   &$\Sigma_{c}^{+}(1^{4}P_{\lambda})-pole$   &$0$\\
              &$\Sigma_{c}^{+}(2S)-pole$    &$\times~(\text{w})$   &   &  &  &$\Sigma_{c}^{+}(2S)-pole$    &$\times~(\text{w})$   &   &\\
              &$\text{total}$   &$-0.94$   &$\text{total}$   &$1.90$  &  &$\text{total}$   &$7.38$   &$\text{total}$   &$-1.11$\\
\specialrule{0em}{2pt}{2pt}
\multirow{1}*{\shortstack{$\Omega_{cc}^{+}\to\Sigma_{c}^{+}\pi^{0}$}} &$\Lambda_{c}^{+}(1S)-pole$  &$4.40$  &$\Lambda_{c}^{+}(1^{2}P_{\lambda})-pole$  &$-0.044$  &\multirow{1}*{\shortstack{$\Omega_{cc}^{+}\to\Lambda_{c}^{+}\pi^{0}$}}   &$\Lambda_{c}^{+}(1S)-pole$   &$\times~(\text{s})$   &$\Lambda_{c}^{+}(1^{2}P_{\lambda})-pole$   &$\times~(\text{s})$\\
                          &$\Lambda_{c}^{+}(2S)-pole$   &$2.64$   &$\Sigma_{c}^{+}(1^{2}P_{\lambda})-pole$    &$\times~(\text{s})$  &  &$\Lambda_{c}^{+}(2S)-pole$   &$\times~(\text{s})$   &$\Sigma_{c}^{+}(1^{2}P_{\lambda})-pole$    &$0$\\
                          &$\Sigma_{c}^{+}(1S)-pole$    &$\times~(\text{w})$   &$\Sigma_{c}^{+}(1^{4}P_{\lambda})-pole$   &$0$  &  &$\Sigma_{c}^{+}(1S)-pole$    &$\times~(\text{w})$   &$\Sigma_{c}^{+}(1^{4}P_{\lambda})-pole$   &$0$\\
                          &$\Sigma_{c}^{+}(2S)-pole$    &$\times~(\text{w})$   &   &  &  &$\Sigma_{c}^{+}(2S)-pole$    &$\times~(\text{w})$   &   &\\
                          &$\text{total}$   &$7.04$   &$\text{total}$   &$-0.044$  &  &$\text{total}$   &$0$   &$\text{total}$   &$0$\\
\specialrule{0em}{2pt}{2pt}
\multirow{1}*{\shortstack{$\Omega_{cc}^{+}\to\Sigma_{c}^{0}\pi^{+}$}}   &$\Lambda_{c}^{+}(1S)-pole$   &$4.40$   &$\Lambda_{c}^{+}(1^{2}P_{\lambda})-pole$   &$-0.044$  &\multirow{1}*{\shortstack{$\Omega_{cc}^{+}\to\Sigma_{c}^{++}\pi^{-}$}}   &$\Lambda_{c}^{+}(1S)-pole$   &$-4.40$   &$\Lambda_{c}^{+}(1^{2}P_{\lambda})-pole$   &$0.044$\\
                         &$\Lambda_{c}^{+}(2S)-pole$   &$2.64$   &$\Sigma_{c}^{+}(1^{2}P_{\lambda})-pole$    &$0$  &  &$\Lambda_{c}^{+}(2S)-pole$   &$-2.64$   &$\Sigma_{c}^{+}(1^{2}P_{\lambda})-pole$    &$0$\\
                         &$\Sigma_{c}^{+}(1S)-pole$    &$\times~(\text{w})$   &$\Sigma_{c}^{+}(1^{4}P_{\lambda})-pole$   &$0$  &  &$\Sigma_{c}^{+}(1S)-pole$    &$\times~(\text{w})$   &$\Sigma_{c}^{+}(1^{4}P_{\lambda})-pole$   &$0$\\
                         &$\Sigma_{c}^{+}(2S)-pole$    &$\times~(\text{w})$   &   &  &  &$\Sigma_{c}^{+}(2S)-pole$    &$\times~(\text{w})$   &   &\\
                         &$\text{total}$   &$7.04$   &$\text{total}$   &$-0.044$  &  &$\text{total}$   &$-7.04$   &$\text{total}$   &$0.044$\\
\specialrule{0em}{2pt}{2pt}
\multirow{1}*{\shortstack{$\Omega_{cc}^{+}\to\Sigma_{c}^{+}\eta$}}   &$C^{\prime}-diagram$   &$0$   &$C^{\prime}-diagram$   &$0$  &\multirow{1}*{\shortstack{$\Omega_{cc}^{+}\to\Lambda_{c}^{+}\eta$}}   &$C^{\prime}-diagram$   &$-0.068$   &$C^{\prime}-diagram$   &$0.37$\\
                        &$\Omega_{cc}^{+}(1S)-pole$   &$\times~(\text{w})$   &$\Omega_{cc}^{+}(1^{2}P_\rho)-pole$   &$\times~(\text{s})$  &  &$\Omega_{cc}^{+}(1S)-pole$   &$-2.16$   &$\Omega_{cc}^{+}(1^{2}P_\rho)-pole$   &$\times~(\text{s})$\\
                        &$\Omega_{cc}^{+}(2S)-pole$   &$\times~(\text{w})$   &$\Omega_{cc}^{+}(1^{2}P_\lambda)-pole$   &$0$  &  &$\Omega_{cc}^{+}(2S)-pole$   &$-0.16$   &$\Omega_{cc}^{+}(1^{2}P_\lambda)-pole$   &$-0.077$\\
                        &$\Lambda_{c}^{+}(1S)-pole$   &$\times~(\text{s})$   &$\Omega_{cc}^{+}(1^{4}P_\lambda)-pole$   &$\times~(\text{w})$  &  &$\Lambda_{c}^{+}(1S)-pole$   &$\times~(\text{s})$   &$\Omega_{cc}^{+}(1^{4}P_\lambda)-pole$   &$1.66$\\
                        &$\Lambda_{c}^{+}(2S)-pole$   &$\times~(\text{s})$   &$\Lambda_{c}^{+}(1^{2}P_{\lambda})-pole$   &$\times~(\text{s})$  &  &$\Lambda_{c}^{+}(2S)-pole$   &$\times~(\text{s})$   &$\Lambda_{c}^{+}(1^{2}P_{\lambda})-pole$   &$\times~(\text{s})$\\
                        &$\Sigma_{c}^{+}(1S)-pole$    &$\times~(\text{w})$   &$\Sigma_{c}^{+}(1^{2}P_{\lambda})-pole$    &$0$  &  &$\Sigma_{c}^{+}(1S)-pole$    &$\times~(\text{w})$   &$\Sigma_{c}^{+}(1^{2}P_{\lambda})-pole$    &$\times~(\text{s})$\\
                        &$\Sigma_{c}^{+}(2S)-pole$    &$\times~(\text{w})$   &$\Sigma_{c}^{+}(1^{4}P_{\lambda})-pole$    &$0$  &  &$\Sigma_{c}^{+}(2S)-pole$    &$\times~(\text{w})$   &$\Sigma_{c}^{+}(1^{4}P_{\lambda})-pole$    &$0$\\
                        &$\text{total}$   &$0$   &$\text{total}$   &$0$  &  &$\text{total}$   &$-2.39$   &$\text{total}$   &$1.95$\\
\specialrule{0em}{2pt}{2pt}
\multirow{1}*{\shortstack{$\Omega_{cc}^{+}\to\Sigma_{c}^{+}\eta^{\prime}$}}   &$C^{\prime}-diagram$   &$0$   &$C^{\prime}-diagram$   &$0$  &\multirow{1}*{\shortstack{$\Omega_{cc}^{+}\to\Lambda_{c}^{+}\eta^{\prime}$}}   &$C^{\prime}-diagram$   &$0.082$   &$C^{\prime}-diagram$   &$-0.49$\\
              &$\Omega_{cc}^{+}(1S)-pole$   &$\times~(\text{w})$   &$\Omega_{cc}^{+}(1^{2}P_\rho)-pole$   &$\times~(\text{s})$  &  &$\Omega_{cc}^{+}(1S)-pole$   &$2.86$   &$\Omega_{cc}^{+}(1^{2}P_\rho)-pole$   &$\times~(\text{s})$\\
              &$\Omega_{cc}^{+}(2S)-pole$   &$\times~(\text{w})$   &$\Omega_{cc}^{+}(1^{2}P_\lambda)-pole$   &$0$  &  &$\Omega_{cc}^{+}(2S)-pole$   &$0.14$   &$\Omega_{cc}^{+}(1^{2}P_\lambda)-pole$   &$0.093$\\
              &$\Lambda_{c}^{+}(1S)-pole$   &$\times~(\text{s})$   &$\Omega_{cc}^{+}(1^{4}P_\lambda)-pole$   &$\times~(\text{w})$  &  &$\Lambda_{c}^{+}(1S)-pole$   &$\times~(\text{s})$   &$\Omega_{cc}^{+}(1^{4}P_\lambda)-pole$   &$-1.87$\\
              &$\Lambda_{c}^{+}(2S)-pole$   &$\times~(\text{s})$   &$\Lambda_{c}^{+}(1^{2}P_{\lambda})-pole$   &$\times~(\text{s})$  &  &$\Lambda_{c}^{+}(2S)-pole$   &$\times~(\text{s})$   &$\Lambda_{c}^{+}(1^{2}P_{\lambda})-pole$   &$\times~(\text{s})$\\
              &$\Sigma_{c}^{+}(1S)-pole$    &$\times~(\text{w})$   &$\Sigma_{c}^{+}(1^{2}P_{\lambda})-pole$    &$0$  &  &$\Sigma_{c}^{+}(1S)-pole$    &$\times~(\text{w})$   &$\Sigma_{c}^{+}(1^{2}P_{\lambda})-pole$    &$\times~(\text{s})$\\
              &$\Sigma_{c}^{+}(2S)-pole$    &$\times~(\text{w})$   &$\Sigma_{c}^{+}(1^{4}P_{\lambda})-pole$    &$0$  &  &$\Sigma_{c}^{+}(2S)-pole$    &$\times~(\text{w})$   &$\Sigma_{c}^{+}(1^{4}P_{\lambda})-pole$    &$0$\\
              &$\text{total}$   &$0$   &$\text{total}$   &$0$  &  &$\text{total}$   &$3.08$   &$\text{total}$   &$-2.27$\\
\bottomrule[0.5pt]
\bottomrule[1pt]
\end{tabular*}
\end{table*}

In Table~\ref{tab:PhysicalObservablesComparison}, we compare our results of factorizable and nonfactorizable amplitudes of $\text{PV}$ and $\text{PC}$ amplitudes of CF decays with other theoretical works~\cite{Dhir:2018twm,Gutsche:2018msz,Cheng:2020wmk,Liu:2023dvg,Hu:2024uia}, in which particularly the nonfactorizable contributions from $W$-exchange diagrams were estimated by CCQM~\cite{Gutsche:2018msz}, or by pole model with relevant nonperturbative parameters being derived from MIT bag model~\cite{Cheng:2020wmk} or flavor symmetry~\cite{Dhir:2018twm,Liu:2023dvg}. In addition, the authors of Ref.~\cite{Hu:2024uia} considered long-distance contributions arising from final-state-interaction (FSI) effects. In this table, the decay amplitudes $A$ and $B$ are defined by
\begin{equation}
\mathcal{M}[\mathcal{B}_{i}\to\mathcal{B}_{f}+P]=\bar{u}_{f}(A-B\gamma_{5})u_{i},
\end{equation}
and they can convert to $\mathcal{M}_{\text{PV,PC}}$ by
\begin{equation}
\begin{split}
\mathcal{M}_{\text{PV}}=&\frac{1}{4\pi}\sqrt{\frac{m_{f}}{{\pi}E_{f}E_{P}}}\sqrt{\frac{E_{f}+m_{f}}{2m_{f}}}\chi_{f}^{\dagger}\chi_{i}A,\\
\mathcal{M}_{\text{PC}}=&\frac{1}{4\pi}\sqrt{\frac{m_{f}}{{\pi}E_{f}E_{P}}}\sqrt{\frac{E_{f}+m_{f}}{2m_{f}}}\chi_{f}^{\dagger}\frac{\pmb{\sigma}\cdot\pmb{P}_{f}}{E_{f}+m_{f}}\chi_{i}B.
\end{split}
\end{equation}
Clearly, the decay mode $\Omega_{cc}^{+}\to\Omega_{c}^{0}\pi^{+}$ receives contribution only from $T$-diagram and is therefor purely factorizable. Our computed PV and PC amplitudes are comparable to those of Refs.~\cite{Dhir:2018twm,Cheng:2020wmk,Liu:2023dvg}, expect that the PC amplitude is almost a factor of 2 larger than that of Ref.~\cite{Liu:2023dvg}. The decay mode $\Omega_{cc}^{+}\to\Xi_{c}^{+}\bar{K}^{0}$ receives substantial contributions from nonfactorizable amplitudes. For both the PV or the PC amplitudes, the factorizable and nonfactorizable parts interfere destructively. This is consistent with the results of Refs.~\cite{Gutsche:2018msz,Cheng:2020wmk}. In contrast, the decay mode $\Omega_{cc}^{+}\to\Xi_{c}^{\prime+}\bar{K}^{0}$ has a small nonfactorizable contribution in pole model. However, a theoretical study suggested that the FSI effects may play a significant role in nonfactorizable amplitude of $\Omega_{cc}^{+}\to\Xi_{c}^{\prime+}\bar{K}^{0}$ decay~\cite{Hu:2024uia}. More theoretical analyses are required to verify the nonfactorizable contributions.

\begin{table*}[htbp]\centering
\caption{The comparison of factorizable and nonfactorizable amplitudes (in the units of $10^{-2}G_{F}~\text{GeV}^{-2}$), as well as the branching fractions and asymmetry parameters $\alpha$ of CF decays by various theoretical works. It should be mention that the lifetime of $\Omega_{cc}^{+}$ is chosen as $\tau_{\Omega_{cc}^{+}}=128~\text{fs}$ in this table.}
\label{tab:PhysicalObservablesComparison}
\renewcommand\arraystretch{1.05}
\begin{threeparttable}
\begin{tabular*}{162mm}{ll@{\extracolsep{\fill}}cccccccc}
\toprule[1pt]
\toprule[0.5pt]
\multicolumn{2}{c}{$\Omega_{cc}^{+}\to\Omega_{c}^{0}\pi^{+}~\text{process}$}  &$A^{\text{fac}}$ &$A^{\text{nf}}$ &$A^{\text{tot}}$  &$B^{\text{fac}}$ &$B^{\text{nf}}$ &$B^{\text{tot}}$  &$\mathcal{B}$ &$\alpha$\\
\midrule[0.5pt]
This work&  &$4.78$ &$0$ &$4.78$  &$-71.52$ &$0$ &$-71.52$  &$4.41\%$  &$-0.70$\\
Ref.~\cite{Cheng:2020wmk}&  &$5.71$ &$0$ &$5.71$  &$-67.48$ &$0$ &$-67.48$  &$3.96\%$ &$-0.83$\\
\multirow{2}*{Ref.~\cite{Dhir:2018twm}$^{[a]}$}
&{\footnotesize NRQM}  &$6.25$ &$0$ &$6.25$  &$-81.75$ &$0$ &$-81.75$  &$6.02\%$ &$-0.77$\\
&{\footnotesize HQET}  &$7.32$ &$0$ &$7.32$  &$-95.32$ &$0$ &$-95.32$  &$8.19\%$ &$-0.77$\\
Ref.~\cite{Liu:2023dvg}&  &$-4.11$ &$0$ &$-4.11$  &$32.96$ &$0$ &$32.96$  &$1.33\%$ &$-0.96$\\
Ref.~\cite{Hu:2024uia}&  &$-7.39^{+0.07}_{-0.07}$ &$0$ &$-7.39^{+0.07}_{-0.07}$  &$61.60^{+0.57}_{-0.57}$ &$0$ &$61.60^{+0.57}_{-0.57}$  &$(6.81^{+0.13}_{-0.13})\%$  &$-0.96$\\
\midrule[0.5pt]
\multicolumn{2}{c}{$\Omega_{cc}^{+}\to\Xi_{c}^{+}\bar{K}^{0}~\text{process}$}  &$$ &$$ &$$  &$$ &$$ &$$  &$$ &$$\\
\midrule[0.5pt]
This work&  &$-8.45$ &$19.05$ &$10.60$  &$22.03$ &$-75.88$ &$-53.85$  &$6.11\%$  &$-1.00$\\
Ref.~\cite{Gutsche:2018msz}&  &$-4.02$ &$12.17$ &$8.15$  &$6.20$ &$-19.23$ &$-13.02$  &$1.98\%$ &$-0.54$\\
Ref.~\cite{Cheng:2020wmk}&  &$2.62$ &$-8.90$ &$-6.28$  &$-5.29$ &$13.40$ &$8.11$  &$1.15\%$ &$-0.45$\\
\multirow{4}*{Ref.~\cite{Dhir:2018twm}$^{[a]}$}
&{\footnotesize NRQM+FI}  &$3.43$ &$0$ &$3.43$  &$-8.13$ &$-25.26$ &$-33.39$  &$1.47\%$ &$-0.84$\\
&{\footnotesize HQET+FI}  &$5.58$ &$0$ &$5.58$  &$-11.55$ &$-25.26$ &$-36.81$  &$2.26\%$ &$-0.97$\\
&{\footnotesize NRQM+FD}  &$3.43$ &$0$ &$3.43$  &$-8.13$ &$-53.00$ &$-61.13$  &$4.14\%$ &$-0.54$\\
&{\footnotesize HQET+FD}  &$5.58$ &$0$ &$5.58$  &$-11.55$ &$-53.00$ &$-64.56$  &$5.13\%$ &$-0.75$\\
\multirow{2}*{Ref.~\cite{Hu:2024uia}}&  &\multirow{2}*{$0.12^{+0.00}_{-0.00}$}  &$(2.41^{+1.13}_{-0.98}$  &$(2.53^{+1.13}_{-0.98}$  &\multirow{2}*{$-0.96^{+0.01}_{-0.01}$}  &$(3.36^{+1.24}_{-1.23}$  &$(2.40^{+1.24}_{-1.23}$  &\multirow{2}*{$(1.33^{+1.56}_{-0.89})\%$}  &\multirow{2}*{$-0.54^{+0.01}_{-0.03}$}\\
&  &  &$-4.31^{+1.79}_{-2.14}i)$  &$-4.31^{+1.79}_{-2.14}i)$  &  &$+11.10^{+4.39}_{-5.21}i)$  &$+11.10^{+4.39}_{-5.21}i)$  &  &\\
\midrule[0.5pt]
\multicolumn{2}{c}{$\Omega_{cc}^{+}\to\Xi_{c}^{\prime+}\bar{K}^{0}~\text{process}$}  &$$ &$$ &$$  &$$ &$$ &$$  &$$ &$$\\
\midrule[0.5pt]
This work&  &$4.64$ &$0$ &$4.64$  &$-59.24$ &$0$ &$-59.24$  &$3.21\%$  &$-0.77$\\
Ref.~\cite{Gutsche:2018msz}&  &$2.26$ &$-0.11$ &$2.14$  &$-17.34$ &$0.69$ &$-16.64$  &$0.31\%$ &$-0.97$\\
Ref.~\cite{Cheng:2020wmk}&  &$-1.68$ &$-0.04$ &$-1.72$  &$17.44$ &$0.06$ &$17.50$  &$0.29\%$ &$-0.88$\\
\multirow{2}*{Ref.~\cite{Dhir:2018twm}$^{[a]}$}
&{\footnotesize NRQM}  &$-2.15$ &$0$ &$-2.15$  &$26.9$ &$0$ &$26.9$  &$0.67\%$ &$-0.78$\\
&{\footnotesize HQET}  &$-2.96$ &$0$ &$-2.96$  &$37.6$ &$0$ &$37.6$  &$1.30\%$ &$-0.77$\\
\multirow{2}*{Ref.~\cite{Hu:2024uia}}  &  &\multirow{2}*{$0.08^{+0.00}_{-0.00}$} &$(0.99^{+0.53}_{-0.44}$ &$(1.10^{+0.53}_{-0.44}$  &\multirow{2}*{$-0.68^{+0.01}_{-0.01}$} &$(5.86^{+2.08}_{-2.05}$ &$(4.90^{+2.08}_{-2.05}$  &\multirow{2}*{$(0.27^{+0.28}_{-0.16})\%$}  &\multirow{2}*{$-0.42^{+0.02}_{-0.04}$}\\
&  &  &$-1.63^{+0.64}_{-0.73}i)$  &$-1.63^{+0.64}_{-0.73}i)$  &  &$+8.35^{+3.00}_{-3.39}i)$  &$+8.35^{+3.00}_{-3.39}i)$  &  &\\
\bottomrule[0.5pt]
\bottomrule[1pt]
\end{tabular*}
\begin{tablenotes}
\footnotesize
\item[a] ``NRQM" and ``HQET" represent that the factorizable amplitudes are obtained by nonrelativistic quark model and heavy quark effective theory, respectively, and ``FI" and ``FD" represent that the pole amplitudes are obtained without or with considering the flavor-dependent effects, respectively.
\end{tablenotes}
\end{threeparttable}
\end{table*}

Moving forward, we present a systematic investigation of the branching fractions and up-down decay asymmetry parameters $\alpha$ for two-body nonleptonic decays of $\Omega_{cc}^{+}$. The decay width can be obtained by
\begin{equation}
\Gamma(\mathcal{B}_{i}{\to}\mathcal{B}_{f}P)=\frac{8\pi^{2}}{2J_{i}+1}\frac{kE_{f}E_{P}}{m_{i}}\sum_{s_{z}^{f},s_{z}^{i}}\Big{(}\vert\mathcal{M}_{\text{PC}}^{s_{z}^{f},s_{z}^{i}}\vert^{2}+\vert\mathcal{M}_{\text{PV}}^{s_{z}^{f},s_{z}^{i}}\vert^{2}\Big{)},
\label{eq:Br}
\end{equation}
where $m_{i}$ is the mass of particle $\mathcal{B}_{i}$, $E_{f}$ and $E_{P}$ are the energies of $\mathcal{B}_{f}$ and $P$, respectively, and $J$ is the total spin of initial state $\mathcal{B}_{i}$. At the same time, we also evaluate the asymmetry parameter $\alpha$, defined by
\begin{equation}
\alpha=\frac{2\text{Re}[(\mathcal{M}_{\text{PV}}^{-1/2,-1/2})^{*}\mathcal{M}_{\text{PC}}^{-1/2,-1/2}]}{\vert\mathcal{M}_{\text{PC}}^{-1/2,-1/2}\vert^{2}+\vert\mathcal{M}_{\text{PV}}^{-1/2,-1/2}\vert^{2}}.
\label{eq:alpha}
\end{equation}
We present our computed results together with those from other theoretical studies for a comprehensive comparison in Table~\ref{tab:PhysicalObservables}. In general, the CF modes exhibit larger branching fractions, whereas the SCS and DCS modes are suppresses by CKM matrix elements. Our results indicate that the branching fractions of the CF decays $\Omega_{cc}^{+}\to\Omega_{c}^{0}\pi^{+}$ and $\Xi_{c}^{(\prime)+}\pi^{+}$ reach up to a few percent. Interestingly, several SCS channels---$\Omega_{cc}^{+}\to\Omega_{c}^{0}K^{+}$, $\Sigma_{cc}^{++}K^{-}$, $\Xi_{c}^{\prime+}\eta$, and $\Lambda_{c}^{+}\bar{K}^{0}$---can also reach the few-percent level. Particularly for the $\Omega_{cc}^{+}\to\Omega_{c}^{0}K^{+}$ decay, owing to the large factorizable amplitude and constructive pole contributions, the branching fraction is comparable to that of the CF decays. Besides, the purely $W$-exchanged process $\Omega_{cc}^{+}\to\Sigma_{c}^{++}K^{-}$ has large nonfactorizable contributions from $\Xi_{c}^{+}(nS)$ poles, which is different to the result of Ref.~\cite{Dhir:2018twm}. These decay channels can serve as the discovery channels for the double-charmed baryon $\Omega_{cc}^{+}$ in future experiment like LHCb and Belle II.

\begin{table*}[htbp]\centering
\caption{The branching fractions and up-down decay asymmetry parameters $\alpha$ of two-body nonleptonic decays of double-charmed baryon $\Omega_{cc}^{+}$. It should be noted that the lifetime of $\Omega_{cc}^{+}$ is adopted as $128~\text{fs}$ in this work, and $300~\text{fs}$ in Ref.~\cite{Dhir:2018twm}, and $206~\text{fs}$ in Ref.~\cite{Hu:2024uia}.}
\label{tab:PhysicalObservables}
\renewcommand\arraystretch{1.05}
\begin{tabular*}{162mm}{l@{\extracolsep{\fill}}cccccc}
\toprule[1pt]
\toprule[0.5pt]
&\multicolumn{2}{c}{This work}  &\multicolumn{2}{c}{Ref.~\cite{Dhir:2018twm}} &\multicolumn{2}{c}{Ref.~\cite{Hu:2024uia}}\\
&$\mathcal{B}$  &$\alpha$    &$\mathcal{B}$  &$\alpha$    &$\mathcal{B}$  &$\alpha$\\
\midrule[0.5pt]
CF decays&&\\
$\quad\Omega_{cc}^{+}\to\Omega_{c}^{0}\pi^{+}$          &$4.41\%$  &$-0.70$    
&$(12.6,17.2)\%$ &$(-0.794,-0.800)$
&$(6.81^{+0.13}_{-0.13})\%$  &$-0.96$\\
$\quad\Omega_{cc}^{+}\to\Xi_{c}^{+}\bar{K}^{0}$         &$6.11\%$  &$-1.00$    
&$(3.1,4.9)\%$ &$(-0.855,-0.982)$
&$(1.33^{+1.56}_{-0.89})\%$  &$-0.54^{+0.01}_{-0.03}$\\
$\quad\Omega_{cc}^{+}\to\Xi_{c}^{\prime+}\bar{K}^{0}$   &$3.21\%$  &$-0.77$    
&$(1.4,2.7)\%$ &$(-0.791,-0.789)$
&$(2.74^{+2.83}_{-1.66})\times10^{-3}$  &$-0.42^{+0.02}_{-0.04}$\\
\midrule[0.5pt]
SCS decays&&\\
$\quad\Omega_{cc}^{+}\to\Xi_{c}^{0}\pi^{+}$               &$0.45\times10^{-3}$  &$-0.90$    
&$(1.6,3.7)\times10^{-3}$ &$(0.640,0.198)$
&$(2.99^{+0.34}_{-0.26})\times10^{-3}$  &$-0.34^{+0.07}_{-0.07}$  \\
$\quad\Omega_{cc}^{+}\to\Xi_{c}^{\prime0}\pi^{+}$         &$4.80\times10^{-3}$  &$0.26$    
&$(1.6,3.9)\times10^{-3}$ &$(-0.934,-0.871)$
&$(0.96^{+0.19}_{-0.10})\times10^{-3}$  &$-0.74^{+0.37}_{-0.19}$  \\
$\quad\Omega_{cc}^{+}\to\Omega_{c}^{0}K^{+}$              &$5.09\%$  &$-0.41$    
&$(1.2,1.5)\%$ &$(-0.759,-0.780)$
&$(5.03^{+0.35}_{-0.29})\times10^{-3}$  &$-0.98^{+0.00}_{-0.00}$  \\
$\quad\Omega_{cc}^{+}\to\Sigma_{c}^{++}K^{-}$             &$1.04\%$  &$-0.013$    
&$8.0\times10^{-4}$ &$-$
&$(1.67^{+1.91}_{-1.07})\times10^{-4}$  &$0.33^{+0.18}_{-0.24}$  \\
$\quad\Omega_{cc}^{+}\to\Xi_{c}^{+}\eta$                  &$1.01\times10^{-3}$  &$-0.35$    
&$(0.86,1.7)\times10^{-3}$ &$(-0.997,-0.970)$
&&\\
$\quad\Omega_{cc}^{+}\to\Xi_{c}^{\prime+}\eta$            &$1.62\%$  &$-0.24$    
&$(2.0,3.0)\times10^{-3}$ &$(-0.516,-0.578)$
&&\\
$\quad\Omega_{cc}^{+}\to\Xi_{c}^{+}\eta^{\prime}$         &$1.76\times10^{-3}$  &$0.074$    
&$(8.4,0.14)\times10^{-3}$ &$(0.995,0.892)$
&&\\
$\quad\Omega_{cc}^{+}\to\Xi_{c}^{\prime+}\eta^{\prime}$   &$0.66\times10^{-3}$  &$-0.98$    
&$(4.8,7.8)\times10^{-5}$ &$(-0.856,-0.906)$
&&\\
$\quad\Omega_{cc}^{+}\to\Xi_{c}^{+}\pi^{0}$               &$0.11\times10^{-3}$  &$-0.90$    
&$(4.2,5.3)\times10^{-4}$ &$(0.893,0.992)$
&$(2.65^{+3.03}_{-1.75})\times10^{-4}$  &$0.28^{+0.08}_{-0.07}$\\
$\quad\Omega_{cc}^{+}\to\Xi_{c}^{\prime+}\pi^{0}$         &$2.96\times10^{-3}$  &$0.16$    
&$(0.43,1.3)\times10^{-4}$ &$(-0.320,-0.972)$
&$(3.01^{+3.72}_{-2.05})\times10^{-4}$  &$-0.24^{+0.07}_{-0.07}$\\
$\quad\Omega_{cc}^{+}\to\Sigma_{c}^{+}\bar{K}^{0}$        &$5.20\times10^{-3}$  &$-0.013$    
&$4.0\times10^{-4}$ &$-$
&$(2.09^{+2.43}_{-1.39})\times10^{-4}$  &$-0.37^{+0.04}_{-0.08}$\\
$\quad\Omega_{cc}^{+}\to\Lambda_{c}^{+}\bar{K}^{0}$       &$8.83\times10^{-3}$  &$-0.93$    
&$4.0\times10^{-3}$ &$-$
&$(3.93^{+4.59}_{-2.57})\times10^{-4}$  &$0.34^{+0.15}_{-0.20}$\\
\midrule[0.5pt]
DCS decays&&\\
$\quad\Omega_{cc}^{+}\to\Xi_{c}^{0}K^{+}$                 &$1.47\times10^{-4}$  &$-0.79$    
&$(1.3,3.3)\times10^{-4}$ &$(0.100,-0.162)$
&$(0.17^{+0.01}_{-0.01})\times10^{-3}$  &$-0.50^{+0.01}_{-0.01}$\\
$\quad\Omega_{cc}^{+}\to\Xi_{c}^{\prime0}K^{+}$           &$1.01\times10^{-3}$  &$-0.37$    
&$(3.3,6.0)\times10^{-4}$ &$(-0.684,-0.711)$
&$(0.13^{+0.01}_{-0.00})\times10^{-3}$  &$-0.85^{+0.10}_{-0.08}$\\
$\quad\Omega_{cc}^{+}\to\Xi_{c}^{+}K^{0}$                 &$0.71\times10^{-4}$  &$-0.79$    
&$(3.2,6.1)\times10^{-5}$ &$(0.962,0.634)$
&$(2.10^{+2.05}_{-1.27})\times10^{-5}$  &$-0.32^{+0.08}_{-0.07}$\\
$\quad\Omega_{cc}^{+}\to\Xi_{c}^{\prime+}K^{0}$           &$7.72\times10^{-4}$  &$-0.29$    
&$(0.85,1.4)\times10^{-4}$ &$(-0.564,-0.615)$
&$(1.32^{+1.70}_{-0.89})\times10^{-5}$  &$0.25^{+0.17}_{-0.25}$\\
$\quad\Omega_{cc}^{+}\to\Sigma_{c}^{+}\pi^{0}$            &$8.46\times10^{-4}$  &$-0.012$    
&$3.7\times10^{-5}$ &$-$
&$(0.88^{+1.05}_{-0.59})\times10^{-5}$  &$0.93^{+0.02}_{-0.09}$\\
$\quad\Omega_{cc}^{+}\to\Lambda_{c}^{+}\pi^{0}$           &$0$  &$-$    
&&
&$(0.43^{+0.46}_{-0.27})\times10^{-5}$  &$0.17^{+0.26}_{-0.30}$\\
$\quad\Omega_{cc}^{+}\to\Sigma_{c}^{0}\pi^{+}$            &$8.46\times10^{-4}$  &$-0.012$    
&$3.7\times10^{-5}$ &$-$
&$(1.34^{+1.55}_{-0.90})\times10^{-5}$  &$0.99^{+0.01}_{-0.01}$\\
$\quad\Omega_{cc}^{+}\to\Sigma_{c}^{++}\pi^{-}$           &$8.46\times10^{-4}$  &$-0.012$    
&$3.7\times10^{-5}$ &$-$
&$(0.69^{+0.83}_{-0.45})\times10^{-5}$  &$0.49^{+0.15}_{-0.24}$\\
$\quad\Omega_{cc}^{+}\to\Sigma_{c}^{+}\eta$               &$0$  &$-$    
&&
&&\\
$\quad\Omega_{cc}^{+}\to\Lambda_{c}^{+}\eta$              &$1.78\times10^{-4}$  &$-0.98$    
&$1.2\times10^{-4}$ &$-$
&&\\
$\quad\Omega_{cc}^{+}\to\Sigma_{c}^{+}\eta^{\prime}$      &$0$  &$-$    
&&  
&&\\
$\quad\Omega_{cc}^{+}\to\Lambda_{c}^{+}\eta^{\prime}$     &$2.29\times10^{-4}$  &$-0.96$    
&$4.2\times10^{-7}$ &$-$
&&\\
\bottomrule[0.5pt]
\bottomrule[1pt]
\end{tabular*}
\end{table*}

According to previous theoretical studies of meson~\cite{Li:1996cj,Cheng:2002wu,Cheng:2004ru,Lu:2005mx,Cheng:2010ry,Wang:2025mdn} and baryon~\cite{Chen:2002jr,Yu:2017zst,Jiang:2018oak,Ke:2020uks,Jia:2024pyb,Cao:2025kvs,Hu:2024uia,Hu:2025pjg,Shang:2026knt,Feng:2026soj} nonleptonic decays, the FSI effects, as an important source of long-distance contribution, may be non-negligible. Particularly for our concerned $\Omega_{cc}\to\mathcal{B}_{c}P$ processes, the FSI effects have been systematically considered in Ref.~\cite{Hu:2024uia}. In their theoretical framework, the factorizable amplitudes arising from $T$ and $C$ diagrams were computed using the factorizable method, while the nonfactorizable long-distance contributions were evaluated by considering the rescattering processes $\Omega_{cc}\to\mathcal{B}_{c}^{\prime}P^{\prime}(V)\to\mathcal{B}_{c}P$. For the channels that are contributed by $T$ diagram, the factorizable contributions are dominated, and the branching fractions of long-distance contributions are suppressed by $10^{-2}\sim10^{-1}$ in magnitude, while for other channels lacking $T$ diagram, the long-distance contributions are dominated. In our computation, the decay amplitudes of $\Omega_{cc}^{+}\to\Lambda_{c}^{+}\pi^{0}$ and $\Sigma_{c}^{+}\eta^{(\prime)}$ are vanishing in the framework of pole model. However, after considering the long-distance contributions from FSI effects, they predicted that these channels will have considerable branching fractions with an order of magnitude of $10^{-5}$~\cite{Hu:2024uia}. Future experimental measurements of these channels will contribute to reveal the long-distance contributions in weak decay mechanisms of charmed baryons.

\section{Conclusion}
\label{sec4}

The discovery of double-charmed baryons is a forefront research topic in heavy-flavor physics. As their ground states are prohibited from decaying via strong and electromagnetic interactions while can only decay via weak interaction, a comprehensive theoretical investigation of their weak decay modes is essential for guiding future experimental searches. However, due to the nonperturbative nature of low-energy QCD, and the complex dynamics in weak decays particularly the nonfactorizable contributions, the theoretical study of nonleptonic decays remains challenging.

In this paper, we present a theoretical study of the two-body nonleptonic decays $\Omega_{cc}^{+}\to\mathcal{B}_{c}+P$, where $\mathcal{B}_{c}$ denotes a single-charmed baryon and $P$ denotes light-flavor pseudoscalar meson such as $\pi$, $K$ or $\eta^{(\prime)}$. The relevant decay amplitudes are calculated within the framework of nonrelativistic quark model, and in particular the nonfactorizable amplitudes contributed from $W$-exchange diagrams are evaluated under pole model assumption. Moreover, in order to reduce the dependence of decay amplitudes on phenomenological parameters, especially those in the baryon wave functions, we adopted the charmed-baryon wave functions obtained from a nonrelativistic quark potential model.

With the obtained decay amplitudes, we future investigate the corresponding branching fractions. In our calculation, the branching fractions of CF decays $\Omega_{cc}^{+}\to\Omega_{c}^{0}\pi^{+}$ and $\Xi_{c}^{(\prime)+}\pi^{+}$, together with several SCS decays $\Omega_{cc}^{+}\to\Omega_{c}^{0}K^{+}$, $\Sigma_{cc}^{++}K^{-}$, $\Xi_{c}^{\prime+}\eta$ and $\Lambda_{c}^{+}\bar{K}^{0}$, can reach up to a few percents. Notably, the branching fraction of SCS mode $\Omega_{cc}^{+}\to\Omega_{c}^{0}K^{+}$ is comparable to the ones of CF decays, owing to large constructive pole contributions. These decay channels can serve as the discovery channels for the double-charmed baryon $\Omega_{cc}^{+}$ in future experiment such as LHCb and Belle II.

\section*{ACKNOWLEDGMENTS}

This work is supported by the National Natural Science Foundation of China under Grant No. 12447155, and by the Postdoctoral Fellowship Program of CPSF under Grant No. 2025M773368 and under Grant No. GZC20240056.

\begin{widetext}
\appendix

\section{The spin-flavor matrix elements in $W$-emission processes}
\label{app01}

\setcounter{equation}{0}
\renewcommand{\theequation}{A.\arabic{equation}}

\setcounter{table}{0}
\renewcommand{\thetable}{A\Roman{table}}

In this appendix, the spin-flavor matrix elements in $W$-emission processes are presented. Before calculating the concerned values, we give the spin-flavor wave functions of concerned charmed baryons. The spin-flavor wave functions of $S$-wave double-charmed baryons are chosen as~\cite{Perez-Marcial:1989sch}
\begin{equation}
\begin{split}
\vert\mathcal{B}_{ccq}\rangle=-\frac{1}{\sqrt{3}}\big{[}ccq\chi_{S}+(13)+(23)\big{]},
\end{split}
\end{equation}
and that of $S$-wave single-charmed baryons are chosen as
\begin{equation}
\begin{split}
\vert\Omega_{c}^{0}\rangle=&\frac{1}{\sqrt{3}}\big{[}ssc\chi_{S}+(13)+(23)\big{]},~~
\vert\Lambda_{c}^{+}\rangle=\frac{1}{\sqrt{6}}\big{[}(udc-duc)\chi_{A}+(13)+(23)\big{]},\\
\vert\Xi_{c}^{+}\rangle=&\frac{1}{\sqrt{6}}\big{[}(usc-suc)\chi_{A}+(13)+(23)\big{]},~~
\vert\Xi_{c}^{\prime+}\rangle=\frac{1}{\sqrt{6}}\big{[}(usc+suc)\chi_{S}+(13)+(23)\big{]},\\
\vert\Xi_{c}^{0}\rangle=&\frac{1}{\sqrt{6}}\big{[}(dsc-sdc)\chi_{A}+(13)+(23)\big{]},~~
\vert\Xi_{c}^{\prime0}\rangle=\frac{1}{\sqrt{6}}\big{[}(dsc+sdc)\chi_{S}+(13)+(23)\big{]},\\
\vert\Sigma_{c}^{++}\rangle=&\frac{1}{\sqrt{3}}\big{[}uuc\chi_{S}+(13)+(23)\big{]},~~
\vert\Sigma_{c}^{+}\rangle=\frac{1}{\sqrt{6}}\big{[}(udc+duc)\chi_{S}+(13)+(23)\big{]},\\
\vert\Sigma_{c}^{0}\rangle=&\frac{1}{\sqrt{3}}\big{[}ddc\chi_{S}+(13)+(23)\big{]}.
\end{split}
\end{equation}
Here, we have
\begin{equation}
\begin{split}
abc\chi_{S}^{\downarrow}=&\frac{1}{\sqrt{6}}(-2a^{\downarrow}b^{\downarrow}c^{\uparrow}+a^{\uparrow}b^{\downarrow}c^{\downarrow}+a^{\downarrow}b^{\uparrow}c^{\downarrow}),\\
abc\chi_{A}^{\downarrow}=&\frac{1}{\sqrt{2}}(a^{\uparrow}b^{\downarrow}c^{\downarrow}-a^{\downarrow}b^{\uparrow}c^{\downarrow}),
\end{split}
\end{equation}
for the baryons of $\vert{S,S_{z}}\rangle=\vert{1/2,-1/2}\rangle$, and
\begin{equation}
\begin{split}
abc\chi_{S}^{\uparrow}=&\frac{1}{\sqrt{6}}(2a^{\uparrow}b^{\uparrow}c^{\downarrow}-a^{\uparrow}b^{\downarrow}c^{\uparrow}-a^{\downarrow}b^{\uparrow}c^{\uparrow}),\\
abc\chi_{A}^{\uparrow}=&\frac{1}{\sqrt{2}}(a^{\uparrow}b^{\downarrow}c^{\uparrow}-a^{\downarrow}b^{\uparrow}c^{\uparrow}),
\end{split}
\end{equation}
for that of $\vert{S,S_{z}}\rangle=\vert{1/2,+1/2}\rangle$. Besides, the spin wave function of pseudoscalar meson is adopted as $\chi_{P}=\frac{1}{\sqrt{2}}(\uparrow\downarrow-\downarrow\uparrow)$. For calculating the spin matrix element, the particle-hole conjugation $\langle{j,-m}\vert\to{(-1)^{j+m}}\vert{j,m}\rangle$ is useful. This relation makes the antiquark spin transform as $\langle\bar{\uparrow}\vert{\to}\vert{\downarrow}\rangle$ and $\langle\bar{\downarrow}\vert\to -\vert{\uparrow}\rangle$.

The spin-flavor matrix elements $\langle{\mathcal{B}_{f}}{\downarrow}\vert\hat{\mathcal{O}}\vert{\mathcal{B}_{i}}{\downarrow}\rangle$ associated with $W$-emission processes are presented in Tables~\ref{tab:Tfactor}-\ref{tab:Cprimfactor}, which contribute to the PC and PV amplitudes in the $T$, $C$ and $C^{\prime}$ diagrams, respectively. The values of $\langle{\mathcal{B}_{f}}{\uparrow}\vert\hat{\mathcal{O}}\vert{\mathcal{B}_{i}}{\uparrow}\rangle$, $\langle{\mathcal{B}_{f}}{\uparrow}\vert\hat{\mathcal{O}}\vert{\mathcal{B}_{i}}{\downarrow}\rangle$ and $\langle{\mathcal{B}_{f}}{\downarrow}\vert\hat{\mathcal{O}}\vert{\mathcal{B}_{i}}{\uparrow}\rangle$ can also be obtained.

\begin{table*}[htbp]
\centering
\caption{The spin-flavor matrix elements of PC (up panel) and PV (bottom panel) amplitudes for $T$ diagrams. For simplicity, the spin and flavor wave functions of pseudoscalar meson is omitted here.}
\label{tab:Tfactor}
\renewcommand\arraystretch{1.05}
\begin{tabular*}{100mm}{c@{\extracolsep{\fill}}ccc}
\toprule[1pt]
\toprule[0.5pt]
\specialrule{0em}{0.5pt}{0.5pt}
matrix elements   &$\langle{\Omega_{c}^{0}}{\downarrow}\vert\hat{\mathcal{O}}\vert{\Omega_{cc}^{+}}{\downarrow}\rangle$  &$\langle{\Xi_{c}^{0}}{\downarrow}\vert\hat{\mathcal{O}}\vert{\Omega_{cc}^{+}}{\downarrow}\rangle$   &$\langle{\Xi_{c}^{\prime0}}{\downarrow}\vert\hat{\mathcal{O}}\vert{\Omega_{cc}^{+}}{\downarrow}\rangle$\\
\specialrule{0em}{0.5pt}{0.5pt}
\midrule[0.5pt]
$\langle{s_{z,1}^{\prime}}\vert{I}\vert{s_{z,1}}\rangle\langle{s_{z,5}\bar{s}_{z,4}}\vert{\sigma_{z}}\vert{0}\rangle$  &$0$  &$0$  &$0$\\
$\langle{s_{z,1}^{\prime}}\vert{\sigma_{z}}\vert{s_{z,1}}\rangle\langle{s_{z,5}\bar{s}_{z,4}}\vert{I}\vert{0}\rangle$  &$\frac{5\sqrt{2}}{9}$  &$-\frac{\sqrt{3}}{9}$  &$\frac{5}{9}$\\
$(\langle{s_{z,1}^{\prime}}\vert{\pmb{\sigma}}\vert{s_{z,1}}\rangle\times\langle{s_{z,5}\bar{s}_{z,4}}\vert{\pmb{\sigma}}\vert{0}\rangle)_{z}$  &$0$  &$0$  &$0$\\
\specialrule{0em}{1pt}{1pt}
$\langle{s_{z,5}}\vert{I}\vert{s_{z,1}}\rangle\langle{s_{z,1}^{\prime}\bar{s}_{z,4}}\vert{\sigma_{z}}\vert{0}\rangle$  &$\frac{5\sqrt{2}}{18}$  &$-\frac{\sqrt{3}}{18}$  &$\frac{5}{18}$\\
$\langle{s_{z,5}}\vert{\sigma_{z}}\vert{s_{z,1}}\rangle\langle{s_{z,1}^{\prime}\bar{s}_{z,4}}\vert{I}\vert{0}\rangle$  &$\frac{5\sqrt{2}}{18}$  &$-\frac{\sqrt{3}}{18}$  &$\frac{5}{18}$\\
$(\langle{s_{z,5}}\vert{\pmb{\sigma}}\vert{s_{z,1}}\rangle\times\langle{s_{z,1}^{\prime}\bar{s}_{z,4}}\vert{\pmb{\sigma}}\vert{0}\rangle)_{z}$  &$-\frac{5\sqrt{2}i}{9}$  &$\frac{\sqrt{3}i}{9}$  &$-\frac{5i}{9}$\\
\specialrule{0em}{0.5pt}{0.5pt}
\midrule[0.5pt]
\specialrule{0em}{0.5pt}{0.5pt}
$\langle{s_{z,1}^{\prime}}\vert{I}\vert{s_{z,1}}\rangle\langle{s_{z,5}\bar{s}_{z,4}}\vert{I}\vert{0}\rangle$  &$-\frac{\sqrt{2}}{3}$  &$\frac{\sqrt{3}}{3}$  &$-\frac{1}{3}$\\
$\langle{s_{z,1}^{\prime}}\vert{\sigma_{x}}\vert{s_{z,1}}\rangle\langle{s_{z,5}\bar{s}_{z,4}}\vert{\sigma_{x}}\vert{0}\rangle$  &$0$  &$0$  &$0$\\
$\langle{s_{z,1}^{\prime}}\vert{\sigma_{y}}\vert{s_{z,1}}\rangle\langle{s_{z,5}\bar{s}_{z,4}}\vert{\sigma_{y}}\vert{0}\rangle$  &$0$  &$0$  &$0$\\
$\langle{s_{z,1}^{\prime}}\vert{\sigma_{z}}\vert{s_{z,1}}\rangle\langle{s_{z,5}\bar{s}_{z,4}}\vert{\sigma_{z}}\vert{0}\rangle$  &$0$  &$0$  &$0$\\
\specialrule{0em}{1pt}{1pt}
$\langle{s_{z,5}}\vert{I}\vert{s_{z,1}}\rangle\langle{s_{z,1}^{\prime}\bar{s}_{z,4}}\vert{I}\vert{0}\rangle$  &$-\frac{\sqrt{2}}{6}$  &$\frac{\sqrt{3}}{6}$  &$-\frac{1}{6}$\\
$\langle{s_{z,5}}\vert{\sigma_{x}}\vert{s_{z,1}}\rangle\langle{s_{z,1}^{\prime}\bar{s}_{z,4}}\vert{\sigma_{x}}\vert{0}\rangle$  &$-\frac{\sqrt{2}}{6}$  &$\frac{\sqrt{3}}{6}$  &$-\frac{1}{6}$\\
$\langle{s_{z,5}}\vert{\sigma_{y}}\vert{s_{z,1}}\rangle\langle{s_{z,1}^{\prime}\bar{s}_{z,4}}\vert{\sigma_{y}}\vert{0}\rangle$  &$-\frac{\sqrt{2}}{6}$  &$\frac{\sqrt{3}}{6}$  &$-\frac{1}{6}$\\
$\langle{s_{z,5}}\vert{\sigma_{z}}\vert{s_{z,1}}\rangle\langle{s_{z,1}^{\prime}\bar{s}_{z,4}}\vert{\sigma_{z}}\vert{0}\rangle$  &$-\frac{\sqrt{2}}{6}$  &$\frac{\sqrt{3}}{6}$  &$-\frac{1}{6}$\\
\bottomrule[0.5pt]
\bottomrule[1pt]
\end{tabular*}
\end{table*}

\begin{table*}[htbp]
\centering
\caption{Same as Table~\ref{tab:Tfactor} but for $C$ diagrams.}
\label{tab:Cfactor}
\renewcommand\arraystretch{1.05}
\begin{tabular*}{100mm}{c@{\extracolsep{\fill}}cc}
\toprule[1pt]
\toprule[0.5pt]
\specialrule{0em}{0.5pt}{0.5pt}
matrix elements   &$\langle{\Xi_{c}^{+}}{\downarrow}\vert\hat{\mathcal{O}}\vert{\Omega_{cc}^{+}}{\downarrow}\rangle$   &$\langle{\Xi_{c}^{\prime+}}{\downarrow}\vert\hat{\mathcal{O}}\vert{\Omega_{cc}^{+}}{\downarrow}\rangle$\\
\specialrule{0em}{0.5pt}{0.5pt}
\midrule[0.5pt]
$\langle{s_{z,1}^{\prime}}\vert{I}\vert{s_{z,1}}\rangle\langle{s_{z,5}\bar{s}_{z,4}}\vert{\sigma_{z}}\vert{0}\rangle$   &$-\frac{\sqrt{3}}{18}$  &$\frac{5}{18}$\\
$\langle{s_{z,1}^{\prime}}\vert{\sigma_{z}}\vert{s_{z,1}}\rangle\langle{s_{z,5}\bar{s}_{z,4}}\vert{I}\vert{0}\rangle$  &$-\frac{\sqrt{3}}{18}$  &$\frac{5}{18}$\\
$(\langle{s_{z,1}^{\prime}}\vert{\pmb{\sigma}}\vert{s_{z,1}}\rangle\times\langle{s_{z,5}\bar{s}_{z,4}}\vert{\pmb{\sigma}}\vert{0}\rangle)_{z}$  &$\frac{\sqrt{3}i}{9}$  &$-\frac{5i}{9}$\\
\specialrule{0em}{1pt}{1pt}
$\langle{s_{z,5}}\vert{I}\vert{s_{z,1}}\rangle\langle{s_{z,1}^{\prime}\bar{s}_{z,4}}\vert{\sigma_{z}}\vert{0}\rangle$  &$0$  &$0$\\
$\langle{s_{z,5}}\vert{\sigma_{z}}\vert{s_{z,1}}\rangle\langle{s_{z,1}^{\prime}\bar{s}_{z,4}}\vert{I}\vert{0}\rangle$  &$-\frac{\sqrt{3}}{9}$  &$\frac{5}{9}$\\
$(\langle{s_{z,5}}\vert{\pmb{\sigma}}\vert{s_{z,1}}\rangle\times\langle{s_{z,1}^{\prime}\bar{s}_{z,4}}\vert{\pmb{\sigma}}\vert{0}\rangle)_{z}$  &$0$  &$0$\\
\specialrule{0em}{0.5pt}{0.5pt}
\midrule[0.5pt]
\specialrule{0em}{0.5pt}{0.5pt}
$\langle{s_{z,1}^{\prime}}\vert{I}\vert{s_{z,1}}\rangle\langle{s_{z,5}\bar{s}_{z,4}}\vert{I}\vert{0}\rangle$  &$\frac{\sqrt{3}}{6}$  &$-\frac{1}{6}$\\
$\langle{s_{z,1}^{\prime}}\vert{\sigma_{x}}\vert{s_{z,1}}\rangle\langle{s_{z,5}\bar{s}_{z,4}}\vert{\sigma_{x}}\vert{0}\rangle$  &$\frac{\sqrt{3}}{6}$  &$-\frac{1}{6}$\\
$\langle{s_{z,1}^{\prime}}\vert{\sigma_{y}}\vert{s_{z,1}}\rangle\langle{s_{z,5}\bar{s}_{z,4}}\vert{\sigma_{y}}\vert{0}\rangle$  &$\frac{\sqrt{3}}{6}$  &$-\frac{1}{6}$\\
$\langle{s_{z,1}^{\prime}}\vert{\sigma_{z}}\vert{s_{z,1}}\rangle\langle{s_{z,5}\bar{s}_{z,4}}\vert{\sigma_{z}}\vert{0}\rangle$  &$\frac{\sqrt{3}}{6}$  &$-\frac{1}{6}$\\
\specialrule{0em}{1pt}{1pt}
$\langle{s_{z,5}}\vert{I}\vert{s_{z,1}}\rangle\langle{s_{z,1}^{\prime}\bar{s}_{z,4}}\vert{I}\vert{0}\rangle$  &$\frac{\sqrt{3}}{3}$  &$-\frac{1}{3}$\\
$\langle{s_{z,5}}\vert{\sigma_{x}}\vert{s_{z,1}}\rangle\langle{s_{z,1}^{\prime}\bar{s}_{z,4}}\vert{\sigma_{x}}\vert{0}\rangle$  &$0$  &$0$\\
$\langle{s_{z,5}}\vert{\sigma_{y}}\vert{s_{z,1}}\rangle\langle{s_{z,1}^{\prime}\bar{s}_{z,4}}\vert{\sigma_{y}}\vert{0}\rangle$  &$0$  &$0$\\
$\langle{s_{z,5}}\vert{\sigma_{z}}\vert{s_{z,1}}\rangle\langle{s_{z,1}^{\prime}\bar{s}_{z,4}}\vert{\sigma_{z}}\vert{0}\rangle$  &$0$  &$0$\\
\bottomrule[0.5pt]
\bottomrule[1pt]
\end{tabular*}
\end{table*}

\begin{table*}[htbp]
\centering
\caption{Same as Table~\ref{tab:Tfactor} but for $C^{\prime}$ diagrams.}
\label{tab:Cprimfactor}
\renewcommand\arraystretch{1.05}
\begin{tabular*}{120mm}{c@{\extracolsep{\fill}}cccc}
\toprule[1pt]
\toprule[0.5pt]
\specialrule{0em}{0.5pt}{0.5pt}
matrix elements &$\langle{\Xi_{c}^{+}}{\downarrow}\vert\hat{\mathcal{O}}\vert{\Omega_{cc}^{+}}{\downarrow}\rangle$  &$\langle{\Xi_{c}^{\prime+}}{\downarrow}\vert\hat{\mathcal{O}}\vert{\Omega_{cc}^{+}}{\downarrow}\rangle$  &$\langle{\Sigma_{c}^{+}}{\downarrow}\vert\hat{\mathcal{O}}\vert{\Omega_{cc}^{+}}{\downarrow}\rangle$  &$\langle{\Lambda_{c}^{+}}{\downarrow}\vert\hat{\mathcal{O}}\vert{\Omega_{cc}^{+}}{\downarrow}\rangle$\\
\specialrule{0em}{0.5pt}{0.5pt}
\midrule[0.5pt]
$\langle{s_{z,1}^{\prime}}\vert{I}\vert{s_{z,1}}\rangle\langle{s_{z,5}\bar{s}_{z,4}}\vert{\sigma_{z}}\vert{0}\rangle$  &$-\frac{\sqrt{3}}{36}$  &$-\frac{1}{36}$  &$-\frac{1}{36}$  &$-\frac{\sqrt{3}}{36}$\\
$\langle{s_{z,1}^{\prime}}\vert{\sigma_{z}}\vert{s_{z,1}}\rangle\langle{s_{z,5}\bar{s}_{z,4}}\vert{I}\vert{0}\rangle$  &$\frac{\sqrt{3}}{36}$  &$\frac{5}{36}$  &$\frac{5}{36}$  &$\frac{\sqrt{3}}{36}$\\
$(\langle{s_{z,1}^{\prime}}\vert{\pmb{\sigma}}\vert{s_{z,1}}\rangle\times\langle{s_{z,5}\bar{s}_{z,4}}\vert{\pmb{\sigma}}\vert{0}\rangle)_{z}$  &$\frac{\sqrt{3}i}{18}$  &$-\frac{i}{6}$  &$-\frac{i}{6}$  &$\frac{\sqrt{3}i}{18}$\\
\specialrule{0em}{1pt}{1pt}
$\langle{s_{z,5}}\vert{I}\vert{s_{z,1}}\rangle\langle{s_{z,1}^{\prime}\bar{s}_{z,4}}\vert{\sigma_{z}}\vert{0}\rangle$  &$\frac{\sqrt{3}}{36}$  &$-\frac{1}{36}$  &$-\frac{1}{36}$  &$\frac{\sqrt{3}}{36}$\\
$\langle{s_{z,5}}\vert{\sigma_{z}}\vert{s_{z,1}}\rangle\langle{s_{z,1}^{\prime}\bar{s}_{z,4}}\vert{I}\vert{0}\rangle$  &$-\frac{\sqrt{3}}{36}$  &$\frac{5}{36}$  &$\frac{5}{36}$  &$-\frac{\sqrt{3}}{36}$\\
$(\langle{s_{z,5}}\vert{\pmb{\sigma}}\vert{s_{z,1}}\rangle\times\langle{s_{z,1}^{\prime}\bar{s}_{z,4}}\vert{\pmb{\sigma}}\vert{0}\rangle)_{z}$  &$-\frac{\sqrt{3}i}{18}$  &$-\frac{i}{6}$  &$-\frac{i}{6}$  &$-\frac{\sqrt{3}i}{18}$\\
\specialrule{0em}{0.5pt}{0.5pt}
\midrule[0.5pt]
\specialrule{0em}{0.5pt}{0.5pt}
$\langle{s_{z,1}^{\prime}}\vert{I}\vert{s_{z,1}}\rangle\langle{s_{z,5}\bar{s}_{z,4}}\vert{I}\vert{0}\rangle$  &$-\frac{\sqrt{3}}{12}$  &$-\frac{1}{12}$  &$-\frac{1}{12}$  &$-\frac{\sqrt{3}}{12}$\\
$\langle{s_{z,1}^{\prime}}\vert{\sigma_{x}}\vert{s_{z,1}}\rangle\langle{s_{z,5}\bar{s}_{z,4}}\vert{\sigma_{x}}\vert{0}\rangle$  &$\frac{\sqrt{3}}{12}$  &$-\frac{1}{36}$  &$-\frac{1}{36}$  &$\frac{\sqrt{3}}{12}$\\
$\langle{s_{z,1}^{\prime}}\vert{\sigma_{y}}\vert{s_{z,1}}\rangle\langle{s_{z,5}\bar{s}_{z,4}}\vert{\sigma_{y}}\vert{0}\rangle$  &$\frac{\sqrt{3}}{12}$  &$-\frac{1}{36}$  &$-\frac{1}{36}$  &$\frac{\sqrt{3}}{12}$\\
$\langle{s_{z,1}^{\prime}}\vert{\sigma_{z}}\vert{s_{z,1}}\rangle\langle{s_{z,5}\bar{s}_{z,4}}\vert{\sigma_{z}}\vert{0}\rangle$  &$\frac{\sqrt{3}}{12}$  &$-\frac{1}{36}$  &$-\frac{1}{36}$  &$\frac{\sqrt{3}}{12}$\\
\specialrule{0em}{1pt}{1pt}
$\langle{s_{z,5}}\vert{I}\vert{s_{z,1}}\rangle\langle{s_{z,1}^{\prime}\bar{s}_{z,4}}\vert{I}\vert{0}\rangle$  &$\frac{\sqrt{3}}{12}$  &$-\frac{1}{12}$  &$-\frac{1}{12}$  &$\frac{\sqrt{3}}{12}$\\
$\langle{s_{z,5}}\vert{\sigma_{x}}\vert{s_{z,1}}\rangle\langle{s_{z,1}^{\prime}\bar{s}_{z,4}}\vert{\sigma_{x}}\vert{0}\rangle$  &$-\frac{\sqrt{3}}{12}$  &$-\frac{1}{36}$  &$-\frac{1}{36}$  &$-\frac{\sqrt{3}}{12}$\\
$\langle{s_{z,5}}\vert{\sigma_{y}}\vert{s_{z,1}}\rangle\langle{s_{z,1}^{\prime}\bar{s}_{z,4}}\vert{\sigma_{y}}\vert{0}\rangle$  &$-\frac{\sqrt{3}}{12}$  &$-\frac{1}{36}$  &$-\frac{1}{36}$  &$-\frac{\sqrt{3}}{12}$\\
$\langle{s_{z,5}}\vert{\sigma_{z}}\vert{s_{z,1}}\rangle\langle{s_{z,1}^{\prime}\bar{s}_{z,4}}\vert{\sigma_{z}}\vert{0}\rangle$  &$-\frac{\sqrt{3}}{12}$  &$-\frac{1}{36}$  &$-\frac{1}{36}$  &$-\frac{\sqrt{3}}{12}$\\
\bottomrule[0.5pt]
\bottomrule[1pt]
\end{tabular*}
\end{table*}

\section{The spin-flavor matrix elements in $W$-exchange processes}
\label{app02}

\setcounter{equation}{0}
\renewcommand{\theequation}{B.\arabic{equation}}

\setcounter{table}{0}
\renewcommand{\thetable}{B\Roman{table}}

In this appendix, the spin-flavor matrix elements in $W$-exchange processes are presented. For the evaluation of these spin-flavor matrix elements, we employed the following spin-flavor wave functions for excited baryons. For $\lambda$-mode excited charmed baryons with total spin $\vert{S,S_{z}}\rangle=\vert{1/2,\pm1/2}\rangle$, the spin-flavor wave functions $\vert{\mathcal{B}}\rangle_{{}^{2}{\lambda}}$ are taken to be the same as those of the $S$-wave states. In addition, for $\lambda$-mode excited double-charmed baryons with total spin $\vert{S,S_{z}}\rangle=\vert{3/2,\pm1/2}\rangle$, we adopted as
\begin{equation}
\vert\mathcal{B}_{ccq}\rangle_{{}^{4}\lambda}=-\frac{1}{\sqrt{3}}\big{[}ccq\chi_{s}+(13)+(23)\big{]},
\end{equation}
while for $\lambda$-mode excited single-charmed baryons we adopted as
\begin{equation}
\begin{split}
\vert\Omega_{c}^{0}\rangle_{{}^{4}\lambda}=&\frac{1}{\sqrt{3}}\big{[}ssc\chi_{s}+(13)+(23)\big{]},~~
\vert\Sigma_{c}^{++}\rangle_{{}^{4}\lambda}=\frac{1}{\sqrt{3}}\big{[}uuc\chi_{s}+(13)+(23)\big{]}\\
\vert\Sigma_{c}^{+}\rangle_{{}^{4}\lambda}=&\frac{1}{\sqrt{6}}\big{[}(udc+duc)\chi_{s}+(13)+(23)\big{]},~~
\vert\Sigma_{c}^{0}\rangle_{{}^{4}\lambda}=\frac{1}{\sqrt{3}}\big{[}ddc\chi_{s}+(13)+(23)\big{]}\\
\vert\Xi_{c}^{\prime+}\rangle_{{}^{4}\lambda}=&\frac{1}{\sqrt{6}}\big{[}(usc+suc)\chi_{s}+(13)+(23)\big{]},
\vert\Xi_{c}^{\prime0}\rangle_{{}^{4}\lambda}=\frac{1}{\sqrt{6}}\big{[}(dsc+sdc)\chi_{s}+(13)+(23)\big{]},
\end{split}
\end{equation}
where
\begin{equation}
\begin{split}
abc\chi_{s}^{\downarrow}=&\frac{1}{\sqrt{3}}(a^{\downarrow}b^{\downarrow}c^{\uparrow}+a^{\uparrow}b^{\downarrow}c^{\downarrow}+a^{\downarrow}b^{\uparrow}c^{\downarrow}),\\
abc\chi_{s}^{\uparrow}=&\frac{1}{\sqrt{3}}(a^{\downarrow}b^{\uparrow}c^{\uparrow}+a^{\uparrow}b^{\uparrow}c^{\downarrow}+a^{\uparrow}b^{\downarrow}c^{\uparrow}).
\end{split}
\end{equation}
Besides, the spin-flavor wave functions of $\rho$-mode excited double-charmed baryons with total spin $\vert{S,S_{z}}\rangle=\vert{1/2,\pm1/2}\rangle$ can be written as
\begin{equation}
\vert\mathcal{B}_{ccq}\rangle_{{}^{2}\rho}=-\frac{1}{\sqrt{3}}\big{[}ccq\chi_{A}+(13)+(23)\big{]}.
\end{equation}

With the above preparations, the spin-flavor matrix elements for PC and PV amplitudes of $W$-exchange diagrams can be obtained. The concerned values are summarized in Tables~\ref{tab:polefactorPC} and \ref{tab:polefactorPV} for PC and PV amplitudes, respectively. Besides, the matrix elements $\langle{\mathcal{B}^{\prime}{\downarrow}}\vert\hat{I}_{P}\sigma_{z}\vert{\mathcal{B}{\downarrow}}\rangle$ of strong interaction vertices are presented in Tables~\ref{tab:gA} and \ref{tab:gA2}.

\begin{table*}[htbp]
\centering
\caption{The spin-flavor matrix elements for PC amplitudes of $W$-exchange diagrams with $\mathcal{B}_{cc}^{+}=(\Xi_{cc}^{+},\Omega_{cc}^{+})$.}
\label{tab:polefactorPC}
\renewcommand\arraystretch{1.05}
\begin{tabular*}{110mm}{c@{\extracolsep{\fill}}cccc}
\toprule[1pt]
\toprule[0.5pt]
matrix elements &$\langle{\Xi_{c}^{+}}{\downarrow}\vert\hat{\mathcal{O}}\vert{\mathcal{B}_{cc}^{+}}{\downarrow}\rangle$  &$\langle{\Xi_{c}^{\prime+}}{\downarrow}\vert\hat{\mathcal{O}}\vert{\mathcal{B}_{cc}^{+}}{\downarrow}\rangle$  &$\langle{\Lambda_{c}^{+}}{\downarrow}\vert\hat{\mathcal{O}}\vert{\mathcal{B}_{cc}^{+}}{\downarrow}\rangle$  &$\langle{\Sigma_{c}^{+}}{\downarrow}\vert\hat{\mathcal{O}}\vert{\mathcal{B}_{cc}^{+}}{\downarrow}\rangle$\\
\midrule[0.5pt]
$\hat{\alpha}_{1}^{(-)}\hat{\beta}_{2}^{(+)}(1-\pmb{\sigma}_{1}\cdot\pmb{\sigma}_{2})$  &$\sqrt{\frac{2}{3}}$  &$0$  &$\sqrt{\frac{2}{3}}$  &$0$\\
$\hat{\alpha}_{1}^{(-)}\hat{\beta}_{2}^{(+)}(1-\tilde{\pmb{\sigma}}_{1}\cdot\tilde{\pmb{\sigma}}_{2})$  &$\sqrt{\frac{2}{3}}$  &$0$  &$\sqrt{\frac{2}{3}}$  &$0$\\
\bottomrule[0.5pt]
\bottomrule[1pt]
\end{tabular*}
\end{table*}

\begin{table*}[htbp]
\centering
\caption{The spin-flavor matrix elements for PV amplitudes of $W$-exchange diagrams, in which $\mathcal{B}_{cc}^{+}=(\Xi_{cc}^{+},\Omega_{cc}^{+})$, and the subscripts $\rho$ and $\lambda$ represent $\rho$-mode and $\lambda$-mode excitations respectively.}
\label{tab:polefactorPV}
\renewcommand\arraystretch{1.05}
\begin{tabular*}{160mm}{c@{\extracolsep{\fill}}cccccc}
\toprule[1pt]
\toprule[0.5pt]
matrix elements 
&$\langle{\Xi_{c}^{+}}{\downarrow}\vert\hat{\mathcal{O}}\vert{\mathcal{B}_{cc}^{+}}{\downarrow}\rangle_{{}^{2}\lambda}$  &$\langle{\Xi_{c}^{+}}{\downarrow}\vert\hat{\mathcal{O}}\vert{\mathcal{B}_{cc}^{+}}{\downarrow}\rangle_{{}^{2}\rho}$
&$\langle{\Xi_{c}^{+}}{\downarrow}\vert\hat{\mathcal{O}}\vert{\mathcal{B}_{cc}^{+}}{\downarrow}\rangle_{{}^{4}\lambda}$ &$\langle{\Xi_{c}^{\prime+}}{\downarrow}\vert\hat{\mathcal{O}}\vert{\mathcal{B}_{cc}^{+}}{\downarrow}\rangle_{{}^{2}\lambda}$  &$\langle{\Xi_{c}^{\prime+}}{\downarrow}\vert\hat{\mathcal{O}}\vert{\mathcal{B}_{cc}^{+}}{\downarrow}\rangle_{{}^{2}\rho}$
&$\langle{\Xi_{c}^{\prime+}}{\downarrow}\vert\hat{\mathcal{O}}\vert{\mathcal{B}_{cc}^{+}}{\downarrow}\rangle_{{}^{4}\lambda}$\\
$-\hat{\alpha}_{1}^{(-)}\hat{\beta}_{2}^{(+)}(\pmb{\sigma}_{1}-\pmb{\sigma}_{2})_{z}$  &$\frac{1}{3\sqrt{6}}$  &$-\frac{1}{3\sqrt{2}}$  &$-\frac{2}{3\sqrt{3}}$
&$\frac{1}{3\sqrt{2}}$  &$\frac{1}{3\sqrt{6}}$   &$0$\\
$\hat{\alpha}_{1}^{(-)}\hat{\beta}_{2}^{(+)}(\pmb{\sigma}_{1}\times\pmb{\sigma}_{2})_{z}$  &$-\frac{i}{3\sqrt{6}}$  &$\frac{i}{3\sqrt{2}}$  &$\frac{2i}{3\sqrt{3}}$
&$\frac{i}{3\sqrt{2}}$  &$\frac{i}{3\sqrt{6}}$   &$0$\\
$-\hat{\alpha}_{1}^{(-)}\hat{\beta}_{2}^{(+)}(\tilde{\pmb{\sigma}}_{1}-\tilde{\pmb{\sigma}}_{2})_{z}$  &$\frac{1}{3\sqrt{6}}$  &$-\frac{1}{3\sqrt{2}}$  &$-\frac{2}{3\sqrt{3}}$
&$\frac{1}{3\sqrt{2}}$  &$\frac{1}{3\sqrt{6}}$   &$0$\\
$\hat{\alpha}_{1}^{(-)}\hat{\beta}_{2}^{(+)}(\tilde{\pmb{\sigma}}_{1}\times\tilde{\pmb{\sigma}}_{2})_{z}$  &$-\frac{i}{3\sqrt{6}}$  &$\frac{i}{3\sqrt{2}}$  &$\frac{2i}{3\sqrt{3}}$
&$\frac{i}{3\sqrt{2}}$  &$\frac{i}{3\sqrt{6}}$   &$0$\\
\specialrule{0em}{0.5pt}{0.5pt}
\midrule[0.5pt]
\specialrule{0em}{0.5pt}{0.5pt}
matrix elements 
&$\langle{\Lambda_{c}^{+}}{\downarrow}\vert\hat{\mathcal{O}}\vert{\mathcal{B}_{cc}^{+}}{\downarrow}\rangle_{{}^{2}\lambda}$  &$\langle{\Lambda_{c}^{+}}{\downarrow}\vert\hat{\mathcal{O}}\vert{\mathcal{B}_{cc}^{+}}{\downarrow}\rangle_{{}^{2}\rho}$
&$\langle{\Lambda_{c}^{+}}{\downarrow}\vert\hat{\mathcal{O}}\vert{\mathcal{B}_{cc}^{+}}{\downarrow}\rangle_{{}^{4}\lambda}$ &$\langle{\Sigma_{c}^{+}}{\downarrow}\vert\hat{\mathcal{O}}\vert{\mathcal{B}_{cc}^{+}}{\downarrow}\rangle_{{}^{2}\lambda}$  &$\langle{\Sigma_{c}^{+}}{\downarrow}\vert\hat{\mathcal{O}}\vert{\mathcal{B}_{cc}^{+}}{\downarrow}\rangle_{{}^{2}\rho}$
&$\langle{\Sigma_{c}^{+}}{\downarrow}\vert\hat{\mathcal{O}}\vert{\mathcal{B}_{cc}^{+}}{\downarrow}\rangle_{{}^{4}\lambda}$\\
$-\hat{\alpha}_{1}^{(-)}\hat{\beta}_{2}^{(+)}(\pmb{\sigma}_{1}-\pmb{\sigma}_{2})_{z}$  &$\frac{1}{3\sqrt{6}}$  &$-\frac{1}{3\sqrt{2}}$  &$-\frac{2}{3\sqrt{3}}$
&$\frac{1}{3\sqrt{2}}$  &$\frac{1}{3\sqrt{6}}$   &$0$\\
$\hat{\alpha}_{1}^{(-)}\hat{\beta}_{2}^{(+)}(\pmb{\sigma}_{1}\times\pmb{\sigma}_{2})_{z}$  &$-\frac{i}{3\sqrt{6}}$  &$\frac{i}{3\sqrt{2}}$  &$\frac{2i}{3\sqrt{3}}$
&$\frac{i}{3\sqrt{2}}$  &$\frac{i}{3\sqrt{6}}$   &$0$\\
$-\hat{\alpha}_{1}^{(-)}\hat{\beta}_{2}^{(+)}(\tilde{\pmb{\sigma}}_{1}-\tilde{\pmb{\sigma}}_{2})_{z}$  &$\frac{1}{3\sqrt{6}}$  &$-\frac{1}{3\sqrt{2}}$  &$-\frac{2}{3\sqrt{3}}$
&$\frac{1}{3\sqrt{2}}$  &$\frac{1}{3\sqrt{6}}$   &$0$\\
$\hat{\alpha}_{1}^{(-)}\hat{\beta}_{2}^{(+)}(\tilde{\pmb{\sigma}}_{1}\times\tilde{\pmb{\sigma}}_{2})_{z}$  &$-\frac{i}{3\sqrt{6}}$  &$\frac{i}{3\sqrt{2}}$  &$\frac{2i}{3\sqrt{3}}$
&$\frac{i}{3\sqrt{2}}$  &$\frac{i}{3\sqrt{6}}$   &$0$\\
\specialrule{0em}{0.5pt}{0.5pt}
\midrule[0.5pt]
\specialrule{0em}{0.5pt}{0.5pt}
matrix elements 
&${}_{{}^{2}\lambda}\langle{\Xi_{c}^{+}}{\downarrow}\vert\hat{\mathcal{O}}\vert{\mathcal{B}_{cc}^{+}}{\downarrow}\rangle$  &${}_{{}^{2}\lambda}\langle{\Xi_{c}^{\prime+}}{\downarrow}\vert\hat{\mathcal{O}}\vert{\mathcal{B}_{cc}^{+}}{\downarrow}\rangle$
&${}_{{}^{4}\lambda}\langle{\Xi_{c}^{\prime+}}{\downarrow}\vert\hat{\mathcal{O}}\vert{\mathcal{B}_{cc}^{+}}{\downarrow}\rangle$   &${}_{{}^{2}\lambda}\langle{\Lambda_{c}^{+}}{\downarrow}\vert\hat{\mathcal{O}}\vert{\mathcal{B}_{cc}^{+}}{\downarrow}\rangle$  &${}_{{}^{2}\lambda}\langle{\Sigma_{c}^{+}}{\downarrow}\vert\hat{\mathcal{O}}\vert{\mathcal{B}_{cc}^{+}}{\downarrow}\rangle$
&${}_{{}^{4}\lambda}\langle{\Sigma_{c}^{+}}{\downarrow}\vert\hat{\mathcal{O}}\vert{\mathcal{B}_{cc}^{+}}{\downarrow}\rangle$\\
$-\hat{\alpha}_{1}^{(-)}\hat{\beta}_{2}^{(+)}(\pmb{\sigma}_{1}-\pmb{\sigma}_{2})_{z}$  &$\frac{1}{3\sqrt{6}}$  &$\frac{1}{3\sqrt{2}}$  &$\frac{1}{3}$
&$\frac{1}{3\sqrt{6}}$  &$\frac{1}{3\sqrt{2}}$   &$\frac{1}{3}$\\
$\hat{\alpha}_{1}^{(-)}\hat{\beta}_{2}^{(+)}(\pmb{\sigma}_{1}\times\pmb{\sigma}_{2})_{z}$  &$-\frac{i}{3\sqrt{6}}$  &$\frac{i}{3\sqrt{2}}$  &$\frac{i}{3}$
&$-\frac{i}{3\sqrt{6}}$  &$\frac{i}{3\sqrt{2}}$   &$\frac{i}{3}$\\
$-\hat{\alpha}_{1}^{(-)}\hat{\beta}_{2}^{(+)}(\tilde{\pmb{\sigma}}_{1}-\tilde{\pmb{\sigma}}_{2})_{z}$  &$\frac{1}{3\sqrt{6}}$  &$\frac{1}{3\sqrt{2}}$  &$\frac{1}{3}$
&$\frac{1}{3\sqrt{6}}$  &$\frac{1}{3\sqrt{2}}$   &$\frac{1}{3}$\\
$\hat{\alpha}_{1}^{(-)}\hat{\beta}_{2}^{(+)}(\tilde{\pmb{\sigma}}_{1}\times\tilde{\pmb{\sigma}}_{2})_{z}$  &$-\frac{i}{3\sqrt{6}}$  &$\frac{i}{3\sqrt{2}}$  &$\frac{i}{3}$
&$-\frac{i}{3\sqrt{6}}$  &$\frac{i}{3\sqrt{2}}$   &$\frac{i}{3}$\\
\bottomrule[0.5pt]
\bottomrule[1pt]
\end{tabular*}
\end{table*}

\begin{sidewaystable*}[htbp]
\centering
\caption{The values of spin-flavor matrix elements $\langle{\mathcal{B}^{\prime}{\downarrow}}\vert\hat{I}_{P}\sigma_{z}\vert{\mathcal{B}{\downarrow}}\rangle$.}
\label{tab:gA}
\renewcommand\arraystretch{1.05}
\begin{tabular*}{220mm}{c@{\extracolsep{\fill}}ccccccccc}
\toprule[1pt]
\toprule[0.5pt]
matrix elements   &values   &matrix elements   &values   &matrix elements   &values   &matrix elements   &values   &matrix elements   &values\\
\midrule[0.5pt]
$\langle{\Xi_{cc}^{+}{\downarrow}}\vert\hat{I}_{\bar{K}^{0}}\sigma_{z}\vert{\Omega_{cc}^{+}{\downarrow}}\rangle$  &$\frac{1}{3}$  
&${}_{{}^{2}\lambda}\langle{\Xi_{cc}^{+}{\downarrow}}\vert\hat{I}_{\bar{K}^{0}}\sigma_{z}\vert{\Omega_{cc}^{+}{\downarrow}}\rangle$  &$\frac{1}{3}$  
&${}_{{}^{2}\rho}\langle{\Xi_{cc}^{+}{\downarrow}}\vert\hat{I}_{\bar{K}^{0}}\sigma_{z}\vert{\Omega_{cc}^{+}{\downarrow}}\rangle$  &$0$  &${}_{{}^{4}\lambda}\langle{\Xi_{cc}^{+}{\downarrow}}\vert\hat{I}_{\bar{K}^{0}}\sigma_{z}\vert{\Omega_{cc}^{+}{\downarrow}}\rangle$  &$-\frac{2\sqrt{2}}{3}$   &&\\
$\langle{\Omega_{cc}^{+}{\downarrow}}\vert\hat{I}_{\eta}\sigma_{z}\vert{\Omega_{cc}^{+}{\downarrow}}\rangle$  &$-\frac{1}{3}\sin\zeta$  
&${}_{{}^{2}\lambda}\langle{\Omega_{cc}^{+}{\downarrow}}\vert\hat{I}_{\eta}\sigma_{z}\vert{\Omega_{cc}^{+}{\downarrow}}\rangle$  &$-\frac{1}{3}\sin\zeta$  
&${}_{{}^{2}\rho}\langle{\Omega_{cc}^{+}{\downarrow}}\vert\hat{I}_{\eta}\sigma_{z}\vert{\Omega_{cc}^{+}{\downarrow}}\rangle$  &$0$  &${}_{{}^{4}\lambda}\langle{\Omega_{cc}^{+}{\downarrow}}\vert\hat{I}_{\eta}\sigma_{z}\vert{\Omega_{cc}^{+}{\downarrow}}\rangle$  &$\frac{2\sqrt{2}}{3}\sin\zeta$   &&\\
$\langle{\Omega_{cc}^{+}{\downarrow}}\vert\hat{I}_{\eta^{\prime}}\sigma_{z}\vert{\Omega_{cc}^{+}{\downarrow}}\rangle$  &$\frac{1}{3}\cos\zeta$  
&${}_{{}^{2}\lambda}\langle{\Omega_{cc}^{+}{\downarrow}}\vert\hat{I}_{\eta^{\prime}}\sigma_{z}\vert{\Omega_{cc}^{+}{\downarrow}}\rangle$  &$\frac{1}{3}\cos\zeta$  
&${}_{{}^{2}\rho}\langle{\Omega_{cc}^{+}{\downarrow}}\vert\hat{I}_{\eta^{\prime}}\sigma_{z}\vert{\Omega_{cc}^{+}{\downarrow}}\rangle$  &$0$  &${}_{{}^{4}\lambda}\langle{\Omega_{cc}^{+}{\downarrow}}\vert\hat{I}_{\eta^{\prime}}\sigma_{z}\vert{\Omega_{cc}^{+}{\downarrow}}\rangle$  &$-\frac{2\sqrt{2}}{3}\cos\zeta$   &&\\
\specialrule{0em}{0.5pt}{0.5pt}
\midrule[0.5pt]
\specialrule{0em}{0.5pt}{0.5pt}
$\langle{\Omega_{c}^{0}{\downarrow}}\vert\hat{I}_{K^{+}}\sigma_{z}\vert{\Xi_{c}^{+}{\downarrow}}\rangle$  &$\sqrt{\frac{2}{3}}$  
&$\langle{\Omega_{c}^{0}{\downarrow}}\vert\hat{I}_{K^{+}}\sigma_{z}\vert{\Xi_{c}^{+}{\downarrow}}\rangle_{{}^{2}\lambda}$  &$\sqrt{\frac{2}{3}}$  
&$\langle{\Omega_{c}^{0}{\downarrow}}\vert\hat{I}_{K^{+}}\sigma_{z}\vert{\Xi_{c}^{\prime+}{\downarrow}}\rangle$  &$-\frac{2\sqrt{2}}{3}$  
&$\langle{\Omega_{c}^{0}{\downarrow}}\vert\hat{I}_{K^{+}}\sigma_{z}\vert{\Xi_{c}^{\prime+}{\downarrow}}\rangle_{{}^{2}\lambda}$  &$-\frac{2\sqrt{2}}{3}$   &$\langle{\Omega_{c}^{0}{\downarrow}}\vert\hat{I}_{K^{+}}\sigma_{z}\vert{\Xi_{c}^{\prime+}{\downarrow}}\rangle_{{}^{4}\lambda}$  &$\frac{2}{3}$\\
$\langle{\Xi_{c}^{0}{\downarrow}}\vert\hat{I}_{\pi^{+}}\sigma_{z}\vert{\Xi_{c}^{+}{\downarrow}}\rangle$  &$0$  
&$\langle{\Xi_{c}^{0}{\downarrow}}\vert\hat{I}_{\pi^{+}}\sigma_{z}\vert{\Xi_{c}^{+}{\downarrow}}\rangle_{{}^{2}\lambda}$  &$0$  
&$\langle{\Xi_{c}^{0}{\downarrow}}\vert\hat{I}_{\pi^{+}}\sigma_{z}\vert{\Xi_{c}^{\prime+}{\downarrow}}\rangle$  &$\frac{1}{\sqrt{3}}$  
&$\langle{\Xi_{c}^{0}{\downarrow}}\vert\hat{I}_{\pi^{+}}\sigma_{z}\vert{\Xi_{c}^{\prime+}{\downarrow}}\rangle_{{}^{2}\lambda}$  &$\frac{1}{\sqrt{3}}$   &$\langle{\Xi_{c}^{0}{\downarrow}}\vert\hat{I}_{\pi^{+}}\sigma_{z}\vert{\Xi_{c}^{\prime+}{\downarrow}}\rangle_{{}^{4}\lambda}$  &$\sqrt{\frac{2}{3}}$\\
$\langle{\Xi_{c}^{\prime0}{\downarrow}}\vert\hat{I}_{\pi^{+}}\sigma_{z}\vert{\Xi_{c}^{+}{\downarrow}}\rangle$  &$\frac{1}{\sqrt{3}}$  
&$\langle{\Xi_{c}^{\prime0}{\downarrow}}\vert\hat{I}_{\pi^{+}}\sigma_{z}\vert{\Xi_{c}^{+}{\downarrow}}\rangle_{{}^{2}\lambda}$  &$\frac{1}{\sqrt{3}}$  
&$\langle{\Xi_{c}^{\prime0}{\downarrow}}\vert\hat{I}_{\pi^{+}}\sigma_{z}\vert{\Xi_{c}^{\prime+}{\downarrow}}\rangle$  &$-\frac{2}{3}$  
&$\langle{\Xi_{c}^{\prime0}{\downarrow}}\vert\hat{I}_{\pi^{+}}\sigma_{z}\vert{\Xi_{c}^{\prime+}{\downarrow}}\rangle_{{}^{2}\lambda}$  &$-\frac{2}{3}$   &$\langle{\Xi_{c}^{\prime0}{\downarrow}}\vert\hat{I}_{\pi^{+}}\sigma_{z}\vert{\Xi_{c}^{\prime+}{\downarrow}}\rangle_{{}^{4}\lambda}$  &$\frac{\sqrt{2}}{3}$\\
$\langle{\Sigma_{c}^{++}{\downarrow}}\vert\hat{I}_{K^{-}}\sigma_{z}\vert{\Xi_{c}^{+}{\downarrow}}\rangle$  &$-\sqrt{\frac{2}{3}}$  
&$\langle{\Sigma_{c}^{++}{\downarrow}}\vert\hat{I}_{K^{-}}\sigma_{z}\vert{\Xi_{c}^{+}{\downarrow}}\rangle_{{}^{2}\lambda}$  &$-\sqrt{\frac{2}{3}}$  
&$\langle{\Sigma_{c}^{++}{\downarrow}}\vert\hat{I}_{K^{-}}\sigma_{z}\vert{\Xi_{c}^{\prime+}{\downarrow}}\rangle$  &$-\frac{2\sqrt{2}}{3}$  
&$\langle{\Sigma_{c}^{++}{\downarrow}}\vert\hat{I}_{K^{-}}\sigma_{z}\vert{\Xi_{c}^{\prime+}{\downarrow}}\rangle_{{}^{2}\lambda}$  &$-\frac{2\sqrt{2}}{3}$   &$\langle{\Sigma_{c}^{++}{\downarrow}}\vert\hat{I}_{K^{-}}\sigma_{z}\vert{\Xi_{c}^{\prime+}{\downarrow}}\rangle_{{}^{4}\lambda}$  &$\frac{2}{3}$\\
$\langle{\Xi_{c}^{+}{\downarrow}}\vert\hat{I}_{\pi^{0}}\sigma_{z}\vert{\Xi_{c}^{+}{\downarrow}}\rangle$  &$0$  
&$\langle{\Xi_{c}^{+}{\downarrow}}\vert\hat{I}_{\pi^{0}}\sigma_{z}\vert{\Xi_{c}^{+}{\downarrow}}\rangle_{{}^{2}\lambda}$  &$0$  
&$\langle{\Xi_{c}^{+}{\downarrow}}\vert\hat{I}_{\pi^{0}}\sigma_{z}\vert{\Xi_{c}^{\prime+}{\downarrow}}\rangle$  &$\frac{1}{\sqrt{6}}$  
&$\langle{\Xi_{c}^{+}{\downarrow}}\vert\hat{I}_{\pi^{0}}\sigma_{z}\vert{\Xi_{c}^{\prime+}{\downarrow}}\rangle_{{}^{2}\lambda}$  &$\frac{1}{\sqrt{6}}$   &$\langle{\Xi_{c}^{+}{\downarrow}}\vert\hat{I}_{\pi^{0}}\sigma_{z}\vert{\Xi_{c}^{\prime+}{\downarrow}}\rangle_{{}^{4}\lambda}$  &$\frac{1}{\sqrt{3}}$\\
$\langle{\Xi_{c}^{\prime+}{\downarrow}}\vert\hat{I}_{\pi^{0}}\sigma_{z}\vert{\Xi_{c}^{+}{\downarrow}}\rangle$  &$\frac{1}{\sqrt{6}}$  
&$\langle{\Xi_{c}^{\prime+}{\downarrow}}\vert\hat{I}_{\pi^{0}}\sigma_{z}\vert{\Xi_{c}^{+}{\downarrow}}\rangle_{{}^{2}\lambda}$  &$\frac{1}{\sqrt{6}}$  
&$\langle{\Xi_{c}^{\prime+}{\downarrow}}\vert\hat{I}_{\pi^{0}}\sigma_{z}\vert{\Xi_{c}^{\prime+}{\downarrow}}\rangle$  &$-\frac{\sqrt{2}}{3}$  
&$\langle{\Xi_{c}^{\prime+}{\downarrow}}\vert\hat{I}_{\pi^{0}}\sigma_{z}\vert{\Xi_{c}^{\prime+}{\downarrow}}\rangle_{{}^{2}\lambda}$  &$-\frac{\sqrt{2}}{3}$   &$\langle{\Xi_{c}^{\prime+}{\downarrow}}\vert\hat{I}_{\pi^{0}}\sigma_{z}\vert{\Xi_{c}^{\prime+}{\downarrow}}\rangle_{{}^{4}\lambda}$  &$\frac{1}{3}$\\
$\langle{\Lambda_{c}^{+}{\downarrow}}\vert\hat{I}_{\bar{K}^{0}}\sigma_{z}\vert{\Xi_{c}^{+}{\downarrow}}\rangle$  &$0$  
&$\langle{\Lambda_{c}^{+}{\downarrow}}\vert\hat{I}_{\bar{K}^{0}}\sigma_{z}\vert{\Xi_{c}^{+}{\downarrow}}\rangle_{{}^{2}\lambda}$  &$0$  
&$\langle{\Lambda_{c}^{+}{\downarrow}}\vert\hat{I}_{\bar{K}^{0}}\sigma_{z}\vert{\Xi_{c}^{\prime+}{\downarrow}}\rangle$  &$-\frac{1}{\sqrt{3}}$  
&$\langle{\Lambda_{c}^{+}{\downarrow}}\vert\hat{I}_{\bar{K}^{0}}\sigma_{z}\vert{\Xi_{c}^{\prime+}{\downarrow}}\rangle_{{}^{2}\lambda}$  &$-\frac{1}{\sqrt{3}}$   &$\langle{\Lambda_{c}^{+}{\downarrow}}\vert\hat{I}_{\bar{K}^{0}}\sigma_{z}\vert{\Xi_{c}^{\prime+}{\downarrow}}\rangle_{{}^{4}\lambda}$  &$-\sqrt{\frac{2}{3}}$\\
$\langle{\Sigma_{c}^{+}{\downarrow}}\vert\hat{I}_{\bar{K}^{0}}\sigma_{z}\vert{\Xi_{c}^{+}{\downarrow}}\rangle$  &$-\frac{1}{\sqrt{3}}$  
&$\langle{\Sigma_{c}^{+}{\downarrow}}\vert\hat{I}_{\bar{K}^{0}}\sigma_{z}\vert{\Xi_{c}^{+}{\downarrow}}\rangle_{{}^{2}\lambda}$  &$-\frac{1}{\sqrt{3}}$  
&$\langle{\Sigma_{c}^{+}{\downarrow}}\vert\hat{I}_{\bar{K}^{0}}\sigma_{z}\vert{\Xi_{c}^{\prime+}{\downarrow}}\rangle$  &$-\frac{2}{3}$  
&$\langle{\Sigma_{c}^{+}{\downarrow}}\vert\hat{I}_{\bar{K}^{0}}\sigma_{z}\vert{\Xi_{c}^{\prime+}{\downarrow}}\rangle_{{}^{2}\lambda}$  &$-\frac{2}{3}$   &$\langle{\Sigma_{c}^{+}{\downarrow}}\vert\hat{I}_{\bar{K}^{0}}\sigma_{z}\vert{\Xi_{c}^{\prime+}{\downarrow}}\rangle_{{}^{4}\lambda}$  &$\frac{\sqrt{2}}{3}$\\
\specialrule{0em}{0.5pt}{0.5pt}
\midrule[0.5pt]
\specialrule{0em}{0.5pt}{0.5pt}
$\langle{\Sigma_{c}^{++}{\downarrow}}\vert\hat{I}_{\pi^{-}}\sigma_{z}\vert{\Lambda_{c}^{+}{\downarrow}}\rangle$  &$-\sqrt{\frac{2}{3}}$  
&$\langle{\Sigma_{c}^{++}{\downarrow}}\vert\hat{I}_{\pi^{-}}\sigma_{z}\vert{\Lambda_{c}^{+}{\downarrow}}\rangle_{{}^{2}\lambda}$  &$-\sqrt{\frac{2}{3}}$  
&$\langle{\Sigma_{c}^{++}{\downarrow}}\vert\hat{I}_{\pi^{-}}\sigma_{z}\vert{\Sigma_{c}^{+}{\downarrow}}\rangle$  &$-\frac{2\sqrt{2}}{3}$  
&$\langle{\Sigma_{c}^{++}{\downarrow}}\vert\hat{I}_{\pi^{-}}\sigma_{z}\vert{\Sigma_{c}^{+}{\downarrow}}\rangle_{{}^{2}\lambda}$  &$-\frac{2\sqrt{2}}{3}$   &$\langle{\Sigma_{c}^{++}{\downarrow}}\vert\hat{I}_{\pi^{-}}\sigma_{z}\vert{\Sigma_{c}^{+}{\downarrow}}\rangle_{{}^{4}\lambda}$  &$\frac{2}{3}$\\
$\langle{\Sigma_{c}^{0}{\downarrow}}\vert\hat{I}_{\pi^{+}}\sigma_{z}\vert{\Lambda_{c}^{+}{\downarrow}}\rangle$  &$\sqrt{\frac{2}{3}}$  
&$\langle{\Sigma_{c}^{0}{\downarrow}}\vert\hat{I}_{\pi^{+}}\sigma_{z}\vert{\Lambda_{c}^{+}{\downarrow}}\rangle_{{}^{2}\lambda}$  &$\sqrt{\frac{2}{3}}$  
&$\langle{\Sigma_{c}^{0}{\downarrow}}\vert\hat{I}_{\pi^{+}}\sigma_{z}\vert{\Sigma_{c}^{+}{\downarrow}}\rangle$  &$-\frac{2\sqrt{2}}{3}$  
&$\langle{\Sigma_{c}^{0}{\downarrow}}\vert\hat{I}_{\pi^{+}}\sigma_{z}\vert{\Sigma_{c}^{+}{\downarrow}}\rangle_{{}^{2}\lambda}$  &$-\frac{2\sqrt{2}}{3}$   &$\langle{\Sigma_{c}^{0}{\downarrow}}\vert\hat{I}_{\pi^{+}}\sigma_{z}\vert{\Sigma_{c}^{+}{\downarrow}}\rangle_{{}^{4}\lambda}$  &$\frac{2}{3}$\\
$\langle{\Xi_{c}^{0}{\downarrow}}\vert\hat{I}_{K^{+}}\sigma_{z}\vert{\Lambda_{c}^{+}{\downarrow}}\rangle$  &$0$  
&$\langle{\Xi_{c}^{0}{\downarrow}}\vert\hat{I}_{K^{+}}\sigma_{z}\vert{\Lambda_{c}^{+}{\downarrow}}\rangle_{{}^{2}\lambda}$  &$0$  
&$\langle{\Xi_{c}^{0}{\downarrow}}\vert\hat{I}_{K^{+}}\sigma_{z}\vert{\Sigma_{c}^{+}{\downarrow}}\rangle$  &$-\frac{1}{\sqrt{3}}$  
&$\langle{\Xi_{c}^{0}{\downarrow}}\vert\hat{I}_{K^{+}}\sigma_{z}\vert{\Sigma_{c}^{+}{\downarrow}}\rangle_{{}^{2}\lambda}$  &$-\frac{1}{\sqrt{3}}$   &$\langle{\Xi_{c}^{0}{\downarrow}}\vert\hat{I}_{K^{+}}\sigma_{z}\vert{\Sigma_{c}^{+}{\downarrow}}\rangle_{{}^{4}\lambda}$  &$-\sqrt{\frac{2}{3}}$\\
$\langle{\Xi_{c}^{\prime0}{\downarrow}}\vert\hat{I}_{K^{+}}\sigma_{z}\vert{\Lambda_{c}^{+}{\downarrow}}\rangle$  &$\frac{1}{\sqrt{3}}$  
&$\langle{\Xi_{c}^{\prime0}{\downarrow}}\vert\hat{I}_{K^{+}}\sigma_{z}\vert{\Lambda_{c}^{+}{\downarrow}}\rangle_{{}^{2}\lambda}$  &$\frac{1}{\sqrt{3}}$  
&$\langle{\Xi_{c}^{\prime0}{\downarrow}}\vert\hat{I}_{K^{+}}\sigma_{z}\vert{\Sigma_{c}^{+}{\downarrow}}\rangle$  &$-\frac{2}{3}$  
&$\langle{\Xi_{c}^{\prime0}{\downarrow}}\vert\hat{I}_{K^{+}}\sigma_{z}\vert{\Sigma_{c}^{+}{\downarrow}}\rangle_{{}^{2}\lambda}$  &$-\frac{2}{3}$   &$\langle{\Xi_{c}^{\prime0}{\downarrow}}\vert\hat{I}_{K^{+}}\sigma_{z}\vert{\Sigma_{c}^{+}{\downarrow}}\rangle_{{}^{4}\lambda}$  &$\frac{\sqrt{2}}{3}$\\
$\langle{\Lambda_{c}^{+}{\downarrow}}\vert\hat{I}_{\pi^{0}}\sigma_{z}\vert{\Lambda_{c}^{+}{\downarrow}}\rangle$  &$0$  
&$\langle{\Lambda_{c}^{+}{\downarrow}}\vert\hat{I}_{\pi^{0}}\sigma_{z}\vert{\Lambda_{c}^{+}{\downarrow}}\rangle_{{}^{2}\lambda}$  &$0$  
&$\langle{\Lambda_{c}^{+}{\downarrow}}\vert\hat{I}_{\pi^{0}}\sigma_{z}\vert{\Sigma_{c}^{+}{\downarrow}}\rangle$  &$\sqrt{\frac{2}{3}}$  
&$\langle{\Lambda_{c}^{+}{\downarrow}}\vert\hat{I}_{\pi^{0}}\sigma_{z}\vert{\Sigma_{c}^{+}{\downarrow}}\rangle_{{}^{2}\lambda}$  &$\sqrt{\frac{2}{3}}$   &$\langle{\Lambda_{c}^{+}{\downarrow}}\vert\hat{I}_{\pi^{0}}\sigma_{z}\vert{\Sigma_{c}^{+}{\downarrow}}\rangle_{{}^{4}\lambda}$  &$\frac{2}{\sqrt{3}}$\\
$\langle{\Sigma_{c}^{+}{\downarrow}}\vert\hat{I}_{\pi^{0}}\sigma_{z}\vert{\Lambda_{c}^{+}{\downarrow}}\rangle$  &$\sqrt{\frac{2}{3}}$  
&$\langle{\Sigma_{c}^{+}{\downarrow}}\vert\hat{I}_{\pi^{0}}\sigma_{z}\vert{\Lambda_{c}^{+}{\downarrow}}\rangle_{{}^{2}\lambda}$  &$\sqrt{\frac{2}{3}}$  
&$\langle{\Sigma_{c}^{+}{\downarrow}}\vert\hat{I}_{\pi^{0}}\sigma_{z}\vert{\Sigma_{c}^{+}{\downarrow}}\rangle$  &$0$  
&$\langle{\Sigma_{c}^{+}{\downarrow}}\vert\hat{I}_{\pi^{0}}\sigma_{z}\vert{\Sigma_{c}^{+}{\downarrow}}\rangle_{{}^{2}\lambda}$  &$0$   &$\langle{\Sigma_{c}^{+}{\downarrow}}\vert\hat{I}_{\pi^{0}}\sigma_{z}\vert{\Sigma_{c}^{+}{\downarrow}}\rangle_{{}^{4}\lambda}$  &$0$\\
$\langle{\Xi_{c}^{+}{\downarrow}}\vert\hat{I}_{K^{0}}\sigma_{z}\vert{\Lambda_{c}^{+}{\downarrow}}\rangle$  &$0$  
&$\langle{\Xi_{c}^{+}{\downarrow}}\vert\hat{I}_{K^{0}}\sigma_{z}\vert{\Lambda_{c}^{+}{\downarrow}}\rangle_{{}^{2}\lambda}$  &$0$  
&$\langle{\Xi_{c}^{+}{\downarrow}}\vert\hat{I}_{K^{0}}\sigma_{z}\vert{\Sigma_{c}^{+}{\downarrow}}\rangle$  &$-\frac{1}{\sqrt{3}}$  
&$\langle{\Xi_{c}^{+}{\downarrow}}\vert\hat{I}_{K^{0}}\sigma_{z}\vert{\Sigma_{c}^{+}{\downarrow}}\rangle_{{}^{2}\lambda}$  &$-\frac{1}{\sqrt{3}}$   &$\langle{\Xi_{c}^{+}{\downarrow}}\vert\hat{I}_{K^{0}}\sigma_{z}\vert{\Sigma_{c}^{+}{\downarrow}}\rangle_{{}^{4}\lambda}$  &$-\sqrt{\frac{2}{3}}$\\
$\langle{\Xi_{c}^{\prime+}{\downarrow}}\vert\hat{I}_{K^{0}}\sigma_{z}\vert{\Lambda_{c}^{+}{\downarrow}}\rangle$  &$-\frac{1}{\sqrt{3}}$  
&$\langle{\Xi_{c}^{\prime+}{\downarrow}}\vert\hat{I}_{K^{0}}\sigma_{z}\vert{\Lambda_{c}^{+}{\downarrow}}\rangle_{{}^{2}\lambda}$  &$-\frac{1}{\sqrt{3}}$  
&$\langle{\Xi_{c}^{\prime+}{\downarrow}}\vert\hat{I}_{K^{0}}\sigma_{z}\vert{\Sigma_{c}^{+}{\downarrow}}\rangle$  &$-\frac{2}{3}$  
&$\langle{\Xi_{c}^{\prime+}{\downarrow}}\vert\hat{I}_{K^{0}}\sigma_{z}\vert{\Sigma_{c}^{+}{\downarrow}}\rangle_{{}^{2}\lambda}$  &$-\frac{2}{3}$   &$\langle{\Xi_{c}^{\prime+}{\downarrow}}\vert\hat{I}_{K^{0}}\sigma_{z}\vert{\Sigma_{c}^{+}{\downarrow}}\rangle_{{}^{4}\lambda}$  &$\frac{\sqrt{2}}{3}$\\
$\langle{\Lambda_{c}^{+}{\downarrow}}\vert\hat{I}_{\eta}\sigma_{z}\vert{\Lambda_{c}^{+}{\downarrow}}\rangle$  &$0$  
&$\langle{\Lambda_{c}^{+}{\downarrow}}\vert\hat{I}_{\eta}\sigma_{z}\vert{\Lambda_{c}^{+}{\downarrow}}\rangle_{{}^{2}\lambda}$  &$0$  
&$\langle{\Lambda_{c}^{+}{\downarrow}}\vert\hat{I}_{\eta}\sigma_{z}\vert{\Sigma_{c}^{+}{\downarrow}}\rangle$  &$0$  
&$\langle{\Lambda_{c}^{+}{\downarrow}}\vert\hat{I}_{\eta}\sigma_{z}\vert{\Sigma_{c}^{+}{\downarrow}}\rangle_{{}^{2}\lambda}$  &$0$   &$\langle{\Lambda_{c}^{+}{\downarrow}}\vert\hat{I}_{\eta}\sigma_{z}\vert{\Sigma_{c}^{+}{\downarrow}}\rangle_{{}^{4}\lambda}$  &$0$\\
$\langle{\Lambda_{c}^{+}{\downarrow}}\vert\hat{I}_{\eta^{\prime}}\sigma_{z}\vert{\Lambda_{c}^{+}{\downarrow}}\rangle$  &$0$  
&$\langle{\Lambda_{c}^{+}{\downarrow}}\vert\hat{I}_{\eta^{\prime}}\sigma_{z}\vert{\Lambda_{c}^{+}{\downarrow}}\rangle_{{}^{2}\lambda}$  &$0$  
&$\langle{\Lambda_{c}^{+}{\downarrow}}\vert\hat{I}_{\eta^{\prime}}\sigma_{z}\vert{\Sigma_{c}^{+}{\downarrow}}\rangle$  &$0$  
&$\langle{\Lambda_{c}^{+}{\downarrow}}\vert\hat{I}_{\eta^{\prime}}\sigma_{z}\vert{\Sigma_{c}^{+}{\downarrow}}\rangle_{{}^{2}\lambda}$  &$0$   &$\langle{\Lambda_{c}^{+}{\downarrow}}\vert\hat{I}_{\eta^{\prime}}\sigma_{z}\vert{\Sigma_{c}^{+}{\downarrow}}\rangle_{{}^{4}\lambda}$  &$0$\\
$\langle{\Sigma_{c}^{+}{\downarrow}}\vert\hat{I}_{\eta}\sigma_{z}\vert{\Lambda_{c}^{+}{\downarrow}}\rangle$  &$0$  
&$\langle{\Sigma_{c}^{+}{\downarrow}}\vert\hat{I}_{\eta}\sigma_{z}\vert{\Lambda_{c}^{+}{\downarrow}}\rangle_{{}^{2}\lambda}$  &$0$  
&$\langle{\Sigma_{c}^{+}{\downarrow}}\vert\hat{I}_{\eta}\sigma_{z}\vert{\Sigma_{c}^{+}{\downarrow}}\rangle$  &$-\frac{2\sqrt{2}}{3}\cos\zeta$  
&$\langle{\Sigma_{c}^{+}{\downarrow}}\vert\hat{I}_{\eta}\sigma_{z}\vert{\Sigma_{c}^{+}{\downarrow}}\rangle_{{}^{2}\lambda}$  &$-\frac{2\sqrt{2}}{3}\cos\zeta$   &$\langle{\Sigma_{c}^{+}{\downarrow}}\vert\hat{I}_{\eta}\sigma_{z}\vert{\Sigma_{c}^{+}{\downarrow}}\rangle_{{}^{4}\lambda}$  &$\frac{2}{3}\cos\zeta$\\
$\langle{\Sigma_{c}^{+}{\downarrow}}\vert\hat{I}_{\eta^{\prime}}\sigma_{z}\vert{\Lambda_{c}^{+}{\downarrow}}\rangle$  &$0$  
&$\langle{\Sigma_{c}^{+}{\downarrow}}\vert\hat{I}_{\eta^{\prime}}\sigma_{z}\vert{\Lambda_{c}^{+}{\downarrow}}\rangle_{{}^{2}\lambda}$  &$0$  
&$\langle{\Sigma_{c}^{+}{\downarrow}}\vert\hat{I}_{\eta^{\prime}}\sigma_{z}\vert{\Sigma_{c}^{+}{\downarrow}}\rangle$  &$-\frac{2\sqrt{2}}{3}\sin\zeta$  
&$\langle{\Sigma_{c}^{+}{\downarrow}}\vert\hat{I}_{\eta^{\prime}}\sigma_{z}\vert{\Sigma_{c}^{+}{\downarrow}}\rangle_{{}^{2}\lambda}$  &$-\frac{2\sqrt{2}}{3}\sin\zeta$   &$\langle{\Sigma_{c}^{+}{\downarrow}}\vert\hat{I}_{\eta^{\prime}}\sigma_{z}\vert{\Sigma_{c}^{+}{\downarrow}}\rangle_{{}^{4}\lambda}$  &$\frac{2}{3}\sin\zeta$\\
\bottomrule[0.5pt]
\bottomrule[1pt]
\end{tabular*}
\end{sidewaystable*}

\begin{table*}[htbp]
\centering
\caption{Same as Table~\ref{tab:gA} but for $\Omega_{cc}^{+}\to\Xi_{c}^{(\prime)+}\eta^{(\prime)}$ decays, where the first value in braces is corresponding to $E_{1}$ diagram while the second one is corresponding to $E_{2}$ diagram.}
\label{tab:gA2}
\renewcommand\arraystretch{1.05}
\begin{tabular*}{160mm}{c@{\extracolsep{\fill}}ccccc}
\toprule[1pt]
\toprule[0.5pt]
matrix elements   &values   &matrix elements   &values   &matrix elements   &values\\
\midrule[0.5pt]
$\langle{\Xi_{c}^{+}{\downarrow}}\vert\hat{I}_{\eta}\sigma_{z}\vert{\Xi_{c}^{+}{\downarrow}}\rangle$  &$\{0,0\}$  
&$\langle{\Xi_{c}^{+}{\downarrow}}\vert\hat{I}_{\eta}\sigma_{z}\vert{\Xi_{c}^{+}{\downarrow}}\rangle_{{}^{2}\lambda}$  &$\{0,0\}$  &&\\  
$\langle{\Xi_{c}^{+}{\downarrow}}\vert\hat{I}_{\eta}\sigma_{z}\vert{\Xi_{c}^{\prime+}{\downarrow}}\rangle$  &$\{\frac{1}{\sqrt{6}}\cos\zeta,\frac{1}{\sqrt{3}}\sin\zeta\}$  
&$\langle{\Xi_{c}^{+}{\downarrow}}\vert\hat{I}_{\eta}\sigma_{z}\vert{\Xi_{c}^{\prime+}{\downarrow}}\rangle_{{}^{2}\lambda}$  &$\{\frac{1}{\sqrt{6}}\cos\zeta,\frac{1}{\sqrt{3}}\sin\zeta\}$  &$\langle{\Xi_{c}^{+}{\downarrow}}\vert\hat{I}_{\eta}\sigma_{z}\vert{\Xi_{c}^{\prime+}{\downarrow}}\rangle_{{}^{4}\lambda}$  &$\{\frac{1}{\sqrt{3}}\cos\zeta,\sqrt{\frac{2}{3}}\sin\zeta\}$\\
$\langle{\Xi_{c}^{\prime+}{\downarrow}}\vert\hat{I}_{\eta}\sigma_{z}\vert{\Xi_{c}^{+}{\downarrow}}\rangle$  &$\{\frac{1}{\sqrt{6}}\cos\zeta,\frac{1}{\sqrt{3}}\sin\zeta\}$  
&$\langle{\Xi_{c}^{\prime+}{\downarrow}}\vert\hat{I}_{\eta}\sigma_{z}\vert{\Xi_{c}^{+}{\downarrow}}\rangle_{{}^{2}\lambda}$  &$\{\frac{1}{\sqrt{6}}\cos\zeta,\frac{1}{\sqrt{3}}\sin\zeta\}$  &&\\  
$\langle{\Xi_{c}^{\prime+}{\downarrow}}\vert\hat{I}_{\eta}\sigma_{z}\vert{\Xi_{c}^{\prime+}{\downarrow}}\rangle$  &$\{-\frac{\sqrt{2}}{3}\cos\zeta,\frac{2}{3}\sin\zeta\}$  
&$\langle{\Xi_{c}^{\prime+}{\downarrow}}\vert\hat{I}_{\eta}\sigma_{z}\vert{\Xi_{c}^{\prime+}{\downarrow}}\rangle_{{}^{2}\lambda}$  &$\{-\frac{\sqrt{2}}{3}\cos\zeta,\frac{2}{3}\sin\zeta\}$  &$\langle{\Xi_{c}^{\prime+}{\downarrow}}\vert\hat{I}_{\eta}\sigma_{z}\vert{\Xi_{c}^{\prime+}{\downarrow}}\rangle_{{}^{4}\lambda}$  &$\{\frac{1}{3}\cos\zeta,-\frac{\sqrt{2}}{3}\sin\zeta\}$\\
\midrule[0.5pt]
$\langle{\Xi_{c}^{+}{\downarrow}}\vert\hat{I}_{\eta^{\prime}}\sigma_{z}\vert{\Xi_{c}^{+}{\downarrow}}\rangle$  &$\{0,0\}$  
&$\langle{\Xi_{c}^{+}{\downarrow}}\vert\hat{I}_{\eta^{\prime}}\sigma_{z}\vert{\Xi_{c}^{+}{\downarrow}}\rangle_{{}^{2}\lambda}$  &$\{0,0\}$  &&\\  
$\langle{\Xi_{c}^{+}{\downarrow}}\vert\hat{I}_{\eta^{\prime}}\sigma_{z}\vert{\Xi_{c}^{\prime+}{\downarrow}}\rangle$  &$\{\frac{1}{\sqrt{6}}\sin\zeta,-\frac{1}{\sqrt{3}}\cos\zeta\}$  
&$\langle{\Xi_{c}^{+}{\downarrow}}\vert\hat{I}_{\eta^{\prime}}\sigma_{z}\vert{\Xi_{c}^{\prime+}{\downarrow}}\rangle_{{}^{2}\lambda}$  &$\{\frac{1}{\sqrt{6}}\sin\zeta,-\frac{1}{\sqrt{3}}\cos\zeta\}$  &$\langle{\Xi_{c}^{+}{\downarrow}}\vert\hat{I}_{\eta^{\prime}}\sigma_{z}\vert{\Xi_{c}^{\prime+}{\downarrow}}\rangle_{{}^{4}\lambda}$  &$\{\frac{1}{\sqrt{3}}\sin\zeta,-\sqrt{\frac{2}{3}}\cos\zeta\}$\\
$\langle{\Xi_{c}^{\prime+}{\downarrow}}\vert\hat{I}_{\eta^{\prime}}\sigma_{z}\vert{\Xi_{c}^{+}{\downarrow}}\rangle$  &$\{\frac{1}{\sqrt{6}}\sin\zeta,-\frac{1}{\sqrt{3}}\cos\zeta\}$  
&$\langle{\Xi_{c}^{\prime+}{\downarrow}}\vert\hat{I}_{\eta^{\prime}}\sigma_{z}\vert{\Xi_{c}^{+}{\downarrow}}\rangle_{{}^{2}\lambda}$  &$\{\frac{1}{\sqrt{6}}\sin\zeta,-\frac{1}{\sqrt{3}}\cos\zeta\}$  &&\\  
$\langle{\Xi_{c}^{\prime+}{\downarrow}}\vert\hat{I}_{\eta^{\prime}}\sigma_{z}\vert{\Xi_{c}^{\prime+}{\downarrow}}\rangle$  &$\{-\frac{\sqrt{2}}{3}\sin\zeta,-\frac{2}{3}\cos\zeta\}$  
&$\langle{\Xi_{c}^{\prime+}{\downarrow}}\vert\hat{I}_{\eta^{\prime}}\sigma_{z}\vert{\Xi_{c}^{\prime+}{\downarrow}}\rangle_{{}^{2}\lambda}$  &$\{-\frac{\sqrt{2}}{3}\sin\zeta,-\frac{2}{3}\cos\zeta\}$  &$\langle{\Xi_{c}^{\prime+}{\downarrow}}\vert\hat{I}_{\eta^{\prime}}\sigma_{z}\vert{\Xi_{c}^{\prime+}{\downarrow}}\rangle_{{}^{4}\lambda}$  &$\{\frac{1}{3}\sin\zeta,\frac{\sqrt{2}}{3}\cos\zeta\}$\\
\bottomrule[0.5pt]
\bottomrule[1pt]
\end{tabular*}
\end{table*}

\end{widetext}

\end{document}